\documentclass[nobibnotes,superscriptaddress,nofootinbib,onecolumn,article]{revtex4-2}

\usepackage{amsmath,amssymb}
\usepackage{graphicx,float}
\usepackage{bm}
\usepackage{bbold}
\usepackage{amsmath}
\usepackage{slashed}

\newcommand{\dmatm}{\Delta m^2_{31}}
\newcommand{\dmsol}{\Delta m^2_{21}}
\newcommand{\Ml}{\mathbf{M}_{\ell}}
\newcommand{\M}{\mathbf{M}}
\newcommand{\U}{\mathbf{U}}
\newcommand{\V}{\mathbf{V}}
\newcommand{\B}{\mathbf{B}}
\newcommand{\Y}{\mathbf{Y}}

\newcommand{\BR}{{\rm BR}}
\newcommand{\CR}{{\rm CR}}
\newcommand{\GeV}{{\rm GeV}}
\newcommand{\TeV}{{\rm TeV}}

\begin{document}
\title{Minimal inverse-seesaw mechanism with Abelian flavour symmetries}
\author{H. B. C\^amara}
\affiliation{Departamento de F\'{\i}sica and CFTP, Instituto Superior T\'ecnico, Universidade de Lisboa, 1049-001 Lisboa, Portugal}
\author{R. G. Felipe}
\affiliation{ISEL - Instituto Superior de Engenharia de Lisboa, Instituto Polit\'ecnico de Lisboa, Rua Conselheiro Em\'{\i}dio Navarro, 1959-007 Lisboa, Portugal}
\affiliation{Departamento de F\'{\i}sica and CFTP, Instituto Superior T\'ecnico, Universidade de Lisboa, 1049-001 Lisboa, Portugal}
\author{F. R. Joaquim}
\affiliation{Departamento de F\'{\i}sica and CFTP, Instituto Superior T\'ecnico, Universidade de Lisboa, 1049-001 Lisboa, Portugal}
\date{\today}

\begin{abstract}

We study the phenomenology of the minimal $(2,2)$ inverse-seesaw model supplemented with Abelian flavour symmetries. To ensure maximal predictability, we establish the most restrictive flavour patterns which can be realised by those symmetries. This setup requires adding an extra scalar doublet and two complex scalar singlets to the Standard Model, paving the way to implement spontaneous CP violation. It is shown that such CP-violating effects can be successfully communicated to the lepton sector through couplings of the scalar singlets to the new sterile fermions. The Majorana and Dirac CP phases turn out to be related, and the active-sterile neutrino mixing is determined by the active neutrino masses, mixing angles and CP phases. We investigate the constraints imposed on the model by the current experimental limits on lepton flavour-violating decays, especially those on the branching ratio $\BR(\mu\rightarrow e \gamma)$ and the capture rate $\CR(\mu-e,{\rm Au})$. The prospects to further test the framework put forward in this work are also discussed in view of the projected sensitivities of future experimental searches sensitive to the presence of heavy sterile neutrinos. Namely, we investigate at which extent upcoming searches for $\mu\rightarrow e \gamma$, $\mu \rightarrow 3e$ and $\mu-e$ conversion in nuclei will be able to test our model, and how complementary will future high-energy collider and beam-dump experiments be in that task. 
\end{abstract}
\maketitle

\section{Introduction}

The discovery of neutrino oscillations has not only established that neutrinos are massive particles, but also opened the way to a whole new domain of searches for physics beyond the Standard Model (SM). Although such a milestone represents a major achievement in particle physics, our current understanding of neutrino properties leaves unanswered many questions concerning the fundamental properties of those particles, namely the ones related to the origin of their masses. From a theoretical perspective, the seesaw mechanism~\cite{Minkowski:1977sc,GellMann:1980vs,Yanagida:1979as,Schechter:1980gr,Glashow:1979nm,Mohapatra:1979ia} offers an elegant framework for the explanation of neutrino masses and lepton mixing in extensions of the SM. As opposed to the canonical type-I seesaw, where very heavy ``right-handed'' (RH) neutrinos or tiny Yukawa couplings are required to generate small neutrino masses, in the so-called inverse seesaw (ISS)~\cite{Mohapatra:1986aw,PhysRevD.34.1642,GonzalezGarcia:1988rw} neutrino mass suppression is triggered by small lepton-number violating (LNV) mass parameters. In this case, the lightness of neutrinos stems from an approximate lepton-number symmetry which is restored when those parameters are set to zero. Therefore, the ISS provides a natural neutrino-mass generation mechanism in the t'Hooft sense~\cite{tHooft:1980xss}.

A crucial feature of the ISS (not shared by the canonical type I seesaw) is that small Majorana neutrino masses can be generated with RH neutrino masses at the TeV scale (or below) and $\mathcal{O}(1)$ Yukawa coupling parameters. As a result, the mixing between the (active) light neutrinos and the new (sterile) states can be sizeable for sterile neutrino masses lying not far from the electroweak scale. The presence of new neutral fermions interacting with SM leptons and gauge bosons motivates phenomenological studies beyond the SM, making the ISS a perfect theoretical framework to guide new physics probes. In particular, experimental searches for charged lepton flavour violating (cLFV) processes like $\mu\rightarrow e \gamma$~\cite{TheMEG:2016wtm,Baldini:2018nnn}, $\mu\rightarrow e e e$~\cite{Bellgardt:1987du} and $\mu-e$ conversion in heavy nuclei~\cite{Dohmen:1993mp,Honecker:1996zf,Bertl:2006up,Kuno:2013mha,Natori:2014yba} have been studied in the ISS framework~\cite{Deppisch:2005zm,Abada:2014vea,Arganda:2014dta,Abada:2015oba,Abada:2014cca,DeRomeri:2015ipa,Arganda:2015ija,Abada:2018nio} with the purpose of understanding at which extent our current knowledge on those processes is able to constrain the ISS parameter space. Depending on their masses and mixing with the SM degrees of freedom, sterile neutrinos may also lead to interesting signals potentially observable at the Large Hadron Collider (LHC), as well as at other experiments sensitive to new physics effects induced by the presence of those particles~\cite{delAguila:2008cj,Das:2012ze,Deppisch:2015qwa,Antusch:2016ejd,Caputo:2017pit,Bhardwaj:2018lma,Bolton:2019pcu}.

Global analyses of neutrino oscillation data in a three-neutrino mixing scheme~\cite{Capozzi:2020qhw,deSalas:2020pgw,Esteban:2020cvm} imply the existence of at least two massive active neutrinos. Yet, controversial results from current neutrino oscillation experiments like LSND~\cite{Aguilar:2001ty} and MiniBooNE~\cite{Aguilar-Arevalo:2018gpe} may be hinting at the existence of sterile neutrinos with masses in the eV range. To accommodate all data, more general active-sterile neutrino mixing schemes would then be required. This inspired several analyses of oscillation data beyond the three-neutrino paradigm (see, for instance, Refs.~\cite{Abazajian:2012ys,Kopp:2013vaa,Dentler:2018sju,Boser:2019rta,Hagstotz:2020ukm}). As turns out, it is possible to construct several ISS low-scale models that are compatible with neutrino oscillation data and, simultaneously, satisfy all phenomenological constraints. In particular, it has been shown that the minimal ISS realisation corresponds to extending the SM with two RH neutrinos and two sterile singlet fermions~\cite{Abada:2014vea}, to which we will refer as the ISS(2,2) model.

A longstanding and challenging issue in particle physics is the lack of a guiding principle to explain the flavour structure of the SM, i.e., the observed fermion mass spectra and mixing patterns. This {\em flavour puzzle} provides a strong motivation for building models with additional particle content and extended continuous and/or discrete symmetries. Once such symmetries are explicitly or spontaneously broken, they will lead to the required fermion mass and mixing structures. Several frameworks have been put forward to tackle this puzzle~\cite{symmetries}. One of the simplest approaches consists on the implementation of texture zeros in the Yukawa coupling and mass matrices, imposed by continuous U(1) and/or discrete $\mathbb{Z}_N$ transformations~(see, for instance, Refs.~\cite{Grimus:2004hf,Dighe:2009xj,Adhikary:2009kz,Dev:2011jc,Felipe:2014vka,Cebola:2015dwa,Samanta:2015oqa,Kobayashi:2018zpq,Rahat:2018sgs,Nath:2018xih,Barreiros:2018ndn,Barreiros:2018bju,Correia:2019vbn}). In the SM extended with RH neutrinos, the realisation of texture zeros with such symmetries is not compatible with data since, in general, they lead to massless charged leptons, massless neutrinos or vanishing lepton mixing angles~\cite{Correia:2019vbn,Low:2003dz}. This is due to the fact that all fermions in the SM couple to the same Higgs field. Thus, enlarging the Higgs sector is a viable solution to surmount this difficulty, being the two-Higgs doublet model (2HDM)~\cite{Branco:2011iw} the most economical one.

Inspired by the above ideas, in this work we consider the ISS(2,2) within the 2HDM supplemented with Abelian symmetries to ensure maximal predictability, i.e., to impose the most constraining flavour structure, so that the charged-lepton masses and current neutrino data can be accommodated, while fulfilling all relevant phenomenological constraints. This can be realised by adding to the scalar sector of the SM another scalar doublet and two complex scalar singlets which, upon spontaneous symmetry breaking, generate all relevant mass terms required to implement the ISS(2,2). Moreover, we will show that CP can be spontaneously broken by the complex vacuum expectation value (VEV) of one of the singlets, and that such CP violation (CPV) can be communicated to the neutrino sector via neutrino-scalar interactions. 

This paper is organised as follows. In Section~\ref{sec:ISS}, general aspects of the ISS mechanism are reviewed paying special attention to the comparison between the effective and full treatment of neutrino masses and mixing. The most restrictive flavour structures for the mass matrices in the ISS(2,2) framework are then identified in Section~\ref{sec:compdata} by performing a systematic search of all possible texture-zero combinations leading to low-energy neutrino parameters compatible with global analyses of neutrino oscillation data. After setting the successful cases, in Section~\ref{sec:symmetries} we select those which can be realised by Abelian horizontal symmetries. The phenomenological analysis starts in Section~\ref{sec:numass}, where spontaneous CPV (SCPV) is considered and the relation between the Dirac and Majorana phases is established in light of present neutrino data. Predictions for the effective neutrino mass parameter relevant for neutrinoless double beta decay are also discussed in that section. The impact of radiative corrections on light-neutrino masses is analysed in Section~\ref{sec:radcorr}, while the constraints imposed by cLFV decays on the model parameter space are investigated in Section~\ref{sec:pheno}. Possibilities of testing the ISS(2,2) with Abelian flavour symmetries at other experiments as, for instance, the LHC, future colliders, beam-dump experiments and cLFV searches are discussed in Section~\ref{sec:CONST}. Finally, our concluding remarks are presented in Section~\ref{sec:conclusion}. Details regarding the scalar sector and the computations of cLFV decay rates are collected in the appendices.

\section{General aspects of the inverse seesaw mechanism}
\label{sec:ISS}
The ISS mechanism can be implemented by extending the SM particle content with $n_R$ RH neutrinos $\nu_{R}$ and $n_s$ sterile fermion singlets $s$, leading to what we denote as ISS$(n_R,n_s)$.\footnote{We recall that an arbitrary number of gauge singlets can be introduced in the SM without affecting the anomaly cancellation constraints.} In this framework, the generic mass Lagrangian for leptons is given in the flavour basis by
\begin{equation}
\begin{split}
-\mathcal{L}_{\text{mass}} &=\overline{e_{L}}\, \mathbf{M}_{\ell}\, e_{R} + \overline{\nu_L}\, \mathbf{M}_{D} \nu_R+ \overline{\nu_R}\, \mathbf{M}_{R}  s + \frac{1}{2} \overline{s^c} \ \mathbf{M}_{s} s + \text{H.c.}\,,
\end{split}
\label{eq:Lmass}
\end{equation}
where  $\nu_{L}=\left(\nu_{e L}, \nu_{\mu L}, \nu_{\tau L} \right)^T$, $\nu_{R}=\left(\nu_{R1}, ... \ , \nu_{Rn_R} \right)^T$,\, $s=\left(s_{1}, ...\ , s_{n_s} \right)^T$. For a fermion field $\psi$ we have $\psi^c \equiv C \overline{\psi}^T $ with $C$ denoting the charge conjugation matrix. In the above equation,  $\mathbf{M}_{\ell}$ is the $3\times 3$ charged-lepton mass matrix, $\mathbf{M}_{D}$ is a $3\times n_R$ Dirac-type mass matrix, $\mathbf{M}_{R}$ is a $n_R \times n_s$ matrix, and $\mathbf{M}_{s}$ is a LNV $n_s \times n_s$ Majorana mass matrix. The latter can be naturally small in the  t'Hooft~\cite{tHooft:1980xss} sense, since lepton number conservation is restored in the limit where the last term in Eq.~\eqref{eq:Lmass} vanishes. Defining $N_{L}=\left(\nu_{L}, \nu^c_{R}, s \right)^{T}$ of dimension $n_f = 3+n_R+n_s$, we can write $\mathcal{L}_{\text{mass}} $ in the compact form
\begin{equation}
\begin{split}
-\mathcal{L}_{\text{mass}} &=\overline{e_{L}} \mathbf{M}_{\ell}\, e_{R}+\frac12\,\overline{N_{L}^c}\,\bm{\mathcal{M}}\,N_{L}+\text{H.c.}\;\;,\;\; 
\bm{\mathcal{M}}= \begin{pmatrix}
0 & \mathbf{M}_D^{*} & 0 \\
\mathbf{M}_D^{\dagger} & 0 &\mathbf{M}_{R}\\
0 & \mathbf{M}^{T}_{R} & \mathbf{M}_s
\end{pmatrix}\,,
\label{eq:bigm}
\end{split}
\end{equation}
where $\bm{\mathcal{M}}$ is the full $n_f \times n_f$ neutrino mass matrix. The charged-lepton mass matrix is bidiagonalised through the unitary transformations $e_{L}  \rightarrow \mathbf{V}_{L}\, e_{L}, \ e_{R}  \rightarrow \mathbf{V}_{R}\, e_{R}$, so that
\begin{equation}
\begin{aligned}
 \mathbf{V}_{L}^{\dagger} \mathbf{M}_{\ell} \mathbf{V}_{R} = \mathbf{D}_{\ell} = \text{diag}\left(m_e, m_{\mu}, m_{\tau}\right),
\end{aligned}
\label{diagcharg}
\end{equation}
with $m_{e,\mu,\tau}$ denoting the physical charged-lepton masses. For a given $\Ml$, $\mathbf{V}_{L}$ and $\mathbf{V}_{R}$ are determined by diagonalising the Hermitian matrices $\mathbf{H}_{\ell} = \mathbf{M}_{\ell} \mathbf{M}_{\ell}^{\dagger}$ and $\mathbf{H}_{\ell}^\prime = \mathbf{M}_{\ell}^{\dagger} \mathbf{M}_{\ell}$ as
\begin{equation}
\mathbf{V}^{\dagger}_{L} \mathbf{H}_{\ell} \mathbf{V}_{L} = \mathbf{D}^2_{\ell} = \text{diag}\left(m_e^2 , m_{\mu}^2, m_{\tau}^2\right) \; , \; \mathbf{V}^{\dagger}_{R} \mathbf{H}_{\ell}^\prime \mathbf{V}_{R} = \mathbf{D}^2_{\ell} = \text{diag}\left(m_e^2 , m_{\mu}^2, m_{\tau}^2\right).
\end{equation}
The weak-basis states $N_{L,R}$ are related to the mass eigenstates~$(\nu_1,...,\nu_{n_f})^T$ by a $n_f\times n_f$  unitary matrix $\bm{\mathcal{U}}$
\begin{align}
N_L = \bm{\mathcal{U}}\, (\nu_1, \dots \ , \nu_{n_f})^T_L\;,\;
 N_R = N_L^{c} = \bm{\mathcal{U}}^{*}\, (\nu_1, \dots \ , \nu_{n_f})^T_R\,,
 \end{align}
such that the full neutrino mass matrix is diagonalised as
\begin{align}
\bm{\mathcal{U}}^{T} \bm{\mathcal{M}} \ \bm{\mathcal{U}} = \bm{\mathcal{D}}_{\nu} = \text{diag}\,(m_1, \dots \ , m_{n_f})\,,
\label{diagnutr}
\end{align}
where $m_{1,\dots,n_f}$ are the $n_f$ (real and positive) Majorana neutrino masses. Notice that, in general, the light-active (heavy-sterile) neutrino masses are labelled as $m_{1,2,3}$ ($m_{4,...,n_f}$). In the ISS approximation limit ($\mathbf{M}_s, \mathbf{M}_D \ll \mathbf{M}_{R}$), the neutrino mass matrix $\bm{\mathcal{M}}$ of Eq.~\eqref{eq:bigm} can be block-diagonalised by writing it in the form
\begin{equation}
\label{eq:bigmtypeI}
\bm{\mathcal{M}} = \begin{pmatrix}
\def\arraystretch{1.2}
\begin{array}{@{}c|cc@{}}
0 & \mathbf{M}_D^{*} & 0 \\ \hline
\mathbf{M}_D^{\dagger} & 0 &\mathbf{M}_{R}\\
0 & \mathbf{M}^{T}_{R} & \mathbf{M}_s
\end{array}
\end{pmatrix} \equiv \begin{pmatrix}
0 & \mathbf{M}_D^\prime \\
\mathbf{M}_D^{\prime\, T} & \mathbf{M}_R^\prime
\end{pmatrix}.
\end{equation}
The full unitary matrix $\bm{\mathcal{U}}$ of Eq.~\eqref{diagnutr} can then be parametrised as~\cite{Grimus:2000vj}
\begin{equation}
\label{eq:paramU}
\bm{\mathcal{U}} = \begin{pmatrix} 
 \sqrt{\mathbb{1}-\mathbf{F} \mathbf{F}^{\dagger}} &  \mathbf{F} \\
 - \mathbf{F}^{\dagger} & \sqrt{\mathbb{1}-\mathbf{F}^{\dagger} \mathbf{F}}
\end{pmatrix} 
\begin{pmatrix} 
\mathbf{U}_{\nu} &  0 \\
 0 & \mathbf{U}_{s}
\end{pmatrix},
\end{equation}
so that
\begin{equation}
\label{eq:paramU1}
\bm{\mathcal{U}}^T \bm{\mathcal{M}} \ \bm{\mathcal{U}} =
\begin{pmatrix} 
\mathbf{U}_{\nu}^T \mathbf{M}_{\text{eff}} \mathbf{U}_{\nu} & 0 \\
 0 & \mathbf{U}_{s}^T \mathbf{M}_{\text{heavy}} \mathbf{U}_{s}
\end{pmatrix},
\end{equation}
where $\mathbf{U}_{\nu}$ and $\mathbf{U}_{s}$ are $3 \times 3$ and $(n_R+n_s) \times (n_R+n_s)$ unitary matrices, respectively; $\mathbf{M}_{\text{eff}}$ and $\mathbf{M}_{\text{heavy}}$ are the effective light and heavy-neutrino mass matrices. At leading order, $\mathbf{M}_{\text{heavy}} \simeq \mathbf{M}_R^\prime$ yielding $n_R + n_s$ heavy neutrinos. In Eq.~\eqref{eq:paramU}, $\mathbf{F}$ is a $3 \times (n_R + n_s)$ matrix given at first order in the seesaw approximation by
\begin{equation}
\mathbf{F}\simeq \mathbf{M}_D^{\prime *} \left(\mathbf{M}_{R}^{\prime *}\right)^{-1} \simeq \left(0,\ \mathbf{M}_D (\mathbf{M}_R^{\dagger})^{-1}\right).
\label{eq:FISS}
\end{equation}
This leads to the $3\times 3$ effective light-neutrino mass matrix
\begin{equation}
\label{eq:invss}
\mathbf{M}_{\rm eff} = - \mathbf{F}^* \mathbf{M}_R^\prime \mathbf{F}^\dagger = - \mathbf{M}_D^{*}\,\left( \mathbf{M}_{R}\, \mathbf{M}_s^{-1} \mathbf{M}_{R}^{T} \right)^{-1}\mathbf{M}_D^{\dagger}\,,
\end{equation}
which can be diagonalised through a unitary rotation of the active neutrino fields, $\nu_{L}  \rightarrow \mathbf{U}_{\nu}\, \nu_{L}$, satisfying
\begin{equation}
\begin{aligned}
\mathbf{U}_{\nu}^{ T}\, \mathbf{M}_{\rm eff}\, \mathbf{U}_{\nu}  = \mathbf{D}_{\nu} = \text{diag}\left(\tilde{m}_1, \tilde{m}_2, \tilde{m}_3\right),
\end{aligned}
\label{eq:Unudef}
\end{equation}
where $\tilde{m}_{1,2,3}$ are the real and positive light neutrino masses in the ISS approximation. The unitary matrix $\mathbf{U}_{\nu}$ is obtained from the diagonalisation of the Hermitian matrix $\mathbf{H}_{\text{eff}} = \mathbf{M}_{\text{eff}} \mathbf{M}_{\text{eff}}^{\dagger}\,$, 
\begin{equation}
\mathbf{U}^{\dagger}_{\nu} \mathbf{H}_{\text{eff}} \mathbf{U}_{\nu}  = \mathbf{D}^2_{\nu} = \text{diag}\left(\tilde{m}_1^2 , \tilde{m}_{2}^2, \tilde{m}_{3}^2\right),
\end{equation}
yielding the unitary lepton mixing matrix
\begin{equation}
   \mathbf{U}^\prime = \mathbf{V}_{L}^{\dagger} \mathbf{U}_{\nu}\,,
   \label{eq:fullpmns}
\end{equation}
after performing the rotation to the charged-lepton mass basis.

In general, for massive Majorana neutrinos, $ \mathbf{U}^\prime$ can be parametrised by three mixing angles $\theta_{12}$, $\theta_{23}$, and $\theta_{13}$, and three CP-violating phases: the Dirac-type phase $\delta$ and two Majorana-type phases $\alpha_{21}$ and $\alpha_{31}$~\cite{param},
\begin{align}
 \mathbf{U}^\prime  = \begin{pmatrix}  c_{12} c_{13} & s_{12} c_{13} & s_{13}  \\
- s_{12} c_{23} - c_{12} s_{23} s_{13}e^{i\delta} & c_{12} c_{23} - s_{12} s_{23} s_{13} e^{i\delta} & s_{23} c_{13} e^{i\delta} \\ 
  s_{12} s_{23} - c_{12} c_{23} s_{13} e^{i\delta} &- c_{12} s_{23} - s_{12} c_{23} s_{13}e^{i\delta} & c_{23} c_{13} e^{i\delta} \\ 
     \end{pmatrix}
     \begin{pmatrix}  1 & 0 & 0 \\
0 &  e^{i\alpha_{21}} & 0 \\ 
 0 & 0 & e^{i\alpha_{31}} \\ 
     \end{pmatrix},
     \label{eq:Uparam}
\end{align}
where $c_{ij} \equiv \cos\theta_{ij}$ and $s_{ij} \equiv \sin\theta_{ij}$. Several neutrino oscillation experiments have been constraining neutrino mass and mixing parameters, namely $\Delta m_{21}^2= m_2^2 - m_1^2$ , $\Delta m_{31}^2= m_3^2 - m_1^2$, $\theta_{12}$, $\theta_{23}$, $\theta_{13}$ and $\delta$. We present in Table~\ref{tab:dataref} the results obtained from the most recent global fit of neutrino oscillation parameters~\cite{deSalas:2020pgw}. Both mass orderings are considered: normal ordering~(NO) where $m_1< m_2< m_3$, and inverted ordering~(IO) where  $m_3 < m_1< m_2$. Since no experimental information has been obtained so far on CPV Majorana phases, they remain unconstrained in our analysis. Notice that within the three-neutrino paradigm, the lepton sector is described by a total of twelve parameters: three charged lepton masses, three light neutrino masses, three mixing angles and three phases. We also remark that the existence of a massless neutrino is presently not excluded by data. In such case, there is only one physical Majorana phase and, thus, the total number of physical parameters in the lepton sector is reduced to ten. \\
\begin{table}[!t]
\setlength{\tabcolsep}{10pt}
\begin{tabular}{ccc}
\hline
Parameter  & Best Fit $\pm 1 \sigma$ & $3\sigma$ range \\ \hline
$\theta_{12} (^\circ)$ & $34.3\pm1.0$ &  $31.4 \rightarrow 37.4$ \\
$\theta_{23} (^\circ) [\text{NO}]$ & $48.79^{+0.93}_{-1.25}$ &  $ 41.63 \rightarrow 51.32 $ \\
$\theta_{23} (^\circ) [\text{IO}]$ & $48.79^{+1.04}_{-1.30}$  &  $ 41.88 \rightarrow 51.30$ \\
$\theta_{13} (^\circ) [\text{NO}]$ & $8.58^{+0.11}_{-0.15}$ &  $ 8.16 \rightarrow 8.94$\\
$\theta_{13} (^\circ) [\text{IO}]$ & $8.63^{+0.11}_{-0.15}$ &  $8.21 \rightarrow 8.99$ \\
$\delta  (^\circ) [\text{NO}]$ & $216^{+41}_{-25}$ & $144 \rightarrow 360 $ \\
$\delta  (^\circ) [\text{IO}]$ & $277^{+23}_{-24}$ &  $205 \rightarrow 342 $ \\
$\Delta m_{21}^2 \left(\times 10^{-5} \ \text{eV}^2\right)$ & $7.50^{+0.22}_{-0.20}$ &  $ 6.94 \rightarrow 8.14$ \\
$\left|\Delta m_{31}^2\right| \left(\times 10^{-3}  \ \text{eV}^2\right) [\text{NO}]$ & $2.56^{+0.03}_{-0.04}$  & $2.46 \rightarrow 2.65 $ \\
$\left|\Delta m_{31}^2\right| \left(\times 10^{-3} \ \text{eV}^2\right) [\text{IO}]$ & $2.46\pm 0.03$ & $2.37 \rightarrow 2.55$\\
\hline
\end{tabular}
%\end{ruledtabular}
\caption{ Current neutrino data obtained from the global fit of three flavour oscillation parameters~\cite{deSalas:2020pgw}.}
\label{tab:dataref}
\end{table}

We now characterise active and active-sterile mixing considering the full mixing matrix $\bm{\mathcal{U}}$ or, more specifically, the rectangular $3 \times n_f$ matrix $\mathbf{W}_{\alpha j}\equiv \bm{\mathcal{U}}_{\alpha j}$ ($\alpha=e,\mu,\tau$, $j=1,\dots,n_f$) which, according to Eq.~\eqref{eq:paramU}, can be decomposed in the form
\begin{equation}
\mathbf{W} = (\sqrt{\mathbb{1}-\mathbf{F} \mathbf{F}^{\dagger}} \,\U_\nu\,,\, \mathbf{F} \U_s)\equiv \left(\mathbf{W}_{\nu}, \mathbf{W}_s\right),
\label{eq:3x7}
\end{equation}
where $\mathbf{W}_{\nu}$ and $\mathbf{W}_{s}$ are $3\times 3$ and $3 \times (n_R + n_s)$ matrices, respectively. It is clear that $ \mathbf{V}_{L}^{\dagger} \mathbf{W}_{s}$ defines the mixing between the three active neutrinos and the $n_R + n_s$ sterile states in the physical charged-lepton basis. Due to the additional fermion states, active-neutrino mixing is determined by the non-unitary matrix~\cite{FernandezMartinez:2007ms}
\begin{equation}
\mathbf{U} = \mathbf{V}_{L}^{\dagger} \mathbf{W}_{\nu} = \left(\mathbb{1} - \boldsymbol{\eta} \right) \mathbf{U}^\prime ,
\label{eq:nonunitpmns}
\end{equation}
where $\mathbf{U}^\prime$ is the unitary mixing matrix given in Eq.~\eqref{eq:fullpmns} and $\boldsymbol{\eta}$ is an Hermitian matrix encoding deviations from unitarity of $\mathbf{U}$. Expanding Eq.~\eqref{eq:paramU} up to second order in $\mathbf{F}$, one has $\sqrt{\mathbb{1}-\mathbf{F} \mathbf{F}^{\dagger}} \simeq \mathbb{1}- \frac{1}{2}\mathbf{F} \mathbf{F}^{\dagger}$ which, together with Eqs.~\eqref{eq:FISS} and \eqref{eq:nonunitpmns}, leads to
\begin{equation}
\bm{\eta} = \dfrac{1}{2}\mathbf{V}_L^{\dagger} \mathbf{F} \mathbf{F}^{\dagger} \mathbf{V}_L \simeq \dfrac{1}{2} \mathbf{V}_L^{\dagger}\mathbf{M}_D (\mathbf{M}_R^{\dagger})^{-1}  \mathbf{M}_R^{-1} \mathbf{M}_D^{\dagger}\mathbf{V}_L\,.
\end{equation}
Active-sterile neutrino mixing is described by $\mathbf{W}_{s}$ given in Eq.~\eqref{eq:3x7}, which at first order in $\mathbf{F}$ is
\begin{equation}
\mathbf{V}_L^{\dagger}\mathbf{W}_{s} = \mathbf{V}_L^{\dagger}\mathbf{F} \mathbf{U}_{s} \simeq \mathbf{V}_L^{\dagger}\big(0,\ \mathbf{M}_D (\mathbf{M}_R^{\dagger})^{-1}\big)  \mathbf{U}_{s},
\label{eq:VLW}
\end{equation}
in the basis where $\mathbf{M}_{\ell}$ is diagonal. For the analysis that follows, it is convenient to define the following matrices~\cite{Ilakovac:1994kj},
\begin{equation}
\mathbf{B}_{\alpha j} = \sum_{k=1}^{3} (\mathbf{V}_{L}^{*})_{k\alpha} \mathbf{W}_{k j}, \quad \bm{\mathcal{C}}_{i j} = \sum_{k=1}^{3} \mathbf{W}^{*}_{k i} \mathbf{W}_{k j},
\label{eq:BC}
\end{equation}
which obey the equalities
\begin{equation}
\sum_{k=1}^{n_f} \mathbf{B}_{\alpha k} \mathbf{B}_{\beta k}^{*}=\delta_{\alpha \beta}, \quad \sum_{k=1}^{n_f} \mathbf{B}_{\alpha k} \bm{\mathcal{C}}_{k i}= \mathbf{B}_{\alpha i}, \quad \sum_{k=1}^{n_f} \bm{\mathcal{C}}_{i k} \bm{\mathcal{C}}_{j k}^{*}=\bm{\mathcal{C}}_{i j}, \quad \sum_{\alpha=1}^{3} \mathbf{B}_{\alpha i}^{*} \mathbf{B}_{\alpha j}=\bm{\mathcal{C}}_{i j},
\label{eq:equality1}
\end{equation}
\begin{equation}
\sum_{k=1}^{n_f} m_k \bm{\mathcal{C}}_{ik} \bm{\mathcal{C}}_{jk}=0, \quad \sum_{k=1}^{n_f} m_k \mathbf{B}_{\alpha k} \bm{\mathcal{C}}_{ki}^{*}=0, \quad \sum_{k=1}^{n_f} m_k \mathbf{B}_{\alpha k} \mathbf{B}_{\beta k}=0.
\label{eq:equality2}
\end{equation}
Note that the mixing between the light and sterile neutrinos is given by the matrix elements $\mathbf{B}_{\alpha j}$ for $\alpha = e, \mu, \tau$ and $j=4, \dots, n_f$, in the charged-lepton physical basis. Furthermore, the parameters $\boldsymbol{\eta}_{\alpha\beta}$ encoding deviations from unitarity [see Eq.~\eqref{eq:nonunitpmns}] can be expressed in terms of $\mathbf{B}$ through the relation
\begin{equation}
\boldsymbol{\eta}_{\alpha \beta} = \frac{1}{2} \sum_{j=4}^{n_f} \mathbf{B}_{\alpha j} \mathbf{B}^{*}_{\beta j},
\label{eq:epsB}
\end{equation}
where we used the first relation in Eq.~\eqref{eq:equality1} in order to write $\boldsymbol{\eta}_{\alpha \beta}$ solely in terms of the active-sterile mixing.

\section{Maximally-restrictive textures for leptons}
\label{sec:compdata}

In this section, we identify the maximally-restrictive textures for the set of matrices $\left(\mathbf{M}_{\ell},\mathbf{M}_{D},\mathbf{M}_{R},\mathbf{M}_{s}\right)$ compatible with neutrino oscillation data within the minimal ISS(2,2) framework,\footnote{By maximally restrictive we mean that no additional texture zero can be placed into any of the mass matrices while keeping compatibility with the charged-lepton masses and neutrino oscillation data.} where two $\nu_R$ and two $s$ fermion singlets are added to the SM particle content, i.e. $n_R=n_s=2$ and $n_f=7$. Our texture-zero analysis is performed assuming the seesaw approximation given in Eq.~\eqref{eq:invss}. Later on, we will comment on the validity of this approximation when comparing with the results obtained with the full neutrino mass matrix $\bm{\mathcal{M}}$. The identification of the compatible textures is based on a standard $\chi^2$-analysis, using the function
\begin{equation}
\chi^2(x) = \sum_i \frac{\left[\mathcal{P}_i(x) - \mathcal{O}_i\right]^2}{\sigma_i^2},
\end{equation}
where $x$ denotes the input parameters, i.e., the matrix elements of $\mathbf{M}_{\ell}$, $\mathbf{M}_{D}$, $\mathbf{M}_{R}$ and $\mathbf{M}_{s}$; $\mathcal{P}_i(x)$ is the model prediction for a given observable with best-fit (b.f.) value $\mathcal{O}_i$, and $\sigma_i$ denotes its $1\sigma$ experimental uncertainty. In our search for viable sets $(\mathbf{M}_{\ell},\mathbf{M}_{D},\mathbf{M}_{R},\mathbf{M}_{s})$, we require the charged-lepton masses to be at their central values~\cite{Zyla:2020zbs}, such that the $\chi^2$-function is minimised only with respect to the six neutrino observables, namely the two neutrino mass squared differences $\Delta m^2_{21}, \Delta m^2_{31}$, the three mixing angles $\theta_{12}, \theta_{23}, \theta_{13}$ and the Dirac CPV phase $\delta$, using the current data reported in Table~\ref{tab:dataref}~\cite{deSalas:2020pgw}. Notice that, in the ISS(2,2) framework, there is always a massless neutrino ($\tilde{m}_1=0$ for NO or $\tilde{m}_3=0$ for IO).

For a given set of input matrices, we consider compatibility with data if the deviation of each neutrino observable from its experimental value is at most $3\sigma$ at the  $\chi^2$-minimum~\cite{Cebola:2015dwa,Cebola:2016jhz,Felipe:2016sya,Correia:2019vbn}. If this is the case, we also test the compatibility of the textures at $1\sigma$. For the sake of simplicity, we shall use the following sequential notation to label the position of the matrix elements of a given $3\times2$ and $2\times2$ texture $\text{T}$, respectively,
\begin{equation}
\begin{pmatrix} 
1 & 2 \\ 
3 & 4 \\
5 & 6
\end{pmatrix},\quad
\begin{pmatrix} 
1 & 2\\ 
3 & 4 
\end{pmatrix},
\end{equation}
where we denote the position of any vanishing element labelled $i$ with a subscript, i.e., $\text{T}_i$. For instance, in this notation, a matrix with vanishing 11 and 22 elements would be labelled as T$_{14}$.

It is straightforward to show that the ISS formula in Eq.~\eqref{eq:invss} is invariant under the weak-basis permutations
\begin{equation}
\begin{aligned}
\mathbf{M}_{D} \rightarrow \mathbf{M}_{D}\, \mathbf{P}_R\;\;,\;\;
\mathbf{M}_{R} \rightarrow \mathbf{P}_R^T\, \mathbf{M}_{R}\, \mathbf{P}_s\;\;,\;\;
\mathbf{M}_s \rightarrow \mathbf{P}_s^T\, \mathbf{M}_s\, \mathbf{P}_s,
\end{aligned}
\end{equation}
where $\mathbf{P}$ denote the $3\times3$ (or $2\times2$) permutation matrices. Furthermore, for a given  pair $(\mathbf{M}_{\ell} , \mathbf{M}_{\text{eff}})$, the permutations 
\begin{equation}
\begin{aligned}
 \mathbf{M}_{\ell} \rightarrow \mathbf{P}^{T}_{\ell}\, \mathbf{M}_{\ell}\,  \mathbf{P}_{e}\;\;,\;\;
 \mathbf{M}_{\text{eff}} \rightarrow \mathbf{P}^{T}_{\ell}\, \mathbf{M}_{\text{eff}}\,  \mathbf{P}_{\nu},
\end{aligned}
\end{equation}
leave the lepton mixing matrix in Eq.~\eqref{eq:fullpmns} invariant. Therefore, for each weak-basis permutation class, only one representative set of textures needs to be identified.
\begin{table}[!t]
    \begin{minipage}{.25\linewidth}
      \centering
\begin{tabular}{cccc} 
\hline
\multicolumn{4}{c}{$\mathbf{M}_{\ell}=6^{\ell}$}\\
        \hline
		$\mathbf{M}_{D}\;\;$ &$\mathbf{M}_{R}\;\;$ & $\mathbf{M}_{s}\;\;$ & $\mathbf{M}_{\text{eff}}$\\
		\hline
		$\text{T}_{1}$	&$\text{T}_{14}$ & $\text{T}_{23}$ &-\\
			$\text{T}_{4}$	&$\text{T}_{14}$ & $\text{T}_{23}$ &-\\
		$\text{T}_{5}$	&$\text{T}_{14}$ & $\text{T}_{23}$ &-\\
		$\text{T}_{14}$	&$\text{T}_{1}$ & $\text{T}_{23}$ &-\\
			$\text{T}_{16}$	&$\text{T}_{1}$ & $\text{T}_{23}$ &-\\
		$\text{T}_{23}$	&$\text{T}_{1}$ & $\text{T}_{23}$ &-\\
		$\text{T}_{25}$	&$\text{T}_{1}$ & $\text{T}_{23}$ &-\\
			$\text{T}_{36}$	&$\text{T}_{1}$ & $\text{T}_{23}$ &-\\
		$\text{T}_{45}$	&$\text{T}_{1}$ & $\text{T}_{23}$ &-\\
		\hline
	\end{tabular}
	 \end{minipage}%
    \begin{minipage}{.25\linewidth}
      \centering
      \vspace{-1.8cm} 
\begin{tabular}{cccc} 
\hline
\multicolumn{4}{c}{$\mathbf{M}_{\ell}=5^{\ell}_1$}\\
\hline
$\mathbf{M}_{D}\;\;$ &$\mathbf{M}_{R}\;\;$ & $\mathbf{M}_{s}\;\;$ & $\mathbf{M}_{\text{eff}}$\\
\hline
$\text{T}_{13}$ &$\text{T}_{14}$ & $\text{T}_{23}$ &- \\
$\text{T}_{14}$ &$\text{T}_{14}$ & $\text{T}_{23}$ & $1_4^{\nu}$\\
$\text{T}_{16}$ &$\text{T}_{14}$ & $\text{T}_{23}$ & $1_5^{\nu}$\\
$\text{T}_{35}$ &$\text{T}_{14}$ & $\text{T}_{23}$ &-\\
$\text{T}_{45}$ &$\text{T}_{14}$ & $\text{T}_{23}$ & $1_6^{\nu}$\\
\hline
	\end{tabular}
    \end{minipage}%
    \begin{minipage}{.25\linewidth}
      \centering
\begin{tabular}{cccc} 
\hline
\multicolumn{4}{c}{$\mathbf{M}_{\ell}= 4^{\ell}_1$, $4^{\ell}_2$, $4^{\ell}_3$}\\
\hline
$\mathbf{M}_{D}\;\;$ &$\mathbf{M}_{R}\;\;$ & $\mathbf{M}_{s}$ & $\mathbf{M}_{\text{eff}}$\\
\hline
$\text{T}_{124}$	&$\text{T}_{14}$ & $\text{T}_{23}$ & $3_{11}^{\nu}$\\
$\text{T}_{125}$	&$\text{T}_{14}$ & $\text{T}_{23}$ & $3_{11}^{\nu}$\\
$\text{T}_{134}$	&$\text{T}_{14}$ & $\text{T}_{23}$ & $3_{16}^{\nu}$\\
$\text{T}_{136}$	&$\text{T}_{14}$ & $\text{T}_{23}$ & $2_{15}^{\nu}$\\
$\text{T}_{145}$	&$\text{T}_{14}$ & $\text{T}_{23}$ & $2_{14}^{\nu}$\\
$\text{T}_{146}$	&$\text{T}_{14}$ & $\text{T}_{23}$ & $2_{13}^{\nu}$\\
$\text{T}_{156}$	&$\text{T}_{14}$ & $\text{T}_{23}$ & $3_{19}^{\nu}$\\
$\text{T}_{345}$	&$\text{T}_{14}$ & $\text{T}_{23}$ & $3_{16}^{\nu}$\\
$\text{T}_{456}$	&$\text{T}_{14}$ & $\text{T}_{23}$ & $3_{19}^{\nu}$\\
\hline
	\end{tabular}
    \end{minipage} 
        \caption{Maximally-restrictive texture sets for $\mathbf{M}_{\ell}=6^{\ell}$ (left), $5^{\ell}_1$ (centre) and $4^{\ell}_{1,2,3}$ (right).}
        \label{tab:maxtex}
\end{table}
The maximally-restrictive texture zero sets $(\mathbf{M}_{\ell},\mathbf{M}_{D},\mathbf{M}_{R},\mathbf{M}_{s})$ compatible with neutrino oscillation data for NO are presented in Table~\ref{tab:maxtex}. It turns out that these sets of matrices are also viable for IO. Moreover, all the sets are compatible with data at $1\sigma$. The labelling used for the charged-lepton mass matrix and the effective neutrino mass matrix follows Ref.~\cite{Ludl:2014axa} and the corresponding textures are presented in Tables~\ref{tab:notMl} and~\ref{tab:notMnu}, respectively.
\begin{table}[!t]
%	\begin{ruledtabular}
\begin{tabular}{ccc}
$4_1^{\ell} \sim \begin{pmatrix} 
0 & 0 & \times \\ 
0 & \times & 0\\
\times & \times & \times
\end{pmatrix}$ & 
$4_2^{\ell} \sim \begin{pmatrix} 
0 & 0 & \times \\ 
0 & \times &\times\\
\times & 0 & \times
\end{pmatrix}$ &
$4_3^{\ell} \sim \begin{pmatrix} 
0 & 0 & \times \\ 
0 & \times & \times\\
\times &  \times& 0
\end{pmatrix}$ \\
\\
$5_1^{\ell} \sim \begin{pmatrix} 
0 & 0 & \times \\ 
0 & \times & 0 \\
\times & 0 & \times
\end{pmatrix} $ &
$6^{\ell} \sim \begin{pmatrix} 
\times & 0 & 0 \\ 
0 & \times & 0 \\
0 & 0 & \times
\end{pmatrix}$\\
\end{tabular}
%\end{ruledtabular}
\caption{Textures for the charged-lepton mass matrix $\mathbf{M}_{\ell}$.}
\label{tab:notMl}
\end{table}
\begin{table}[!t]
%	\begin{ruledtabular}
	\begin{tabular}{ccc}
	$1_4^{\nu} \sim \begin{pmatrix} 
\times & 0 & \times \\ 
0& \times & \times\\
\times & \times & \times
\end{pmatrix}$ & 
$1_5^{\nu} \sim \begin{pmatrix} 
\times & \times& 0\\ 
\times & \times & \times\\
0 & \times & \times
\end{pmatrix}$ &
$1_6^{\nu} \sim \begin{pmatrix} 
\times & \times& \times \\ 
\times & \times & 0\\
\times & 0 & \times
\end{pmatrix} $
\\
\\
	$2_{13}^{\nu} \sim \begin{pmatrix} 
\times & 0& 0 \\ 
0 & \times & \times\\
0 & \times & \times
\end{pmatrix}$ & 
$2_{14}^{\nu} \sim \begin{pmatrix} 
\times & 0& \times \\ 
0 & \times & 0\\
\times & 0 & \times
\end{pmatrix}$ &
$2_{15}^{\nu} \sim \begin{pmatrix} 
\times & \times& 0 \\ 
\times & \times & 0\\
0 & 0 & \times
\end{pmatrix} $
\\
\\
$	3_{11}^{\nu} \sim \begin{pmatrix} 
0 & 0& 0 \\ 
0 & \times & \times\\
0 & \times & \times
\end{pmatrix} $& 
$3_{16}^{\nu} \sim \begin{pmatrix} 
\times & 0& \times \\ 
0 & 0 & 0\\
\times & 0 & \times
\end{pmatrix}$ &
$3_{19}^{\nu} \sim \begin{pmatrix} 
\times & \times& 0 \\ 
\times & \times &0\\
0& 0 & 0
\end{pmatrix} $
\\
\end{tabular}
%\end{ruledtabular}
\caption{Textures for the effective neutrino mass matrix $\mathbf{M}_{\text{eff}}$.}
\label{tab:notMnu}
\end{table}

\section{Abelian symmetry realisation of compatible textures}
\label{sec:symmetries}

We start this section by specifying the scalar sector of the model. As mentioned before, maximally-restrictive texture zeros in Yukawa coupling matrices cannot be implemented in the SM with Abelian symmetries, since all fermion fields couple to the same Higgs doublet. Hence, to realise such textures, our minimal setup will require the presence of at least two Higgs doublets $\Phi_a$ ($a=1,2$). Furthermore, to avoid bare mass terms in the Lagrangian, we also add two complex scalar fields $S_a$ ($a=1,2$), so that $\M_s$ and $\M_R$ are dynamically generated through couplings of $S_1$ and $S_2$ with $s^{T} C s$ and $\overline{\nu_R}\, s$, respectively. We parameterise $\Phi_a$ and $S_a$ as
\begin{equation}
\Phi_a =\begin{pmatrix}
\phi_a^{+} \\
\phi_a^0
\end{pmatrix}= \frac{1}{\sqrt{2}}  \begin{pmatrix}
 \sqrt{2} \phi_a^{+} \\
 v_a e^{i \theta_a} + \rho_a + i \eta_a
\end{pmatrix} \; , \; S_a = \frac{1}{\sqrt{2}}\left( u_a e^{i \xi_a} + \rho_{a+2} + i \eta_{a+2}\right)\;,\; a=1,2\,,
\label{eq:scalarfield}
\end{equation}
where $v_a$ and $u_a$ are the VEVs of the neutral components of Higgs doublets~$\phi_a^0$ and the scalar singlet fields, respectively. Note that only the phase difference $\theta=\theta_2-\theta_1$ is physical (a more detailed analysis of the scalar sector can be found in Appendix~\ref{sec:ScalarSector}).

Given the minimal fermion and scalar contents described above, the Yukawa Lagrangian relevant for our work is~\footnote{Notice that there are other Yukawa interactions invariant under the SM gauge symmetry, as well as bare mass terms for the neutrino singlets. We do not list all terms here since they will be forbidden by the Abelian symmetries to be considered next.}
\begin{align}
-\mathcal{L}_{\text{Yuk.}} &= \overline{\ell_{L}} \left(\mathbf{Y}_{\ell}^1 \Phi_1 +  \mathbf{Y}^2_{\ell} \Phi_2 \right) e_{R} +\overline{\ell_{L}} \left(\mathbf{Y}^1_{D} \tilde{\Phi}_1 + \mathbf{Y}^2_{D} \tilde{\Phi}_2\right) \nu_{R} \nonumber\\ 
&+ \frac{1}{2}\, \overline{s^c} \left(\mathbf{Y}_{s}^{1} S_1 + \mathbf{Y}_{s}^{2} S_1^{*}\right) s + \overline{\nu_{R}} \left(\mathbf{Y}_{R}^{1} S_2 + \mathbf{Y}_{R}^{2} S_2^{*}\right) s + \text{H.c.}.
\label{eq:Yuk}
\end{align}
Upon spontaneous symmetry breaking, the scalar fields acquire non-zero VEVs and the above Yukawa interactions yield the generic mass Lagrangian of Eq.~\eqref{eq:Lmass} for the ISS(2,2). The corresponding mass matrices are given by 
\begin{equation}
\begin{aligned}
   & \mathbf{M}_{\ell} = \frac{v_1}{\sqrt{2}} \mathbf{Y}_{\ell}^1 + \frac{v_2}{\sqrt{2}} \mathbf{Y}^2_{\ell} \ e^{i \theta}, \ \mathbf{M}_{D} = \frac{v_1}{\sqrt{2}} \mathbf{Y}^1_{D} + \frac{v_2}{\sqrt{2}} \mathbf{Y}^2_{D} \ e^{-i \theta}, \\
    &\mathbf{M}_{s} = \frac{u_1}{\sqrt{2}}  \left(\mathbf{Y}_{s}^{1} \ e^{i \xi_1} +  \mathbf{Y}_{s}^{2} \ e^{-i \xi_1}\right), \ \mathbf{M}_{R} = \frac{u_2}{\sqrt{2}} \left(\mathbf{Y}_{R}^{1} \ e^{i \xi_2} +  \mathbf{Y}_{R}^{2} \ e^{-i \xi_2}\right).
\end{aligned}
\label{eq:massyuk}
\end{equation}

To implement Abelian flavour symmetries, we require the full Lagrangian to be invariant under the field transformations
\begin{equation}
\begin{aligned}
\Phi_a \rightarrow \mathbf{X}_{\Phi_a} \Phi_a, \quad  S_a \rightarrow \mathbf{X}_{S_a} S_a\;,\;
\ell_{L}  \rightarrow \mathbf{X}_{\ell} \ell_{L}, \ e_{R}  \rightarrow \mathbf{X}_{e} e_{R}\;,\;
\nu_{R}  \rightarrow \mathbf{X}_{R} \nu_{R}, \ s  \rightarrow \mathbf{X}_{s} s,
\end{aligned}
\label{eq:transf}
\end{equation}
where, for each field component $F$, $\mathbf{X}_{F}$ denotes a phase of the form $e^{i x_{F}}$. This invariance requirement yields the following constraints on the Yukawa matrices of Eq.~\eqref{eq:Yuk}:
\begin{equation}
\begin{aligned}
    &\mathbf{Y}^a_{\ell} = \mathbf{X}_{\ell}^{\dagger} \mathbf{Y}^a_{\ell} \mathbf{X}_{e} \mathbf{X}_{\Phi_a}, \ \mathbf{Y}^a_{D} = \mathbf{X}_{\ell}^{\dagger} \mathbf{Y}^a_{D} \mathbf{X}_{R} \mathbf{X}^{*}_{\Phi_a}, \\
    & \mathbf{Y}_{s}^1 = \mathbf{X}_{s}^{T} \mathbf{Y}_{s}^1 \mathbf{X}_{s} \mathbf{X}_{S_1}, \ \mathbf{Y}_{s}^2 = \mathbf{X}_{s}^{T} \mathbf{Y}_{s}^2 \mathbf{X}_{s} \mathbf{X}_{S_1}^{*}, \\
    &\mathbf{Y}_{R}^1 = \mathbf{X}_{R}^{\dagger} \mathbf{Y}_{R}^1 \mathbf{X}_{s} \mathbf{X}_{S_2}, \ \mathbf{Y}_{R}^2 = \mathbf{X}_{R}^{\dagger} \mathbf{Y}_{R}^2 \mathbf{X}_{s} \mathbf{X}_{S_2}^*\,,
\end{aligned}
\label{eq:phaserel}
\end{equation}
which can be translated into relations among the various field-transformation phases $x_F$. To implement a texture-zero entry in one of the above Yukawa matrices we require that the corresponding phase relation is not fulfilled. 

We now proceed to identify which of the maximally-restrictive texture sets compatible with neutrino data (see previous section) can be realised by imposing discrete or continuous Abelian symmetries. To this end, we will apply two methods that complement each other, namely the canonical~\cite{Ferreira:2010ir,Serodio:2013gka} and the Smith normal form (SNF)~\cite{Ivanov:2011ae,Ivanov:2013bka} methods. Our methodology follows closely the one employed in Refs.~\cite{Felipe:2016sya,Correia:2019vbn}. We start with the canonical approach applied to the maximally-restrictive textures presented in Table~\ref{tab:maxtex} to reduce the scope of realisable textures before employing the SNF method. We recall that the charged-lepton textures $4_1^{\ell}$ and $4_2^{\ell}$ cannot be realised through Abelian symmetries in the 2HDM~\cite{Correia:2019vbn}. For the remaining cases, we first write all possible decompositions of the mass matrix textures into the corresponding two Yukawa matrices defined in Eq.~\eqref{eq:massyuk}. Afterwards, for a given $(\mathbf{M}_{\ell},\mathbf{M}_{D},\mathbf{M}_{R},\mathbf{M}_{s})$ combination, and for all decompositions of its matrices, we solve the corresponding system of algebraic relations for the field phases (or charges) stemming from Eq.~\eqref{eq:phaserel}. If a solution exists for a set of charges, then that specific $(\mathbf{M}_{\ell},\mathbf{M}_{D},\mathbf{M}_{R},\mathbf{M}_{s})$ is realisable by Abelian symmetries with the fields carrying those charges. In Table~\ref{tab:RealizableDecomp}, we present the realisable mass matrix textures and their corresponding Yukawa decompositions, respectively. Notice that although in some cases two decompositions are possible for a given mass matrix, only one is realisable. We set the ordering for $\mathbf{Y}^{1,2}_{\ell}$ and $\mathbf{Y}^{1,2}_{D}$ as the one given in Table~\ref{tab:RealizableDecomp}. Also, we use the notation $\mathbf{Y}_{R} \equiv \mathbf{Y}_{R}^1$ [see Eq.~\eqref{eq:Yuk}] since $\mathbf{Y}_{R}^2$ is forbidden by the symmetries as we shall see promptly. Hence, since for all realisable cases $\mathbf{M}_{R}$ and $\mathbf{M}_{s}$ are fixed by the textures $\text{T}_{14}$ and $\text{T}_{23}$, respectively, from now on we will refer to each case just through the pair notation $(\mathbf{M}_{\ell},\mathbf{M}_{D})$.
\renewcommand{\arraystretch}{1.5}
\begin{table}[!t]
%\begin{ruledtabular}
\begin{minipage}{.3\linewidth}
\vspace{-9.85cm}
\begin{tabular}{ccccc}
\hline
$\mathbf{M}_{\ell}$ &\quad $\mathbf{M}_{D}$ &\quad $\mathbf{M}_{R}$ &\quad $\mathbf{M}_{s}$ &\quad $\mathbf{M}_{\text{eff}}$\\
\hline
 $5^{\ell}_{1,\text{I}}$ &$\text{T}_{45}$ &$\text{T}_{14}$ & $\text{T}_{23}$ & $1_{6}^{\nu}$ \\ \hline 
 $4^{\ell}_3$ & $\text{T}_{124}$ &$\text{T}_{14}$ & $\text{T}_{23}$ & $3_{11}^{\nu}$\\
  &$\text{T}_{456}$ &$\text{T}_{14}$ & $\text{T}_{23}$ & $3_{19}^{\nu}$\\
  &$\text{T}_{136,\text{I}}$ &$\text{T}_{14}$ & $\text{T}_{23}$ & $2_{15}^{\nu}$\\
  &$\text{T}_{146,\text{I}}$ &$\text{T}_{14}$ & $\text{T}_{23}$ & $2_{13}^{\nu}$\\
  \hline
\end{tabular}
\end{minipage}
\begin{minipage}{.3\linewidth}
\vspace{-8mm}
\begin{tabular}{ccc}
\hline
\setlength{\tabcolsep}{8pt}
 $\mathbf{M}_{\ell}$  & $\mathbf{Y}_{\ell}^{1}$ & $\mathbf{Y}_{\ell}^{2}$\\
\hline
\\
$4^{\ell}_3$ & $\begin{pmatrix} 
0 & 0& \times \\ 
0 & \times & 0 \\
\times & 0 & 0
\end{pmatrix}$ & $\begin{pmatrix} 
0 & 0 & 0 \\ 
0 & 0 & \times \\
0 & \times & 0
\end{pmatrix}$ \\
$5^{\ell}_{1,\text{I}}$ &
$\begin{pmatrix} 
0 & 0& \times \\ 
0 & 0 & 0 \\
\times & 0 & 0
\end{pmatrix}$ &  $\begin{pmatrix} 
0 & 0 & 0 \\ 
0 & \times & 0 \\
0 & 0 & \times
\end{pmatrix} $\\
$5^{\ell}_{1,\text{II}}$ & 
$\begin{pmatrix} 
0 & 0& \times \\ 
0 & \times & 0 \\
\times & 0 & 0
\end{pmatrix}$ & $ \begin{pmatrix} 
0 & 0 &  0 \\ 
0 & 0 &  0 \\
0 & 0 & \times
\end{pmatrix}$ \\
\\
\hline
\hline
 $\mathbf{M}_{R}$  & $\mathbf{Y}_{R}$& \\
\hline
\\
$\text{T}_{14}$& $\begin{pmatrix} 
0 & \times\\ 
\times& 0 
\end{pmatrix} $& \\
\\
\hline
\hline
$\mathbf{M}_{s}$ & $\mathbf{Y}_{s}^{1}$ & $\mathbf{Y}_{s}^{2}$\\
\hline
\\
$\text{T}_{23}$& $\begin{pmatrix} 
\times & 0 \\ 
0 & 0 
\end{pmatrix} $ & $\begin{pmatrix} 
0 & 0  \\ 
0 & \times
\end{pmatrix}$\\
\\
\hline
\end{tabular}
\end{minipage}
\begin{minipage}{.3\linewidth}
\setlength{\tabcolsep}{8pt}
\begin{tabular}{ccc}
\hline
 $\mathbf{M}_{D}$ &$\mathbf{Y}_{D}^{1}$ & $\mathbf{Y}_{D}^{2}$\\
\hline
\\
$\text{T}_{45}$&$\begin{pmatrix} 
\times & 0 \\ 
0 & 0 \\
0 & \times
\end{pmatrix}$  & $\begin{pmatrix} 
0 & \times  \\ 
\times & 0 \\
0 & 0 
\end{pmatrix} $\\
$\text{T}_{124}$& $\begin{pmatrix} 
0 & 0 \\ 
0 & 0 \\
\times & 0 
\end{pmatrix}$  & $\begin{pmatrix} 
0 & 0  \\ 
\times & 0 \\
0 & \times 
\end{pmatrix} $\\
$\text{T}_{456}$& $\begin{pmatrix} 
0 & \times \\ 
\times & 0 \\
 0 & 0 
\end{pmatrix}$  & $\begin{pmatrix} 
\times & 0  \\ 
0 & 0 \\
0 & 0 
\end{pmatrix}$ \\
$\text{T}_{136,\text{I}}$& $\begin{pmatrix} 
0 & 0 \\ 
0 & \times \\
0 & 0 
\end{pmatrix}$  & $\begin{pmatrix} 
0 & \times \\ 
0 & 0 \\
\times & 0 
\end{pmatrix}$ \\
$\text{T}_{136,\text{II}}$& $ \begin{pmatrix} 
0 & \times \\ 
0 & 0\\
0 & 0 
\end{pmatrix}$  & $ \begin{pmatrix} 
0 & 0 \\ 
0 & \times \\
\times & 0 
\end{pmatrix} $\\
$\text{T}_{146,\text{I}}$& $\begin{pmatrix} 
0 &  \times \\ 
0 & 0 \\
\times & 0 
\end{pmatrix}$  & $\begin{pmatrix} 
0 & 0  \\ 
\times & 0 \\
0 & 0 
\end{pmatrix} $\\
$\text{T}_{146,\text{II}}$& $\begin{pmatrix} 
0 &  \times \\ 
\times & 0 \\
0 & 0 
\end{pmatrix} $ & $\begin{pmatrix} 
0 & 0  \\ 
0 & 0 \\
\times & 0 
\end{pmatrix}$ \\
\\
\hline
\end{tabular}
\end{minipage}
%\end{ruledtabular}
\caption{ \label{tab:RealizableDecomp} [Left] Maximally-restrictive texture-zero sets compatible with neutrino oscillation data and realisable through Abelian symmetries. [Centre and right] Decomposition of mass matrices into the Yukawa textures according to Eq.~\eqref{eq:massyuk}.}
\end{table}
\renewcommand{\arraystretch}{1}

Applying the SNF method to the texture sets passing the canonical method test, we find that the minimal Abelian symmetry group $\mathbf{G}$ realising such textures is
\begin{equation}
\begin{aligned}
\mathbf{G} = \mathbb{Z}_{n} \times \left[\text{U}(1)\right]^3,\quad n=2,4.
\end{aligned}
\end{equation}
Irrespective of the type of Yukawa textures, the Lagrangian is invariant under the global continuous symmetry $\text{U}(1)_{Y}$, $Y$ being the SM hypercharge. Since, obviously, that $\text{U}(1)_{Y}$ does not impose any texture zero, the actual flavour symmetry group $\mathbf{G_{\text{F}}}= \mathbf{G}/\text{U}(1)_{Y}$ is
\begin{equation}
\begin{aligned}
\mathbf{G_{\text{F}}}  = \mathbb{Z}_{n} \times \left[\text{U}(1)\right]^2,\quad n=2,4.
\end{aligned}
\end{equation}
Furthermore, the Yukawa Lagrangian \eqref{eq:Yuk} is also invariant under the following $\text{U}(1)$ global symmetry,
\begin{equation}
\begin{aligned}
\ell_{L} \rightarrow e^{i q_1} \,\ell_L, \ e_{R} \rightarrow e^{i q_1} \, e_{R},\
\nu_{R} \rightarrow e^{i q_1} \, \nu_{R},\ S_2 \rightarrow e^{i q_1} \, S_2,
\end{aligned}
\label{eq:forbid}
\end{equation}
with the other fields remaining invariant. Although this symmetry does not impose any texture zero on the mass matrices, it restricts the possible coupling terms that can appear in the Lagrangian. Besides the ones given in Eq.~\eqref{eq:Yuk}, only a bare Majorana mass term of the form $s^{T} C s$ is allowed by this symmetry. Nevertheless, it can be shown that the remaining Abelian symmetries forbid such a bare term. The minimal group that discretises this $\text{U}(1)$ symmetry is $\mathbb{Z}_4$, with $q_1 = \pi/2$. Thus, the actual flavour symmetry group is 
\begin{equation}
\begin{aligned}
\mathbf{G_{\text{F}}} = \text{U}(1) \times \mathbb{Z}_{n} \times \text{U}(1)_{\text{F}},\quad n=2,4.
\end{aligned}
\end{equation}

The maximally restrictive texture sets $\left(5^{\ell}_{1,\text{I}},\text{T}_{45}\right), \left(4^{\ell}_3,\text{T}_{124}\right)$ and $\left(4^{\ell}_3,\text{T}_{456}\right)$ are realised by the flavour symmetry for~$n=2$, while $\left(4^{\ell}_3,\text{T}_{136,\text{I}}\right)$ and $\left(4^{\ell}_3,\text{T}_{146,\text{I}}\right)$ are realised by the Abelian symmetry group for $n=4$. The corresponding~$\text{U}(1)_{\text{F}}$ charges can be determined through the canonical method, while the discrete group charges are obtained resorting to the SNF method. We present in Table~\ref{tab:symcharges}, for each texture set, the Abelian symmetry group that realises the set and the associated transformation charges for each field. In all cases, the full texture decomposition is imposed by the $\text{U}(1)_{\text{F}}$ symmetry alone. The discrete groups, $\mathbb{Z}_{2}$ or $\mathbb{Z}_{4}$, only preserve some of the texture zeros but ultimately fail in imposing them totally. Yet, they are crucial in forbidding the bare Majorana mass term for sterile singlet fermions $s$. In fact, the $\text{U}(1)_{\text{F}}$ charges alone only forbid the diagonal elements of the bare mass term, while the charges of the discrete groups forbid the remaining off diagonal elements. Therefore, the Yukawa Lagrangian remains restricted to the form given in Eq.~\eqref{eq:Yuk} where the term with $\mathbf{Y}_R^2$ is forbidden and $\mathbf{Y}_R \equiv \mathbf{Y}_R^1$. As a final comment, let us note that for the realisable cases the $\text{U}(1)_{\text{F}}$ symmetry can be discretised into a minimal set of charges corresponding to a $\mathbb{Z}_5$ symmetry.
\renewcommand{\arraystretch}{1.5}
\begin{table}[!t]
%\begin{ruledtabular}
\renewcommand{\tabcolsep}{12pt}
\begin{tabular}{ccccccc}\hline
 & & $(5^{\ell}_{1,\text{I}},\text{T}_{45})$ & $(4^{\ell}_3,\text{T}_{124})$ & $(4^{\ell}_3,\text{T}_{456})$ & $(4^{\ell}_3,\text{T}_{136,\text{I}})$ & $(4^{\ell}_3,\text{T}_{146,\text{I}})$\\
 Fields &  $\text{U}(1)$ & $\mathbb{Z}_{2} \times \text{U}(1)_{\text{F}}$ & $\mathbb{Z}_{2} \times \text{U}(1)_{\text{F}}$ & $\mathbb{Z}_{2} \times \text{U}(1)_{\text{F}}$ & $\mathbb{Z}_{4} \times \text{U}(1)_{\text{F}}$ &  $\mathbb{Z}_{4} \times \text{U}(1)_{\text{F}}$ \\ \hline
$\Phi_1$ &   0     & $(1,  1)$    & $(0, -5 )$ & $(1, 1)$  & $(1, 2 )$   & $(0,  1)$           \\ %\hline
$\Phi_2$  &  0    & $(0, - 1)$     & $(1, -3 )$& $(0, - 1)$  & $(0,  1)$           & $(3, 0)$       \\ %\hline
$S_1$      &   0  & $(0, 2 )$     & $(0, -2 )$ & $(0, -2)$ & $(0, -2  )$         & $(0, -2  )$         \\ %\hline
$S_2$  &   $1$      & $(0, 0)$      & $(0, 0)$       & $(1, 0)$      & $(0, 0)$               & $(0, 0)$               \\ %\hline
$\ell_{e_L}$ &  $1$   & $(1, 0)$    & $(0, 0)$      & $(0, 0)$      & $(2, 0)$             & $(2, 0)$             \\ %\hline
$\ell_{\mu_L}$ & $1$  & $(0, 2  )$     & $(1, 2  )$ & $(1, -2  )$& $(1, - 1 )$   & $(1, -  1)$   \\ %\hline
$\ell_{\tau_L}$& $1$  & $(0, -2  )$    & $(0, 4  )$  & $(0, -4  )$ & $(0, -2  )$         & $(0, -2)$         \\ %\hline
$e_R$  &   $1$       & $(1, -3  )$  & $(0, 9  )$  & $(1, -5  )$& $(3, -4  )$& $(0, -3)$         \\ %\hline
$\mu_{R}$ &   $1$    & $(0, 3  )$     & $(1, 7  )$ & $(0, -3  )$ & $(0, -3  )$         & $(1, -2)$  \\ %\hline
$\tau_{R}$ & $1$    & $(0, - 1 )$     & $(0, 5  )$  & $(1, - 1 )$ & $(1, -2  )$  & $(2, - 1)$          \\ %\hline
$\nu_{R_{1}}$& $1$    & $(0,  1 )$      & $(0, - 1 )$  & $(0, -1  )$  & $(0, -1  )$          & $(0, - 1)$          \\ %\hline
$\nu_{R_{2}}$ &$1$    & $(1, -1  )$   & $(1,  1  )$  & $(1,  1 )$  & $(2,1   )$          & $(2,  1)$            \\ %\hline
$s_{1}$  &0 & $(1, - 1 )$   & $(1,1   )$  & $(0,1   )$   & $(2,   1)$          & $(2,  1)$            \\ %\hline
$s_{2}$ & 0 & $(0,1   )$      & $(0, - 1 )$  & $(1, -1  )$ & $(0, - 1 )$          & $(0, - 1)$          \\ \hline
\end{tabular}
%\end{ruledtabular}
\caption{\label{tab:symcharges} Maximally-restrictive texture sets realisable through an Abelian symmetry group. For each texture pair, we provide the $\mathbb{Z}_{n}$ charges $q_n$ such that the transformation phases are $e^{2 \pi i q_n/n}$. The U(1) and U(1)$_{\rm F}$ charges are expressed as multiples of the arbitrary charges $q_1$ and $q_{\rm F}$, respectively.}
\end{table}
\renewcommand{\arraystretch}{1}

\section{Lepton masses, mixing and leptonic CPV}
\label{sec:numass}

In the previous section we have identified which of the maximally-restrictive texture sets can be realised through Abelian symmetries in the ISS(2,2) context. As will become clear later (see Section~\ref{sec:pheno}), throughout the rest of this paper we restrict our phenomenological analysis to the combination $\left(5^{\ell}_{1,\text{I}},\text{T}_{45}\right)$. In this case, the charged-lepton mass matrix can be parameterised as
\begin{equation}
5_1^{\ell} : \mathbf{M}_{\ell} = \begin{pmatrix} 
0 & 0 & a_1\\ 
0 & a_3 & 0 \\
a_2 & 0 & a_4     
\end{pmatrix}, \ a_1^2 = \frac{m^2_{\ell_{2}} m^2_{\ell_{3}}}{a_2^2}, \ a_3^2 =  m^2_{\ell_{1}}, \ a_4^2 = \frac{(a_2^2-m^2_{\ell_{2}}) (m^2_{\ell_{3}}-a_2^2)}{a_2^2}, \ m_{\ell_{2}}<a_2<m_{\ell_{3}},
\end{equation}
where $a_i$ can always be made real by phase field redefinitions, and $m_{\ell_{1,2,3}}$ are the charged lepton masses. Note that the charged-lepton state $\ell_1$ is decoupled from the remaining ones. The unitary matrices $\mathbf{V}_{L,R}^\prime$ that diagonalise the Hermitian matrices $\mathbf{H}_{\ell} = \mathbf{M}_{\ell} \mathbf{M}_{\ell}^{\dagger}$ and $\mathbf{H}_{\ell}^\prime = \mathbf{M}_{\ell}^{\dagger} \mathbf{M}_{\ell} $ are given by
\begin{equation}
\mathbf{H}_{\ell} = \begin{pmatrix} 
a_1^2 & 0 & a_1 a_4\\ 
0 & a_3^2 & 0 \\
a_1 a_4 & 0 & a_2^2 + a_4^2     
\end{pmatrix}\;,\; \mathbf{V}_{L}^\prime = \begin{pmatrix} 
c_L & 0 & s_L\\ 
0 & 1 & 0 \\
- s_L & 0 & c_L
\end{pmatrix} \;,\;
\mathbf{H}_{\ell}^\prime = \begin{pmatrix} 
a_1^2 & 0 & a_2 a_4\\ 
0 & a_3^2 & 0 \\
a_2 a_4 & 0 & a_1^2 + a_4^2     
\end{pmatrix}\;,\; \ \mathbf{V}_{R}^\prime = \begin{pmatrix} 
c_R & 0 & s_R\\ 
0 & 1 & 0 \\
- s_R & 0 & c_R
\end{pmatrix}\,.
\end{equation}
Here, we have used the compact notation $c_{L,R}\equiv \cos \theta_{L,R}$ and $s_{L,R}\equiv \sin \theta_{L,R}$ with the angles $\theta_{L,R}$ given by
\begin{equation}
\tan \left(2 \theta_{L}\right) = \frac{2 m_{\ell_{2}} m_{\ell_{3}} \sqrt{(a_2^2-m_{\ell_{2}}^2)(m_{\ell_{3}}^2-a_2^2)}}{a_2^2(m_{\ell_{2}}^2+m_{\ell_{3}}^2)-2 m_{\ell_{2}}^2 m_{\ell_{3}}^2} \;,\;
\tan \left(2 \theta_{R}\right) = \frac{2 \sqrt{(a_2^2-m_{\ell_{2}}^2)(m_{\ell_{3}}^2-a_2^2)}}{m_{\ell_{2}}^2+m_{\ell_{3}}^2-2 a_2^2}.
\end{equation}
We consider the three distinct cases of $5_1^{\ell_1}$ textures with $\ell_{1}=e,\mu,\tau$, labelled  as $5_1^{e,\mu,\tau}$. Since after the diagonalisation of the charged-lepton mass matrix the unitary matrices $\mathbf{V}_{L,R}$ are such that the correct mass ordering is obtained as in Eq.~\eqref{diagcharg}, we have
\begin{equation}
5_1^{e}: \mathbf{V}_{L,R}=\mathbf{V}_{L,R}^\prime \mathbf{P}_{12}, \quad 5_1^{\mu}: \mathbf{V}_{L,R}=\mathbf{V}_{L,R}^\prime, \quad 5_1^{\tau}: \mathbf{V}_{L,R}=\mathbf{V}_{L,R}^\prime \mathbf{P}_{23},
\label{eq:VLdecoupled}
\end{equation}
with
\begin{equation}
\mathbf{P}_{12} = \begin{pmatrix} 
0 & 1 & 0 \\ 
1 & 0 & 0 \\
0 & 0 & 1 
\end{pmatrix} \;,\;\mathbf{P}_{23} = \begin{pmatrix} 
1 & 0 & 0 \\ 
0 & 0 & 1 \\
0 & 1 & 0
\end{pmatrix}.
\label{eq:P12P13}
\end{equation}
As seen in the previous section, for $\mathbf{M}_\ell$ of the type $5_1^{\ell}$, only the $\mathbf{M}_D$, $\mathbf{M}_R$ and $\mathbf{M}_s$ matrices of the type $\text{T}_{45}$, $\text{T}_{14}$ and $\text{T}_{23}$, respectively, lead to maximally-restricted neutrino mass matrices. In terms of the Yukawa matrices given in Eq.~\eqref{eq:Yuk}, those mass matrices are realised through Abelian symmetries for the decompositions (see Table~\ref{tab:RealizableDecomp}) 
\begin{align}
\mathbf{Y}_{D}^1 =\begin{pmatrix} 
b_1 & 0\\
 0 & 0\\
 0 & b_2
\end{pmatrix}\;,\;
\mathbf{Y}_{D}^2 =\begin{pmatrix} 
 0 & b_3\,e^{i\beta_1}\\
 b_4 & 0\\
 0 & 0
\end{pmatrix} \;,\;
\mathbf{Y}_{R}=\begin{pmatrix} 
0  & d_2\\
d_1 & 0
\end{pmatrix}\;,\;
\mathbf{Y}_{s}^{1}=\begin{pmatrix} 
f_2  & 0\\
0 & 0
\end{pmatrix} \;,\;
\mathbf{Y}_{s}^{2}=\begin{pmatrix} 
0  & 0\\
0 & f_1 \,e^{i\beta_2}
\end{pmatrix}\,.
\label{eq:YYY}
\end{align}
Notice that we have rephased the fields to remove the unphysical phases such that $b_i$, $d_i$ and $f_i$ are real and only the phases $\beta_{1,2}$ remain in all Yukawa couplings of Eq.~\eqref{eq:Yuk}. 

From now on, instead of considering the general case of complex Yukawa couplings, we will consider the scenario in which CP is imposed at the Lagrangian level and, thus, $\beta_{1,2}=0$. As shown in Appendix~\ref{sec:ScalarSector}, the scalar potential of the fields $\Phi$ and $S_{1,2}$, with the soft breaking of the ${\rm U(1)}\times \mathbb{Z}_2\times {\rm U(1)}_{\rm F}$ symmetry, allows for a CP-violating vacuum configuration with~\footnote{Spontaneous CP violation can also be successfully communicated to the lepton sector for the texture combinations $\left(4^{\ell}_{3},\text{T}_{124}\right)$ and~$\left(4^{\ell}_{3},\text{T}_{456}\right)$. This is not the case for the pairs $\left(4^{\ell}_{3},\text{T}_{136,\text{I}}\right)$ and $\left(4^{\ell}_{3},\text{T}_{146,\text{I}}\right)$ for which $\mathbf{M}_D$ has one texture zero per row (see Table~\ref{tab:RealizableDecomp}) and the complex phase $\xi$ can be rephased away.}
\begin{align}
    \langle \phi^0_{1} \rangle = v \cos\beta\;,\;
    \langle \phi^0_{2} \rangle = v \sin\beta\;,\; 
    \langle S_1 \rangle=u_1 e^{i\xi} \;,\; \langle S_2 \rangle=u_2\;,\; \tan\beta = \frac{v_2}{v_1}\,.
    \label{eq:tanb}
\end{align}
Together with Eqs.~\eqref{eq:YYY}, these VEVs lead to the mass matrices
\begin{equation}
\mathbf{M}_D = \begin{pmatrix} 
 m_{D_1} & m_{D_3}\\
 m_{D_4} & 0\\
 0 & m_{D_2}
\end{pmatrix}\;,\;\mathbf{M}_R  =
\begin{pmatrix} 
 0 & M\\
 q M & 0
\end{pmatrix}\;,\;
\mathbf{M}_s = 
\begin{pmatrix} 
p\, \mu_s e^{i \xi} & 0\\
0 & \mu_s e^{-i \xi}\end{pmatrix}\,,
\label{eq:MDMRMS}
\end{equation}
being the matrix elements defined as
\begin{equation}
\begin{aligned}
m_{D_{1,2}}=b_{1,2}\, v \cos\beta\;,\quad m_{D_{3,4}}=b_{3,4}\, v \sin\beta\;,\quad
M=d_2 u_2\,,\quad \mu_s=f_1 u_1\,,\quad q=d_1/d_2\,,\quad p=f_2/f_1\,,
\label{eq:pdef}
\end{aligned}
\end{equation}
where $p$ and $q$ are rescaling adimensional parameters, which will be useful for later discussions. Taking into account Eqs.~\eqref{eq:invss} and \eqref{eq:MDMRMS}, the effective neutrino mass matrix in the original symmetry basis reads
\begin{align}
\mathbf{M}_{\rm eff} = 
e^{-i \xi}\begin{pmatrix} 
\dfrac{y^2}{x}+\dfrac{z^2}{w}e^{2i \xi} & y & ze^{2i \xi}\smallskip\\
y & x &0\smallskip\\
z e^{2i \xi}& 0 & w e^{2i \xi}
\end{pmatrix}\,,\; x=\mu_s\frac{m_{D_4}^2}{M^2}\;,\;
y=\mu_s\frac{m_{D_1}m_{D_4}}{M^2}\;,\;
z=\mu_s\frac{m_{D_2}m_{D_3}}{M^2}\dfrac{p}{q^2}\;,\;
w=\mu_s\frac{m_{D_2}^2}{M^2}\dfrac{p}{q^2}\,,
\label{eq:Mxyzw}
\end{align}
with all parameters real. Performing the rotation to the charged-lepton mass basis with one of the unitary matrices $\mathbf{V}_{L}$ given in Eq.~\eqref{eq:VLdecoupled}, we obtain for the $5_1^{e}$ case:
\begin{align}
\V_L^T\,\M_{\rm eff} \V_L= 
 \begin{pmatrix} 
x & y c_L  &y s_L \\
y c_L & \dfrac{y^2 c_L^2}{x}+\dfrac{(zc_L-ws_L)^2}{w}e^{2i \xi} &\dfrac{y^2 s_L c_L}{x}+\dfrac{ \left(z^2-w^2\right)s_Lc_L+
   w z \cos (2 \theta_L)}{w}e^{2 i \xi}\smallskip\\
y s_L &\dfrac{y^2 s_L c_L}{x}+\dfrac{ \left(z^2-w^2\right)s_Lc_L+
   w z \cos (2 \theta_L)}{w}e^{2 i \xi} 
   &\dfrac{y^2 s_L^2}{x}+\dfrac{(wc_L+zs_L)^2}{w}e^{2i \xi}
\end{pmatrix}\,,
\label{eq:MxyzwVL}
\end{align}
while for $5_1^\mu$ and $5_1^{\tau}$ the permutations $\mathbf{P}_{12}$ and $\mathbf{P}_{12}\mathbf{P}_{23}$ of Eq.~\eqref{eq:P12P13} have to be applied on the left and right. The above matrix must be matched with the one defined in terms of the physical low-energy parameters according to Eqs.~\eqref{eq:Unudef}-\eqref{eq:Uparam}, for which the matrix elements are 
\begin{align}
{\rm NO:}&\;M_{ij}= \left[\mathbf{U}'^\ast \text{diag}\left(0 , \sqrt{\dmsol}, \sqrt{\dmatm}\right)  \mathbf{U}'^\dag\right]_{ij}\,,\label{eq:NOMij}\\
{\rm IO:}&\;M_{ij}= \left[\mathbf{U}'^\ast \text{diag}\left(\sqrt{\left|\dmatm\right|} , \sqrt{\dmsol+\left|\dmatm\right|}, 0\right)  \mathbf{U}'^\dag\right]_{ij}\,,
\label{eq:IOMij}
\end{align}
for both NO and IO neutrino masses. The lepton mixing matrix $\mathbf{U}'$ is parametrised as in Eq.~\eqref{eq:Uparam}, with $\alpha_{31}=0$ and $\alpha_{21}\equiv \alpha$ (since one neutrino is massless in our framework - see Section~\ref{sec:compdata}). 

It is clear from Eq.~\eqref{eq:MxyzwVL} that there are six relevant effective parameters ($x$, $y$, $z$, $w$, $\xi$ and $\theta_L$) that determine neutrino masses and mixings. These are to be compared with seven low-energy physical parameters which define $M_{ij}$, namely three mixing angles $\theta_{ij}$, two neutrino masses and two CPV phases (the Dirac and Majorana phases $\delta$ and $\alpha$, respectively). Thus, there is a relation among the elements of the effective neutrino mass matrix, which results in a correlation between two low-energy parameters. It can be shown that, for the $5_1^{e}$ case, the said relation is
\begin{align}
5_1^{e}: 
\,{\rm arg}\left[M_{11}^{\ast 2}M_{13}^2\frac{{\rm D}_{12}}{{\rm D}_{23}}\right]=0\;,\;
{\rm D}_{ij}=M_{ii}M_{jj}-M_{ij}^2\,,
\label{eq:relsLE}
\end{align}
while the corresponding ones for the $5_1^{\mu}$ and $5_1^{\tau}$ cases can be obtained by performing the index replacements $(11\rightarrow 12,13\rightarrow 23)$ and $(11\rightarrow 13,13\rightarrow 33)$, respectively. For a given set of $\theta_{ij}$ and $\Delta m_{21,31}^2$ values, the above equations establish how the two CP-phases $\delta$ and $\alpha$ are correlated. Moreover, we notice that all parameters in Eq.~\eqref{eq:MxyzwVL} can be expressed in terms of low-energy neutrino observables. Indeed, for $5_1^{e}$ we have
\begin{equation}
\begin{aligned}
5_1^{e}&: \tan\theta_L=\left|\frac{M_{13}}{M_{12}}\right|\;,\;
x=|M_{11}|\,,\,\xi={\rm arg}\left[M_{12}^{\ast}\sqrt{{\rm D}_{12}}\right] \;,\;
|y|^2=|D_{23}|x^2\left(\dfrac{\sqrt{|D_{13}|}c_L^2\pm \sqrt{|D_{12}|}s_L^2}{|D_{13}|c_L^2- |D_{12}|s_L^2}\right)^2\,,\\
&w=\frac{\sqrt{|D_{13}|}c_L^2 \mp  \sqrt{|D_{12}|}s_L^2}{x}\;,\;
z=\frac{(|D_{13}|-|D_{12}|)\sin(2\theta_L)\mp 2\sqrt{|D_{13}D_{12}|}\cos(2\theta_L) }{4x}\,.
\label{eq:xyzwdf}
\end{aligned}
\end{equation}
The equivalent expressions for $5_1^{\mu}$ and $5_1^{\tau}$ can be obtained performing the replacements $(11\rightarrow 22, 13 \leftrightarrow 23)$ and $(11\rightarrow 33,13\rightarrow 23,12\rightarrow 13,23 \rightarrow 12)$, respectively.
\begin{figure}[t!]
	\centering
	\begin{tabular}{ccc}
	\includegraphics[width=0.41\textwidth]{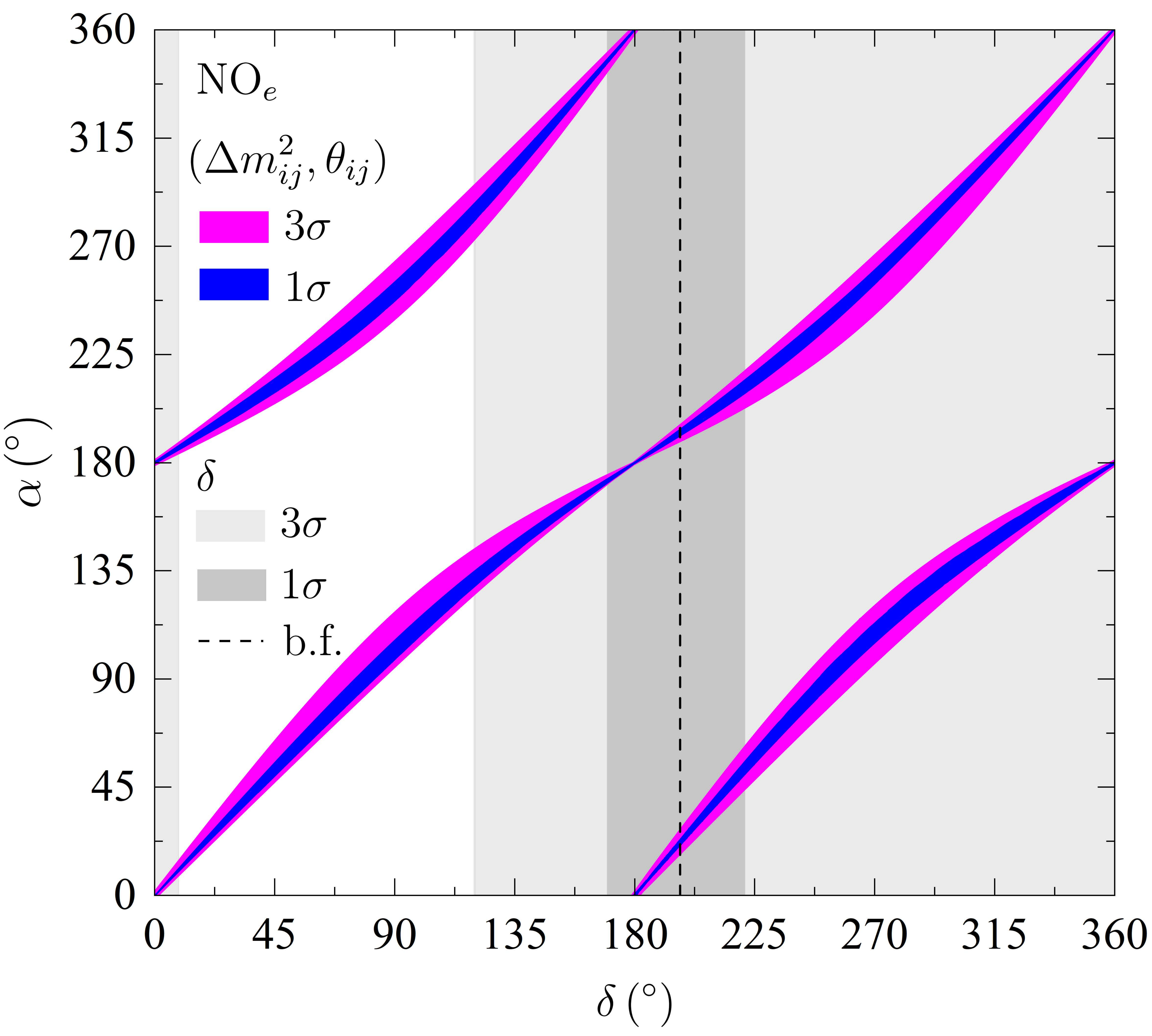}
	&\includegraphics[width=0.41\textwidth]{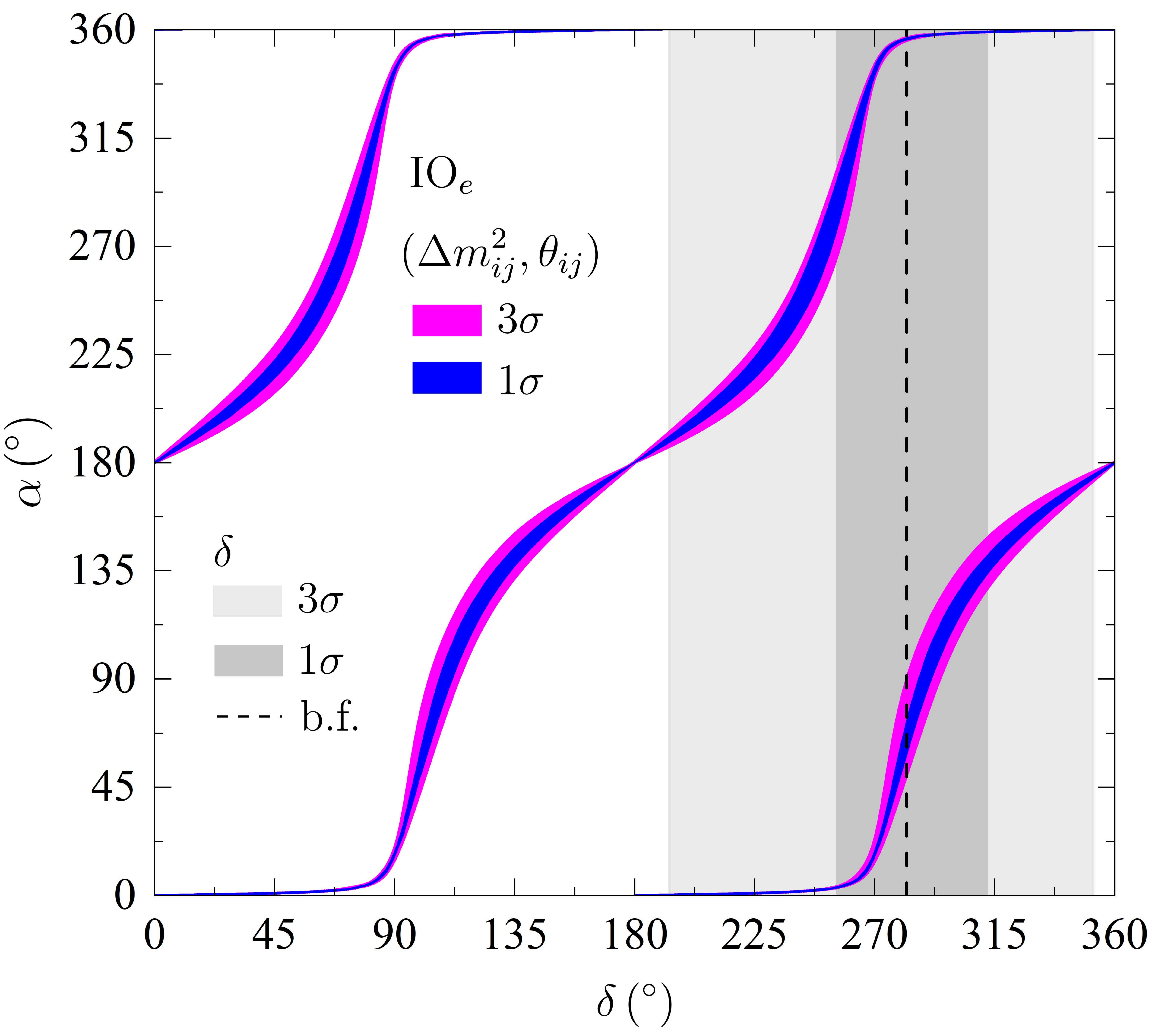}\\
		\includegraphics[width=0.41\textwidth]{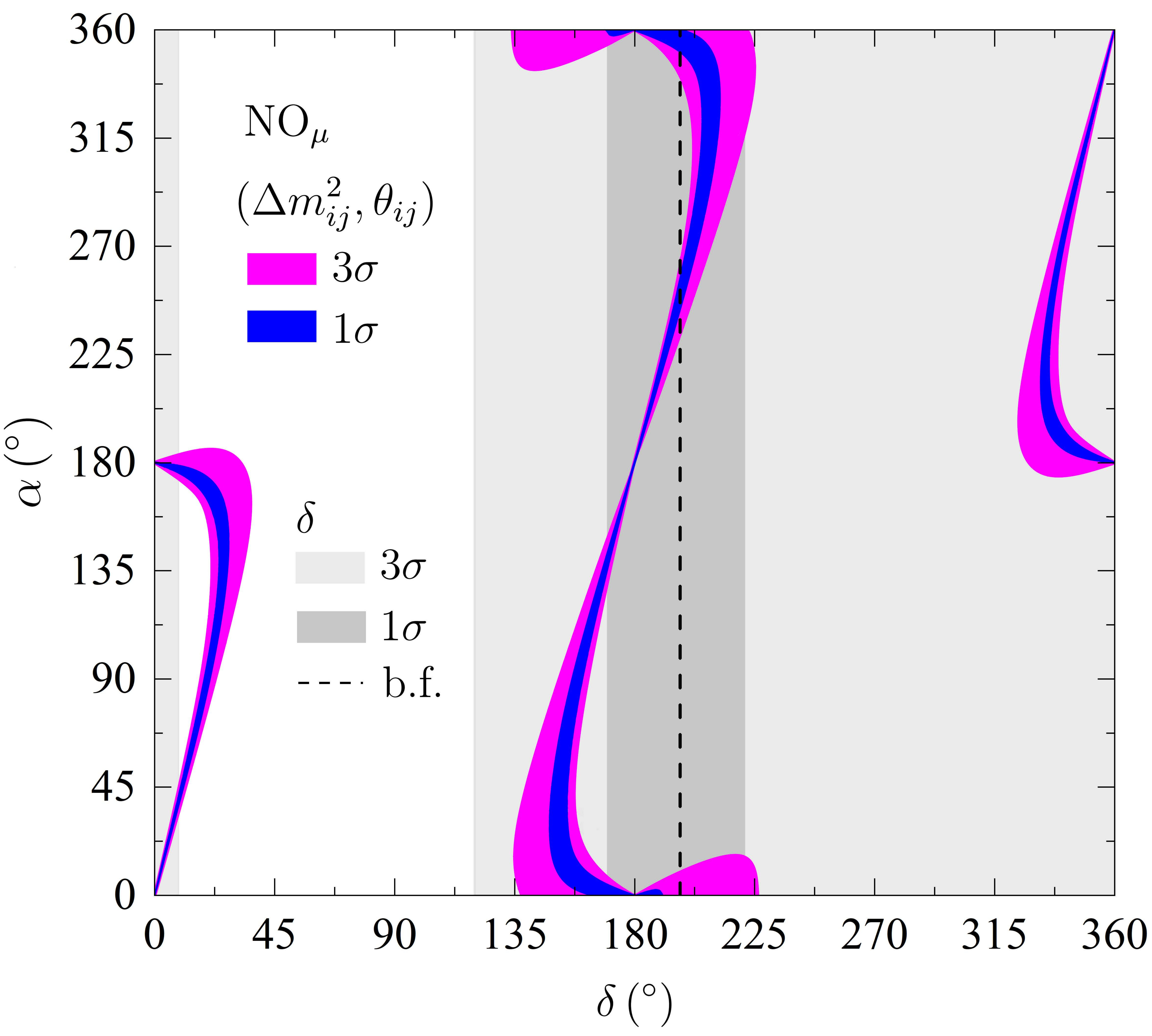}
	&\includegraphics[width=0.41\textwidth]{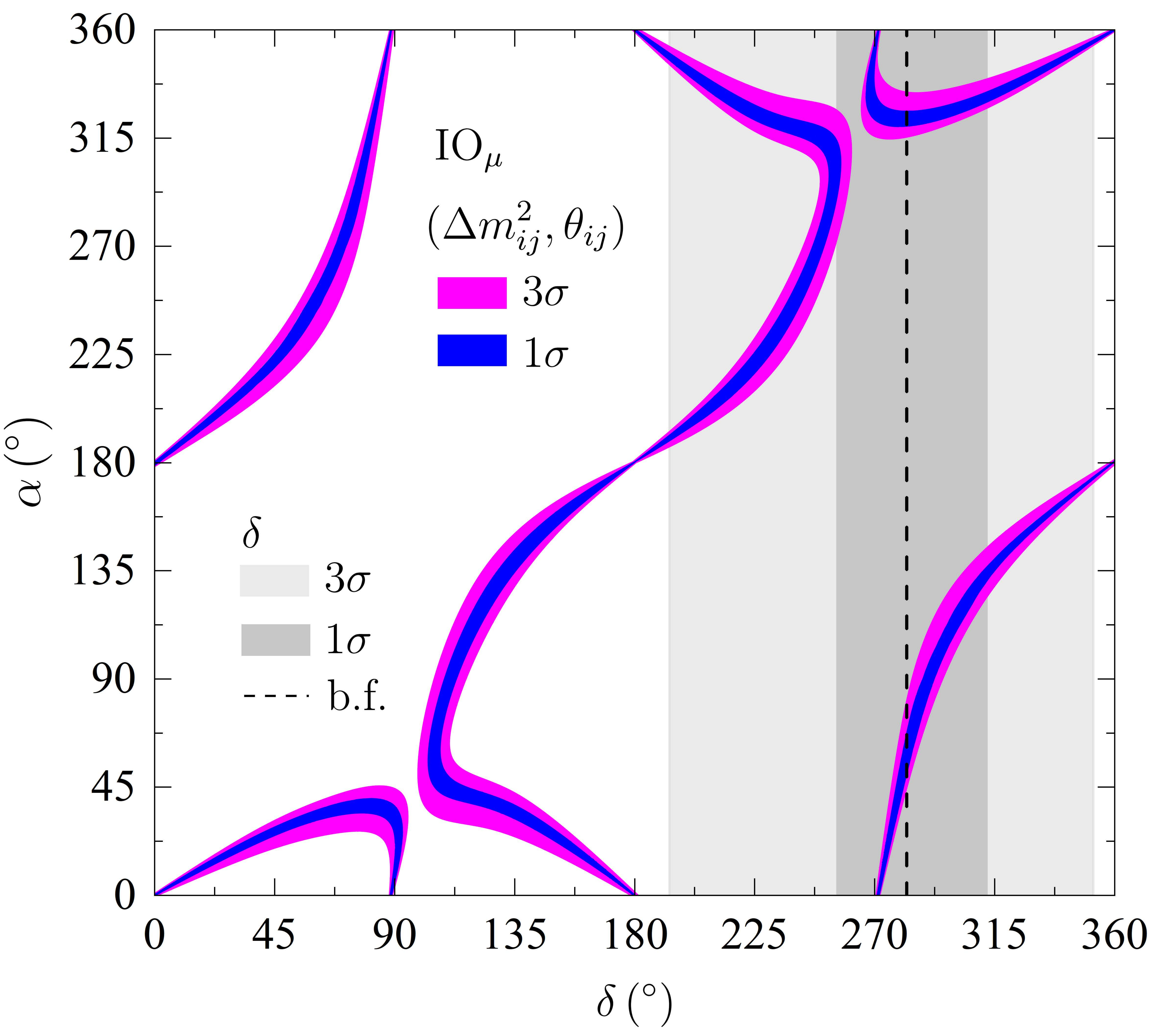}\\
		\includegraphics[width=0.41\textwidth]{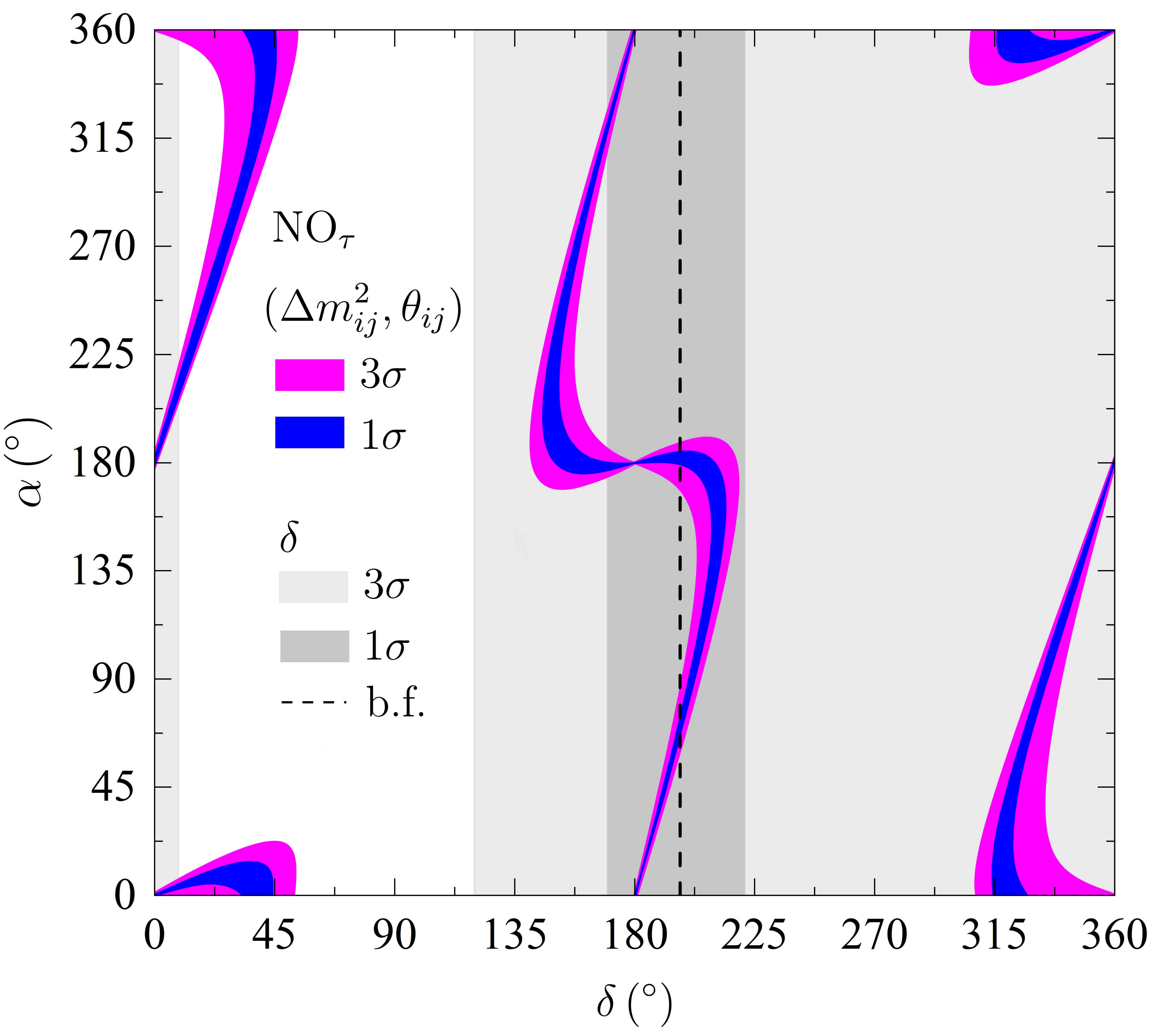}
	&\includegraphics[width=0.41\textwidth]{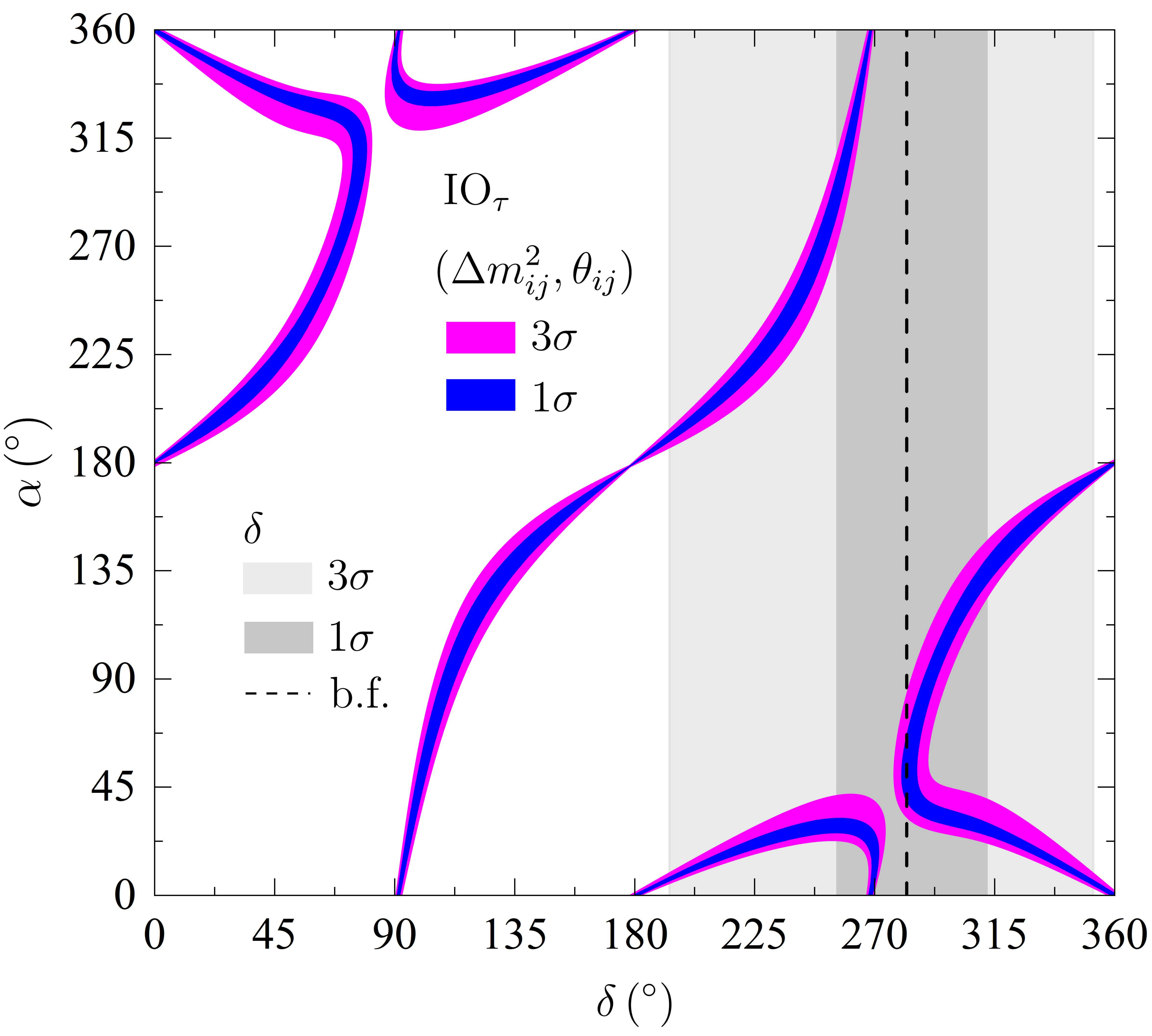}
	\end{tabular}
\caption{Predictions for the Majorana phase $\alpha$ as function of the Dirac CP phase $\delta$ varying the neutrino mixing angles $\theta_{ij}$ and the neutrino mass-squared differences $\dmsol$ and $\dmatm$ in the $1\sigma$ (blue) and $3\sigma$ (magenta) allowed ranges given in Table~\ref{tab:dataref}. The dark (light) grey vertical band marks the $1\sigma$ ($3\sigma$) range for $\delta$ shown in the same table, while the vertical dashed line is at the $\delta$ best-fit value. The left (right) column corresponds to the cases with NO (IO) neutrino mass spectrum and a charged lepton mass matrix of the $5_1^e$ (top), $5_1^\mu$ (middle) and $5_1^\tau$ (bottom) type. Hereafter, these different possibilities will be labelled as NO$_{e,\mu,\tau}$ and IO$_{e,\mu,\tau}$.}
\label{Fig1}
\end{figure} 

In order to establish numerically how $\delta$ and $\alpha$ are related, we vary the mixing angles $\theta_{ij}$ and the neutrino mass-squared differences $\Delta m_{21,31}^2$ within their experimental $1\sigma$ and $3\sigma$ ranges (see Table~\ref{tab:dataref}), while changing $\delta$ from 0 to $2\pi$. Then, for both NO and IO cases, we compute $M_{ij}$ through Eqs.~\eqref{eq:NOMij} and \eqref{eq:IOMij}, respectively. The Majorana phase $\alpha$ is obtained by solving Eq.~\eqref{eq:relsLE} for $5_1^{e,\mu,\tau}$, leading to the results presented in Fig.~\ref{Fig1}, where in the left (right) column we show the allowed regions in the $(\delta,\alpha)$ plane for the NO$_{e,\mu,\tau}$ (IO$_{e,\mu,\tau}$) corresponding to $5_1^{e,\mu,\tau}$ with a NO (IO) neutrino mass spectrum. The blue (magenta) regions were obtained taking the $1\sigma$ ($3\sigma$) intervals for $\theta_{ij}$ and $\Delta m_{21,31}^2$, while the vertical dark (light) grey band marks the current experimentally allowed region for the Dirac CP phase $\delta$ at $1\sigma$ $(3\sigma)$. The results show that there is a strong correlation between $\alpha$ and $\delta$. For both NO and IO mass ordering, the plots exhibit an approximate symmetry under the shift $\delta \rightarrow \delta + \pi$, which is due to the fact that Eq.~\eqref{eq:relsLE} is nearly invariant under that transformation at zeroth order in the smallest mixing angle $\theta_{13}$. Note that the absence of Dirac-type CP violation $(\delta=0,\pi)$ implies $\alpha=k\pi\,(k \in \mathbb{Z})$. This can be confirmed analytically by evaluating the Dirac and Majorana CP weak basis invariants, $\mathcal{J}^{\rm CP}_{\rm Dirac}$ and $\mathcal{J}^{\rm CP}_{\rm Maj}$~\cite{Branco:2011zb}, which are both proportional to $\sin(2\xi)$. Notice also that a future measurement of $\delta$ in the intervals $[45^\circ,135^\circ]$ and $[225^\circ,315^\circ]$ would exclude the NO$_\mu$ and NO$_\tau$ cases since, in these ranges, Eq.~\eqref{eq:relsLE} has no solution.

\begin{figure}[t!]
% 	\centering
% 	\begin{tabular}{lll}
% 	\includegraphics[width=0.31\textwidth]{fig2a.pdf}
% 	&\includegraphics[width=0.2742\textwidth]{fig2b.pdf} &
% 		\includegraphics[width=0.394\textwidth]{fig2c.pdf}
%	\end{tabular}
\begin{tabular}{cc}
	\includegraphics[width=0.43\textwidth]{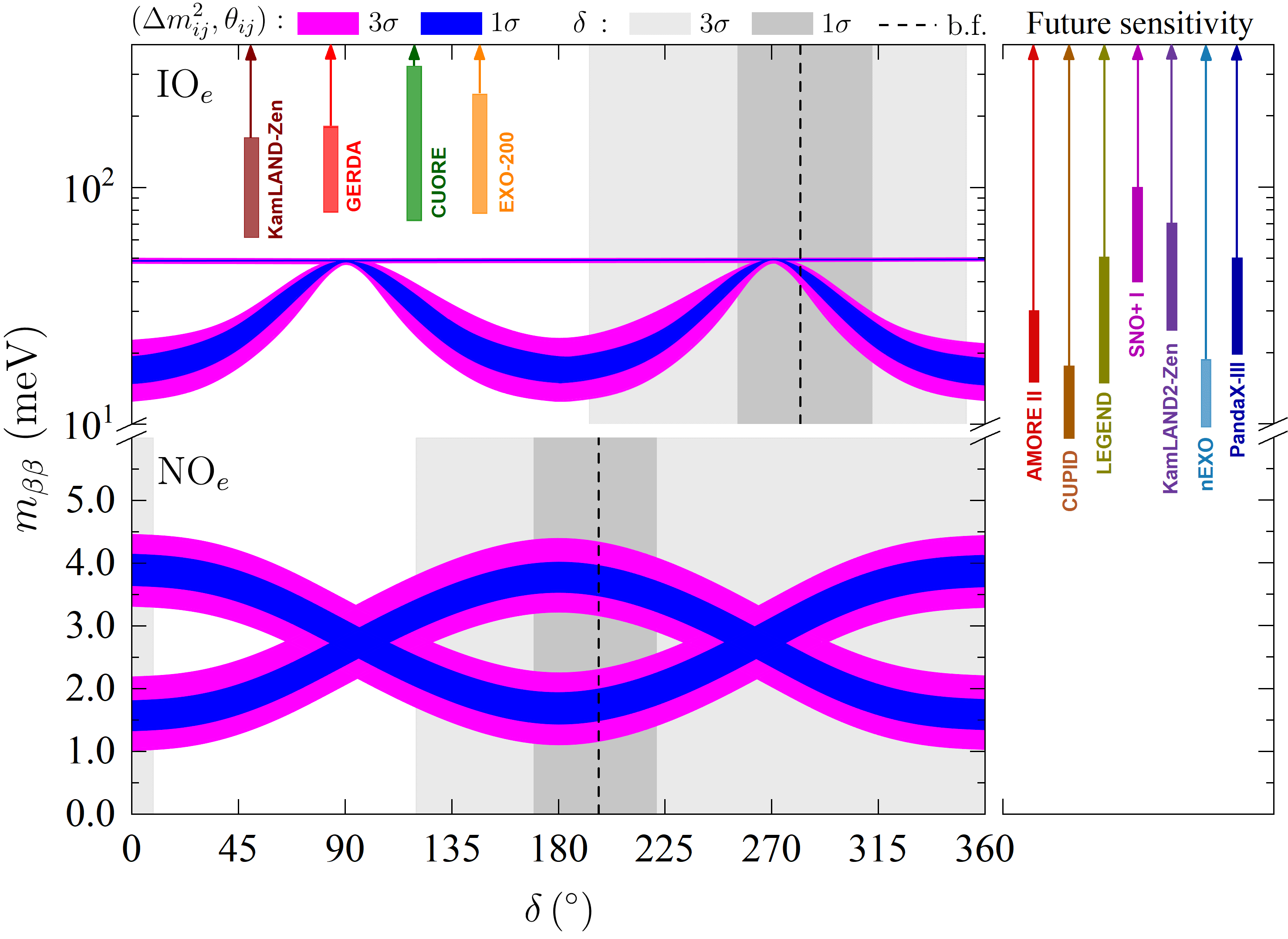}  &\includegraphics[width=0.427\textwidth]{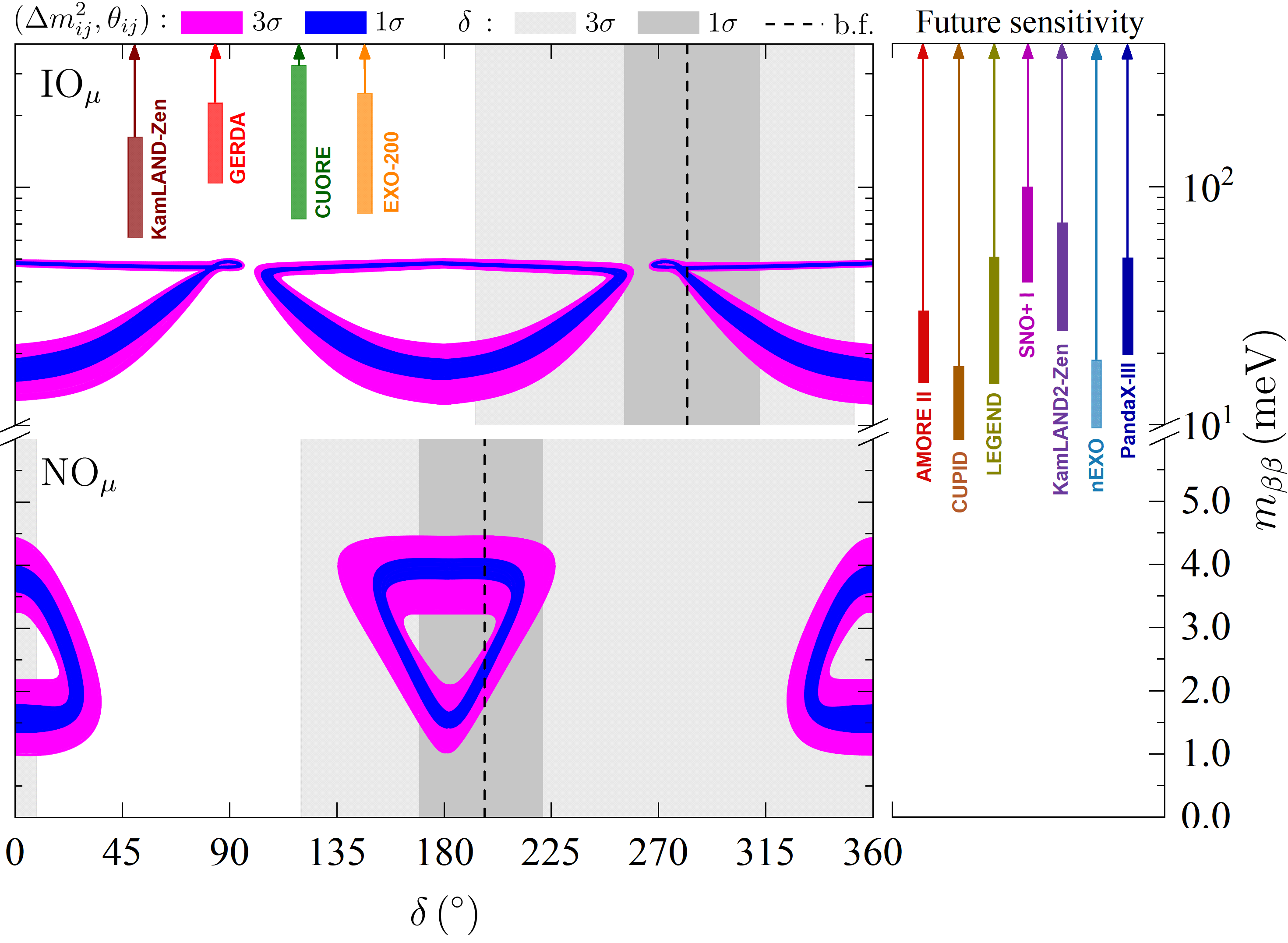}\\
\multicolumn{2}{c}{\includegraphics[width=0.47\textwidth]{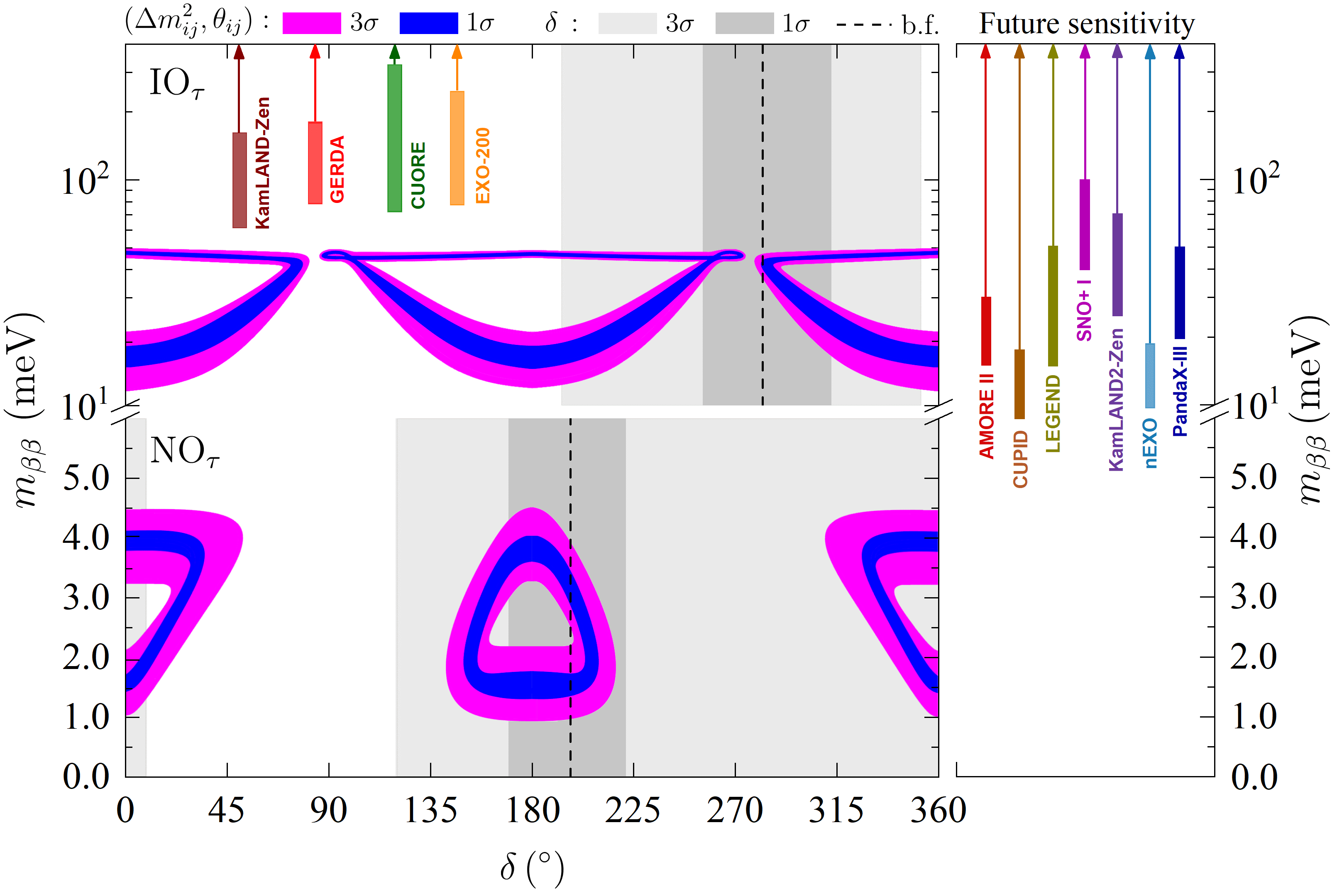}}
	\end{tabular}
\caption{Effective neutrino mass parameter $m_{\beta\beta}$ as function of the Dirac CP-violating phase for the $5_1^e$ (top-left panel), $5_1^\mu$ (top-right panel) and $5_1^\tau$ (bottom) cases (the colour codes are the same as in Fig.~\ref{Fig1}). In each panel we show the NO and IO results in the lower and upper fraction of the vertical scale, respectively. We also show the upper bounds on $m_{\beta\beta}$ reported by the KamLAND-Zen~\cite{KamLAND-Zen:2016pfg}, GERDA~\cite{Agostini:2020xta}, CUORE~\cite{Adams:2019jhp} and EXO-200~\cite{Anton:2019wmi} collaborations (the height of the corresponding rectangles represent the uncertainty in the upper bounds). On the right the sensitivity of several future experiments is shown.}
\label{Fig2}
\end{figure}

The fact that the Majorana and Dirac CP-violating phases are related brings up an interesting connection between neutrinoless double beta decay $(\beta\beta_{0\nu})$ and neutrino oscillations. With one massless neutrino, the $\beta\beta_{0\nu}$ widths are proportional to the effective neutrino mass parameter $m_{\beta\beta}$ that, in the most general case, should only be sensitive to Majorana phases. Namely, for the NO and IO cases~\cite{Joaquim:2003pn} one has
\begin{align}
\text{NO:}\;\;	m_{\beta\beta}=&\left|\sqrt{\dmsol}\,c_ {13}^2s_{12}^2\,e^{-2i\alpha}+\sqrt{\dmatm}\, s_{13}^2\right|\,, \label{NOmbb}\\
\text{IO:}\;\;	m_{\beta\beta}=&\,c_{13}^2\left|\sqrt{\left|\dmatm\right|}\,c_{12}^2+\sqrt{\dmsol+\left|\dmatm\right|}\,s_{12}^2\,e^{-2i\alpha}\right|\label{IOmbb}\,.
\end{align}
Given that $\alpha$ is related to $\delta$ through Eqs.~\eqref{eq:relsLE}, a dependence of $m_{\beta\beta}$ on $\delta$ can be established, as shown in Fig.~\ref{Fig2} for the same cases treated in Fig.~\ref{Fig1}. The present upper limits on $m_{\beta\beta}$ reported by the KamLAND-Zen~\cite{KamLAND-Zen:2016pfg}, GERDA~\cite{Agostini:2020xta}, CUORE~\cite{Adams:2019jhp} and EXO-200~\cite{Anton:2019wmi} collaborations are also shown (the height of the bars reflects the uncertainties in the nuclear matrix elements relevant for the computation of the decay rates). We also show the future $m_{\beta\beta}$ sensitivities planned by the projects AMORE II~\cite{Lee:2020rjh}, CUPID~\cite{Wang:2015raa}, LEGEND~\cite{Abgrall:2017syy}, SNO+~I~\cite{Andringa:2015tza}, KamLAND2-Zen~\cite{KamLAND-Zen:2016pfg}, nEXO~\cite{Albert:2017hjq} and PandaX-III~\cite{Chen:2016qcd}. Our results show that, although $m_{\beta\beta}$ is always below the current bounds (even taking the less conservative limits), several future experiments will be able to probe the whole $m_{\beta\beta}$ for IO neutrino masses. In particular, for the current best-fit values given in Table~\ref{tab:dataref}, we have $m_{\beta\beta}\simeq 40$~meV, which is on the upper IO region. As usual, future experiments with sensitivities of (1.0 - 4.5)~meV will be needed to probe the NO regime (see e.g. Refs.~\cite{Barabash:2019suz} and \cite{Dolinski:2019nrj} for reviews on future prospects of $\beta\beta_{0\nu}$ experiments).

We will now study the properties of the sterile neutrino sector, particularly its spectrum and mixing with active light neutrinos (heavy-light mixing). Although for our phenomenological analysis a full numerical computation will be carried out, we first provide an analytical insight following the discussion in Section~\ref{sec:ISS} -- see Eqs.~\eqref{eq:3x7}-\eqref{eq:BC}. According to the definition of the effective heavy neutrino mass matrix $\mathbf{M}_{\text{heavy}}$ given in Eq.~\eqref{eq:paramU1}, and taking into account $\M_R$ and $\M_s$ from Eq.~\eqref{eq:MDMRMS}, we have for the $4\times4$ unitary matrix $\U_s$:
\begin{equation}
\mathbf{U}_{s} =
\begin{pmatrix} 
- i  c_1 e^{-i \xi/2} &  s_1e^{-i \xi/2} & 0 & 0\\
0 & 0 & - i  c_2e^{i \xi/2} &  s_2e^{i \xi/2} \\
0 & 0 & i  s_2e^{-i \xi/2} &  c_2e^{-i \xi/2} \\
i  s_1 e^{i \xi/2}&  c_1e^{i \xi/2} & 0 & 0
\end{pmatrix}\,,
\label{eq:USmat}
\end{equation}
where $c_{1,2}\equiv \cos \varphi_{1,2}$, $s_{1,2}\equiv \sin \varphi_{1,2}$ and
\begin{equation}
\tan \left(2 \varphi_{1}\right) = \frac{2 M}{\mu_s} \; , \;  \ \tan \left(2 \varphi_{2}\right) = \frac{2 q M}{p \mu_s}\,.
\end{equation}
In the ISS approximation, $\varphi_{1} \simeq \varphi_{2} \simeq \pi/4$ leading to $ c_1 \simeq s_1 \simeq c_2 \simeq s_2 \simeq 1/\sqrt{2}$, which characterises the two pairs of pseudo-Dirac neutrinos with masses~\footnote{The sterile neutrinos are labelled from 4 to 7 in increasing order of their mass. Thus, in the following equations the $-$ (+) sign refers to the lightest (heaviest) neutrino mass, labelled with indices 4 and 6 (5 and 7).}
\begin{align}
\tilde{m}_{4,5} \simeq M \left[\sqrt{1+\left(\frac{\mu_s}{2 M}\right)^2} \mp \frac{\mu_s}{2 M} \right]\,,\quad 
\tilde{m}_{6,7} \simeq q M \left[\sqrt{1+\left(\frac{p \mu_s}{2 q M}\right)^2} \mp \frac{p \mu_s}{2 q M} \right]\,,
\end{align}
and mass differences $\tilde{m}_5-\tilde{m}_4= \mu_s \; , \; \tilde{m}_7-\tilde{m}_6= p \mu_s$. As expected, the degree of degeneracy between the masses is controlled by the small LNV parameter $\mu_s$. 

From Eqs.~\eqref{eq:VLW}, \eqref{eq:VLdecoupled} and \eqref{eq:USmat}, we obtain for the heavy-light neutrino mixing defined in Eq.~\eqref{eq:BC}:
\begin{equation}
\B=
\begin{pmatrix} 
\times & \times &\times  &i \dfrac{m_{D_{4}}}{M} s_1 e^{i \xi/2}  &  \dfrac{m_{D_{4}}}{M} c_1e^{i \xi/2} & 0 & 0\\
\times & \times &\times  &i  \dfrac{m_{D_1}}{M} s_1 c_Le^{i \xi/2} & \dfrac{m_{D_1}}{M} c_1 c_L e^{i \xi/2} & i  \dfrac{(m_{D_{3}} c_L - m_{D_{2}} s_L)}{q M}s_2e^{-i \xi/2}  &   \dfrac{(m_{D_{3}} c_L - m_{D_{2}} s_L)}{q M}c_2e^{-i \xi/2}\\
\times & \times &\times  &i \dfrac{m_{D_1}}{M} s_1 s_L e^{i \xi/2} &  \dfrac{m_{D_1}}{M} c_1 s_L e^{i \xi/2} & i   \dfrac{(m_{D_{3}} s_L + m_{D_{2}} c_L)}{q M} s_2e^{-i \xi/2}&  \dfrac{(m_{D_{3}} s_L + m_{D_{2}} c_L)}{q M}c_2e^{-i \xi/2} 
\end{pmatrix}\,,
\label{eq:Baj}
\end{equation}
in the $5_1^{e}$ case. The corresponding results for $5_1^{\mu}$ and $5_1^{\tau}$ are obtained applying on the left of the above matrix the permutation $\mathbf{P}_{12}$ and $\mathbf{P}_{12}\mathbf{P}_{23}$ of Eq.~\eqref{eq:P12P13}, respectively. Notice that each neutrino in a quasi-Dirac pair couples similarly to each lepton flavour $\alpha$, i.e. $\B_{\alpha 4}\simeq \B_{\alpha 5}$ and $\B_{\alpha 6}\simeq \B_{\alpha 7}$. Since we are working in the charged-lepton mass basis, the above matrix provides an approximation to the mixing among the charged lepton $e_\alpha$ and the sterile neutrinos $\nu_{4-7}$. It turns out that, due to the Abelian symmetries imposed to realise the maximally-restricted textures, the charged-current (CC) flavour mixings for distinct lepton flavours are related by low-energy neutrino parameters. Indeed, for the $5_1^e$ case,
\begin{align}
\frac{\B_{e4}}{\B_{\mu 4}}\simeq \frac{\B_{e5}}{\B_{\mu 5}}
\simeq \frac{x}{y c_L}\;,\quad
\frac{\B_{\tau 4}}{\B_{\mu 4}}\simeq \frac{\B_{\tau 5}}{\B_{\mu 5}} 
\simeq \tan\theta_L\,,\quad
\frac{\B_{\mu 6}}{\B_{\tau 6}}\simeq \frac{\B_{ \mu 7}}{\B_{\tau 7}} \simeq \frac{z-w\tan\theta_L}{w+z \tan\theta_L}\;,\quad \B_{e 6}\simeq \B_{e 7}\simeq 0\,,
\label{eq:Brels}
\end{align}
where all parameters involved depend on the neutrino observables as shown in Eq.~\eqref{eq:xyzwdf}. The corresponding relations for the $5_1^\mu$ and $5_1^\tau$ textures are obtained by performing the replacements ($e \leftrightarrow \mu$) and ($e \leftrightarrow \tau$), respectively. In Table~\ref{tab:BREL} we show the numerical values of some heavy-light mixing ratios using the best-fit values for the low-energy neutrino parameters given in Table~\ref{tab:dataref}.
\begin{table}[!t]
\renewcommand{\arraystretch}{1.5}
\setlength{\tabcolsep}{8pt}
\centering
\begin{tabular}{l|ccc|ccc}   
\hline
 & $\text{NO}_{e}$ & $\text{NO}_{\mu}$ & $\text{NO}_{\tau}$ &  $\text{IO}_{e}$ & $\text{IO}_{\mu}$ & $\text{IO}_{\tau}$ \\ \hline
$\B_{e 4}/\B_{\mu 4} \simeq \B_{e 5}/\B_{\mu 5}$  & 0.21 & 0.17 & 0.17 & 2.73 & 0.21 & 0.41\\ \hline
$\B_{\tau 4}/\B_{\mu 4} \simeq \B_{\tau 5}/\B_{\mu 5}$  & 0.27 & 0.88 & 0.87 & 0.51 & 1.09 & 1.24\\ \hline
$\B_{\tau 4}/\B_{e 4} \simeq \B_{\tau 5}/\B_{e 5}$  & 1.27 & 5.07 & 5.24 & 0.19 & 5.33 & 5.02\\ \hline
$\B_{e 6}/\B_{\mu 6} \simeq \B_{e 7}/\B_{\mu 7}$  & 0 &  $-$ & 0.36 &  0 &  $-$ & 4.96\\ \hline
$\B_{\tau 6}/\B_{\mu 6} \simeq \B_{\tau 7}/\B_{\mu 7}$  & 0.61 & $-$ & 0 & 1.14 & $-$ & 0\\ \hline
$\B_{\tau 6}/\B_{e 6} \simeq \B_{\tau 7}/\B_{e 7}$  & $-$ & 1.64 & 0 & $-$  & 0.23 & 0\\ \hline
\end{tabular}
\caption{Predictions for ratios of heavy-light mixing parameters $\B_{\alpha j}$ computed using Eq.~\eqref{eq:Brels}. The results are shown for the NO$_{e,\mu,\tau}$ and IO$_{e,\mu,\tau}$ cases.}
\label{tab:BREL}
\end{table}

\section{Radiative corrections to neutrino masses}
\label{sec:radcorr}

\begin{figure}[t!]
\centering
\includegraphics[width=0.65\textwidth]{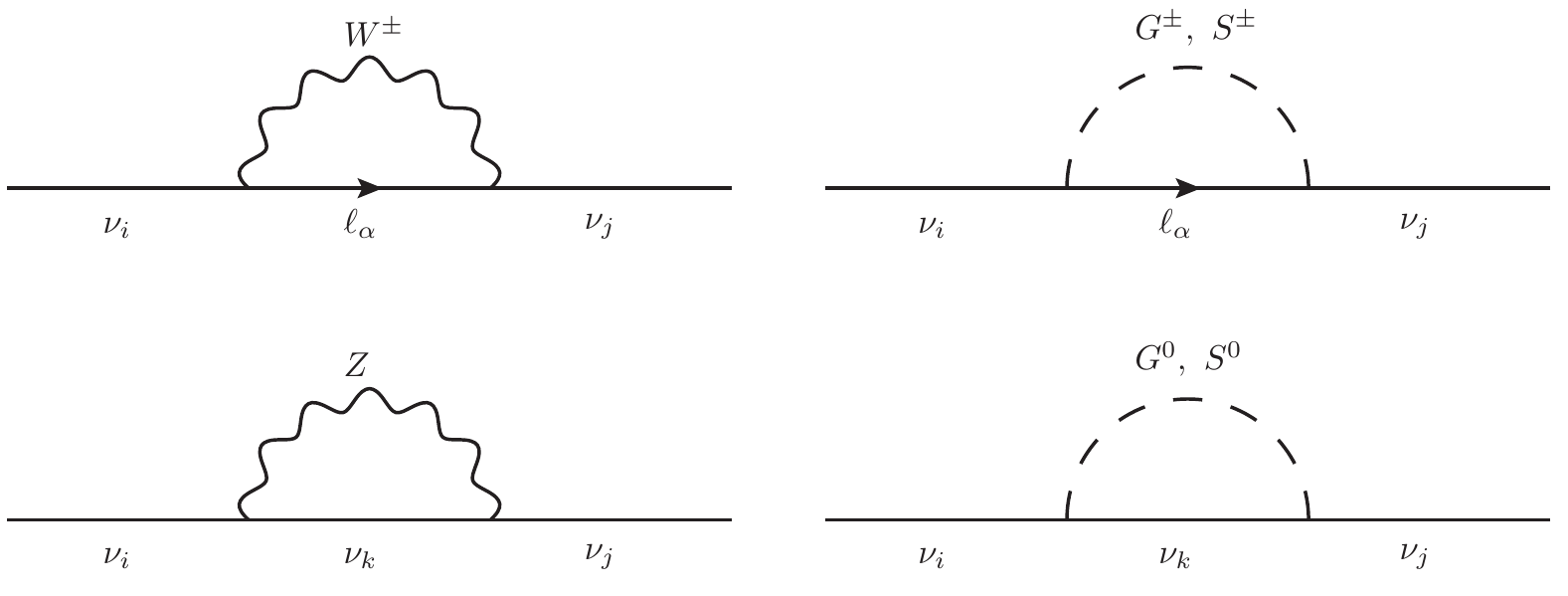}
\caption{Self-energy Feynman diagrams contributing to the radiative corrections to the neutrino masses.}
\label{fig:selfenergy}
\end{figure}

The analysis presented in the previous section was based on the assumption that the (tree-level) ISS approximation for the neutrino mass matrix given in Eq.~\eqref{eq:invss} is valid, i.e., the parameters in Eq.~\eqref{eq:MDMRMS} are such that $\mu_s,m_{D_i} \ll M$. However, the presence of new fermions and scalars may induce relevant corrections to the light-neutrino masses that should not be overlooked. This matter becomes even more important when considering the  high precision achieved in the determination of the oscillation parameters, which is currently at the level of a few per cent for some of those observables (see Table~\ref{tab:dataref}). The one-loop radiative corrections to neutrino masses have been computed in several works~\cite{Pilaftsis:1991ug,Pilaftsis:2002nc,Grimus:2002nk,Dev:2012sg}. Here we revisit the calculation of the one-loop corrections to the light neutrino mass matrix, and adapt it to our case (the one-loop neutrino self-energy diagrams are succinctly presented in Fig.~\ref{fig:selfenergy}). We compute the~$3 \times 3$ one-loop correction matrix $\delta \bm{\mathcal{M}}$ given by~\cite{Pilaftsis:1991ug,Pilaftsis:2002nc}
\begin{equation}
\begin{aligned}
\delta \bm{\mathcal{M}}_{ij} = \mathbf{\Sigma}_{ij}(\slashed{p})\big|_{\slashed{p} = 0} &=  (\delta \bm{\mathcal{M}}_{W^\pm})_{ij} + (\delta \bm{\mathcal{M}}_{G^\pm})_{ij} + \sum_{a=1}^{n_\pm-1} (\delta \bm{\mathcal{M}}_{S^\pm_a})_{ij} \\
& + (\delta \bm{\mathcal{M}}_{Z})_{ij} + (\delta \bm{\mathcal{M}}_{G^0})_{ij} + \sum_{a=1}^{n_0-1} (\delta \bm{\mathcal{M}}_{S^0_a})_{ij} \;,\quad i,j=1,2,3\,,
\label{eq:deltaM}
\end{aligned}
\end{equation}
where $\mathbf{\Sigma}(\slashed{p})$ are the (active) neutrino self-energies, evaluated at $\slashed{p} = 0$ since the tree-level light neutrino masses are extremely small. The number of neutral (charged) scalar mass eigenstates $S^0_a$ ($S^\pm_a$)  is denoted by $n_0-1$ and $(n_\pm-1)$, respectively. First, note that the $W^{\pm}$-boson contribution vanishes since the self-energies are evaluated at $\slashed{p} = 0$. Furthermore, by using the unitarity of the full~$n_f \times n_f$ matrix $\bm{\mathcal{U}}$ defined in Eq.~\eqref{diagnutr}, it can be shown that the $G^{\pm}$ and $S^{\pm}_a$ contributions to $\delta \bm{\mathcal{M}}$ also vanish~\cite{Grimus:2002nk}. Thus, only $\delta \bm{\mathcal{M}}_{Z}$, $\delta \bm{\mathcal{M}}_{G^0}$ and $\delta \bm{\mathcal{M}}_{S^0_a}$ will be relevant. Performing the computation in the basis where the tree-level neutrino mass matrix $\bm{\mathcal{M}}$ given in Eq.~\eqref{diagnutr} is diagonal, we obtain
\begin{align}
&\left(\delta \bm{\mathcal{M}}_{Z}\right)_{ij}= - \frac{\alpha_W}{4 \pi c_W^2} \sum_{k=1}^{n_f} m_k \bm{\mathcal{C}}^{*}_{i k} \bm{\mathcal{C}}_{k j} \ f\left(\frac{m_k^2}{M_Z^2}\right)\;, \\
&\left(\delta \bm{\mathcal{M}}_{G^0}\right)_{i j}= \frac{\alpha_W}{16 \pi} \sum_{k=1}^{n_f} \frac{m_k}{M_W^2} \left( m_i \bm{\mathcal{C}}_{i k}+ m_k \bm{\mathcal{C}}^{*}_{i k}\right) \left(m_j \bm{\mathcal{C}}_{k j}^{*} + m_k \bm{\mathcal{C}}_{k j} \right) \ f\left(\frac{m_k^2}{M_Z^2}\right)\,,\\
&\big(\delta \bm{\mathcal{M}}_{S^0_a}\big)_{i j}= - \frac{1}{4 \pi^2} \sum_{k=1}^{n_f} m_k \left(\mathbf{\Delta}_{\nu}^a\right)^{*}_{i k} \left(\mathbf{\Delta}_{\nu}^a\right)^{*}_{k j} \ f\left(\frac{m_k^2}{m_{S^0_a}^2}\right).
\end{align}
The couplings $\bm{\mathcal{C}}_{ij}$ are defined in Eq.~\eqref{eq:BC}, $m_k$ are the tree-level neutrino masses, and $m_{S^0_a}$ are the neutral scalar masses. The Yukawa couplings matrices $\mathbf{\Delta}_{\nu}^a$ appear in the interaction terms between neutrinos and neutral scalars $S^0_a$ -- see Eq.~\eqref{eq:genericS0} of Appendix~\ref{sec:Allint}. The loop function in the above expressions is
\begin{equation}
f(x)=-\frac{x \ln x}{1-x}-1.
\end{equation}

In our framework, the neutral scalar degrees of freedom originate from the two Higgs doublets $\Phi_{1,2}$ and the two scalar singlets $S_{1,2}$ (see Table~\ref{tab:symcharges}), which results in a total of seven scalar mass eigenstates ($S_{1-7}$) (see Appendix~\ref{sec:ScalarSector} for more details on the scalar sector). In the exact alignment limit with no mixing among scalar doublets and singlets, and following the same argument as before regarding the unitarity of $\bm{\mathcal{U}}$~\cite{Grimus:2002nk}, it can be shown that the $S^0_{3,4,6,7}$ contributions to $\delta \bm{\mathcal{M}}$ vanish. Hence, only the (SM Higgs boson) $H^0$, $R$ and $I$ will be relevant, being their contributions given by
\begin{align}
\left(\delta \bm{\mathcal{M}}_{H^0}\right)_{i j}&= -\frac{\alpha_W}{16 \pi}\sum_{k=1}^{n_f} \frac{m_k}{M_W^2} \left( m_i \bm{\mathcal{C}}_{i k}+ m_k \bm{\mathcal{C}}^{*}_{i k}\right) \left(m_j \bm{\mathcal{C}}_{k j}^{*} + m_k \bm{\mathcal{C}}_{k j} \right) \ f\left(\frac{m_k^2}{m_{H^0}^2}\right)\,,
\label{eq:deltaH0} \\
\left(\delta \bm{\mathcal{M}}_{R}\right)_{i j}&= -\frac{\alpha_W}{16 \pi}\sum_{k=1}^{n_f} \frac{m_k}{M_W^2} \left[\left(\mathbf{N}_{\nu}^\dagger\right)_{i k} + \left(\mathbf{N}_{\nu}^*\right)_{i k}\right] \left[\left(\mathbf{N}_{\nu}^\dagger\right)_{k j} + \left(\mathbf{N}_{\nu}^*\right)_{k j}\right] \ f\left(\frac{m_k^2}{m_{R}^2}\right)\,
\label{eq:deltaR}, \\
\left(\delta \bm{\mathcal{M}}_{I}\right)_{i j}&= \frac{\alpha_W}{16 \pi}\sum_{k=1}^{n_f} \frac{m_k}{M_W^2} \left[\left(\mathbf{N}_{\nu}^\dagger\right)_{i k} + \left(\mathbf{N}_{\nu}^*\right)_{i k}\right] \left[\left(\mathbf{N}_{\nu}^\dagger\right)_{k j} + \left(\mathbf{N}_{\nu}^*\right)_{k j}\right] \ f\left(\frac{m_k^2}{m_{I}^2}\right)\,,
\label{eq:deltaI}
\end{align}
where the matrix $\mathbf{N}_{\nu}$ is defined in Eq.~\eqref{eq:Nnu}. Notice that, assuming $m_{R} = m_{I}$, the $R$ and $I$ contributions cancel each other. This is what we will assume in our analysis (we refer the reader to Ref.~\cite{Aeikens:2020acn} for a model where these neutral scalar contributions are taken into account). As for divergences, those coming from the $G^0$ and $H^0$ loops cancel each other, likewise for the $R$ and $I$ scalars. The $Z$-boson contribution is finite by itself due to the first relation in Eq.~\eqref{eq:equality2}. 

We now wish to evaluate how the results of the previous section are affected when performing a numerical analysis to compute neutrino mass and mixing observables, including the loop corrections discussed above. In particular, we are interested in
\begin{itemize}
 \item Comparing the tree-level light neutrino parameters obtained using the seesaw-approximated $\M_{\rm eff}$ of Eq.~\eqref{eq:invss} with those stemming from the full neutrino mass matrix $\bm{\mathcal{M}}$ in Eq.~\eqref{eq:bigm}. This will not only provide an insight about how the results are affected by neglecting higher-order terms in the seesaw expansion, but will also set limits on the parameters, above which the approximation holds up to a certain precision. As a reference observable we choose the neutrino mass-squared difference $\dmatm$, which turns out to be the most sensitive one to such corrections. To quantify the effect of considering the ISS approximation at lowest order, we define the parameter
    \begin{align}
        \Delta_{\rm ISS} \equiv \frac{|\Delta m^2_{31}-\Delta \tilde{m}^2_{31}|}{\Delta \tilde{m}^2_{31}}\,,\quad \Delta\tilde{m}^2_{31}=\tilde{m}^2_3-\tilde{m}^2_1\,,
            \label{eq:DISS}
    \end{align}
where the light-neutrino masses $m_i$ and $\tilde{m}_i$ are determined using Eqs.~\eqref{diagnutr} and \eqref{eq:Unudef}, respectively. For instance, $\Delta_{\rm ISS}=0.1$ indicates that the value of $\Delta m^2_{31}$ determined from $\bm{\mathcal{M}}$ differs from that computed with $\M_{\rm eff}$ by 10\%.

\item Evaluating the impact of the one-loop $\delta \bm{\mathcal{M}}$ on the determination of low-energy neutrino parameters. For that, we will consider the neutrino mass squared differences $\dmsol$ and $\dmatm$, since for the mixing angles and CP phases the corrections are negligible when compared with the current experimental precision. Likewise the previous case, we define the parameter
\begin{align}
        \Delta_{\rm 1L}^{ij} \equiv \frac{|\Delta \widehat{m}^2_{ij}-\Delta m^2_{ij}|}{\Delta m^2_{ij}}\,,\quad \Delta\widehat{m}^2_{ij}=\widehat{m}^2_i-\widehat{m}^2_j\,,
          \label{eq:D1L}
\end{align} 
where $\widehat{m}_i$ are the one-loop corrected neutrino masses and $\Delta m^2_{ij}$ are the tree-level neutrino mass-squared differences computed with the full $\bm{\mathcal{M}}$. Notice that, since we evaluate $\delta \bm{\mathcal{M}}$ in the basis where $\bm{\mathcal{M}}$ is diagonal, the light neutrino masses $\widehat{m}_i$ are determined diagonalising the matrix $\M_{\rm light}=\text{diag}(m_1,m_2,m_3)+\delta \bm{\mathcal{M}}_{ij}$, where $\delta \bm{\mathcal{M}}_{ij}$ is given by Eqs.~\eqref{eq:deltaM}-\eqref{eq:deltaH0} with $i,j=1,2,3$ and $n_f=7$ (total number of neutrino mass eigenstates).
\end{itemize}

For numerical computations in the $5_1^{e,\mu\tau}$ cases discussed in Sections~\ref{sec:symmetries} and \ref{sec:numass}, we consider a benchmark scenario based on the following assumptions. We choose $p=1$ and $q=10$ in Eq.~\eqref{eq:MDMRMS}, implying $m_{6,7} \simeq 10\,m_{4,5}$ and $m_5-m_4\simeq m_7-m_6 \simeq \mu_s$. Regarding the scalar sector, we take $\tan\beta=1$ [see Eq.~\eqref{eq:tanb}] and consider all physical neutral and charged scalar masses to be 1 TeV, except for the SM Higgs boson with mass $m_{H^0}=125$~GeV. Under these premises we span the parameter space in the following way:

\begin{itemize}
 \item The low-energy neutrino parameters are fixed to their best-fit values given in Table~\ref{tab:dataref}, and the effective neutrino mass matrix elements $M_{ij}$ (defined in the ISS approximation) are computed according to Eqs.~\eqref{eq:NOMij} and \eqref{eq:IOMij}, for both NO and IO neutrino mass spectra. Relations~\eqref{eq:relsLE} are then solved to find the predicted value for the Majorana phase $\alpha$, and the parameters in Eq.~\eqref{eq:xyzwdf} are determined. Notice that, although the values of $x$, $y$, $z$ and $w$ are set by low-energy neutrino parameters, the scales $m_{D_i}$, $M$ and $\mu_s$ are not uniquely defined since $\M_{\rm eff}$ in Eq.~\eqref{eq:Mxyzw} is invariant under the rescalings
\begin{align}
 M\rightarrow a M\;,\; \mu_s \rightarrow b \mu_s\;,\; m_{D_i}\rightarrow \frac{a}{\sqrt{b}}\, m_{D_i}\,.
 \label{eq:resc}
\end{align}
Choosing the initial values $M=100$~GeV and $\mu_s=10$~eV, we determine $m_{D_i}$ as shown in Eq.~\eqref{eq:Mxyzw}.  In order to probe a wide range of scales, we vary $M$ and $\mu_s$ in the intervals $[1,10^4]$~GeV and $[1,10^{11}]$~eV. For an arbitrary pair $(M,\mu_s)$, we set the rescaling parameters with respect to the initial values, namely $a=M/100$ and $b=\mu_s/10$. The corresponding $m_{D_i}$ are obtained using Eq.~\eqref{eq:resc}. This procedure leaves invariant the effective neutrino mass matrix (in the ISS approximation) and guarantees that the low-energy parameters are always those corresponding to the experimental best-fit values. Notice that such procedure may lead to very large Dirac masses $m_{D_i}$. Thus, in order to ensure perturbativity of the Yukawa couplings $b_i$ in Eq.~\eqref{eq:YYY}, we require $y_{\rm max}={\rm max}\{m_{D_{1,2}}/v_1,m_{D_{3,4}}/v_2\}\leq 5$. It is important to stress that rescaling $M$, $\mu_s$ and $m_{D_i}$ is the only way to probe the parameter space of our model since ratios among different $m_{D_i}$ are determined by low-energy parameters which are fixed.

\item For each set of $(M,\mu_s,m_{D_i})$, the full $7\times 7$ neutrino mass matrix $\bm{\mathcal{M}}$ is defined using Eqs.~\eqref{eq:bigm} and \eqref{eq:MDMRMS}, and then diagonalised as indicated in Eq.~\eqref{diagnutr} to determine $\bm{\mathcal{U}}$ and $m_{1-7}$. The active neutrino mixing is characterised by the $3\times 3$ (non-unitary) matrix $\U$ of Eq.~\eqref{eq:nonunitpmns}, which is the upper-left $3\times 3$ block of $\bm{\mathcal{U}}$. The non-unitarity effects are parameterised by the matrix $\bm{\eta}$ of Eq.~\eqref{eq:nonunitpmns}. Finally, we compute the one-loop corrections to the light neutrino masses as explained above. 
\end{itemize}

We remark that the CC mixing between the charged lepton with flavour $\alpha=e,\mu,\tau$ and the heavy sterile neutrino $\nu_j\,(j=4-7)$ is set by $\mathbf{B}_{\alpha j}$ as defined in Eq.~\eqref{eq:BC}--see also Eq.~\eqref{eq:Wint}. In practice, $\mathbf{B}_{\alpha j}$ is the $(\alpha,j)$ element of $\bm{\mathcal{U}}$ computed above and defined in the charged-lepton mass basis. Notice also that we will be able to cover wide ranges of $\mathbf{B}_{\alpha j}$ since these elements scale as $m_D/M$. Throughout the remaining of this work, we will use as reference parameters the average mass of the lightest sterile neutrino pair $m_{45}$, a degeneracy parameter $r_N$ and the mixing of the electron with the lightest sterile neutrino $V_{eN}$, defined as
\begin{align}
    m_{45}=\frac{m_4+m_5}{2}\simeq M\;,\quad r_N=\frac{m_5-m_4}{m_{45}}\simeq \frac{\mu_s}{m_{45}}\;,\quad V_{eN}=|\mathbf{B}_{e4}|\simeq \frac{m_{D_4}}{\sqrt{2}\,m_{45}}\,.
    \label{eq:m45VeN}
\end{align}
\begin{figure}[t!]
	\centering
	\begin{tabular}{ll}
	\includegraphics[width=0.47\textwidth]{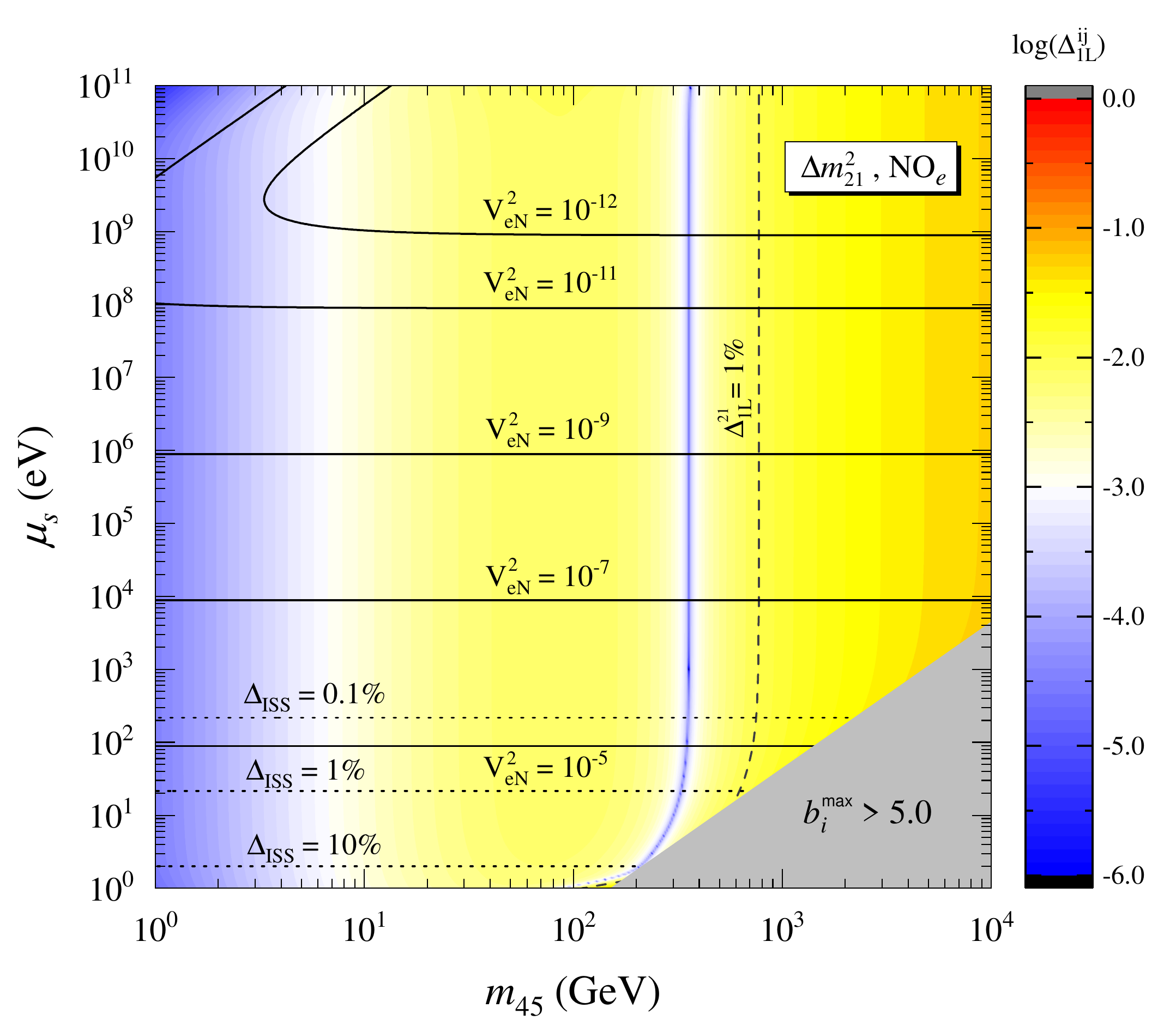}
	&\includegraphics[width=0.4\textwidth]{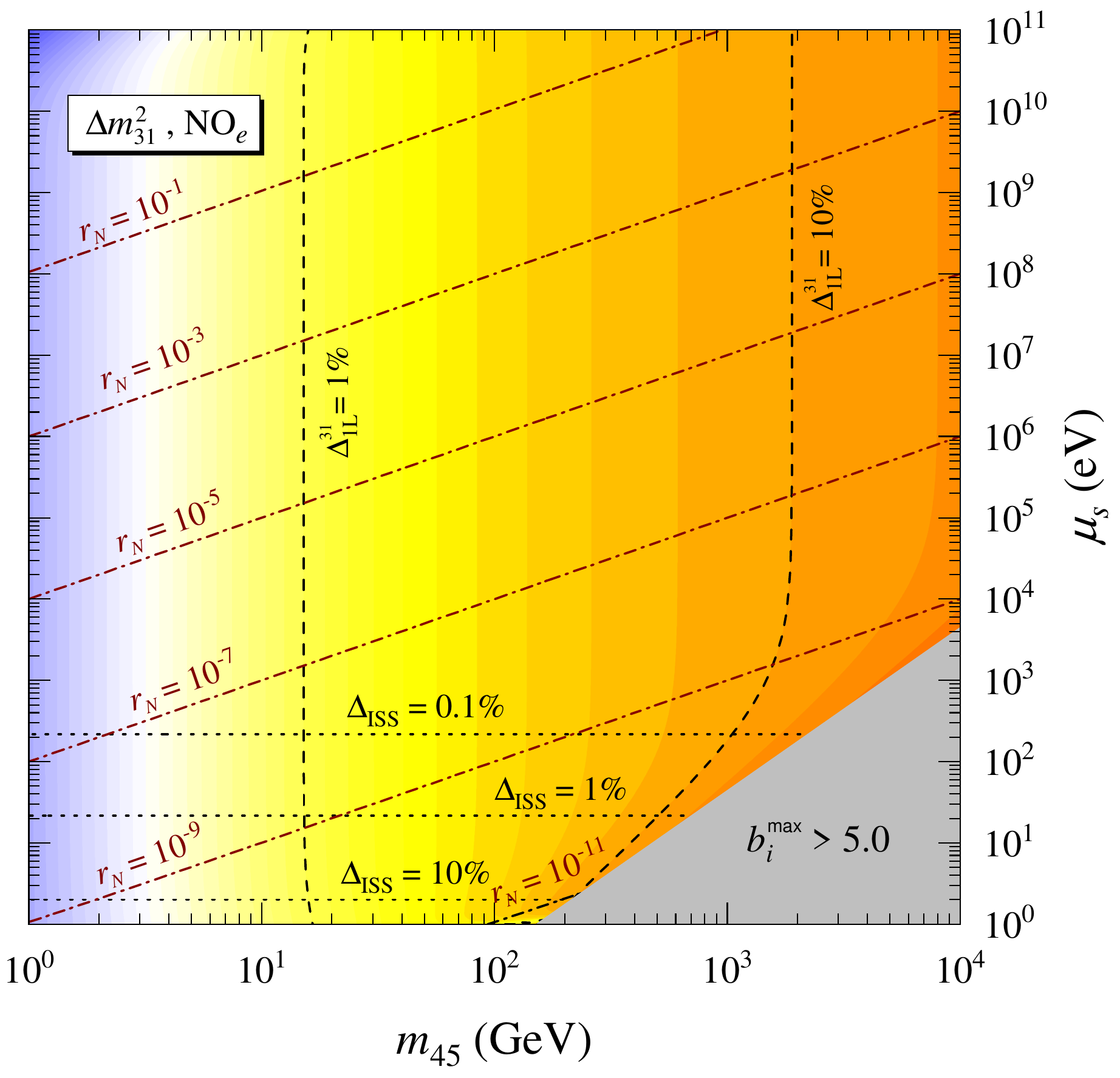}\\
		\includegraphics[width=0.47\textwidth]{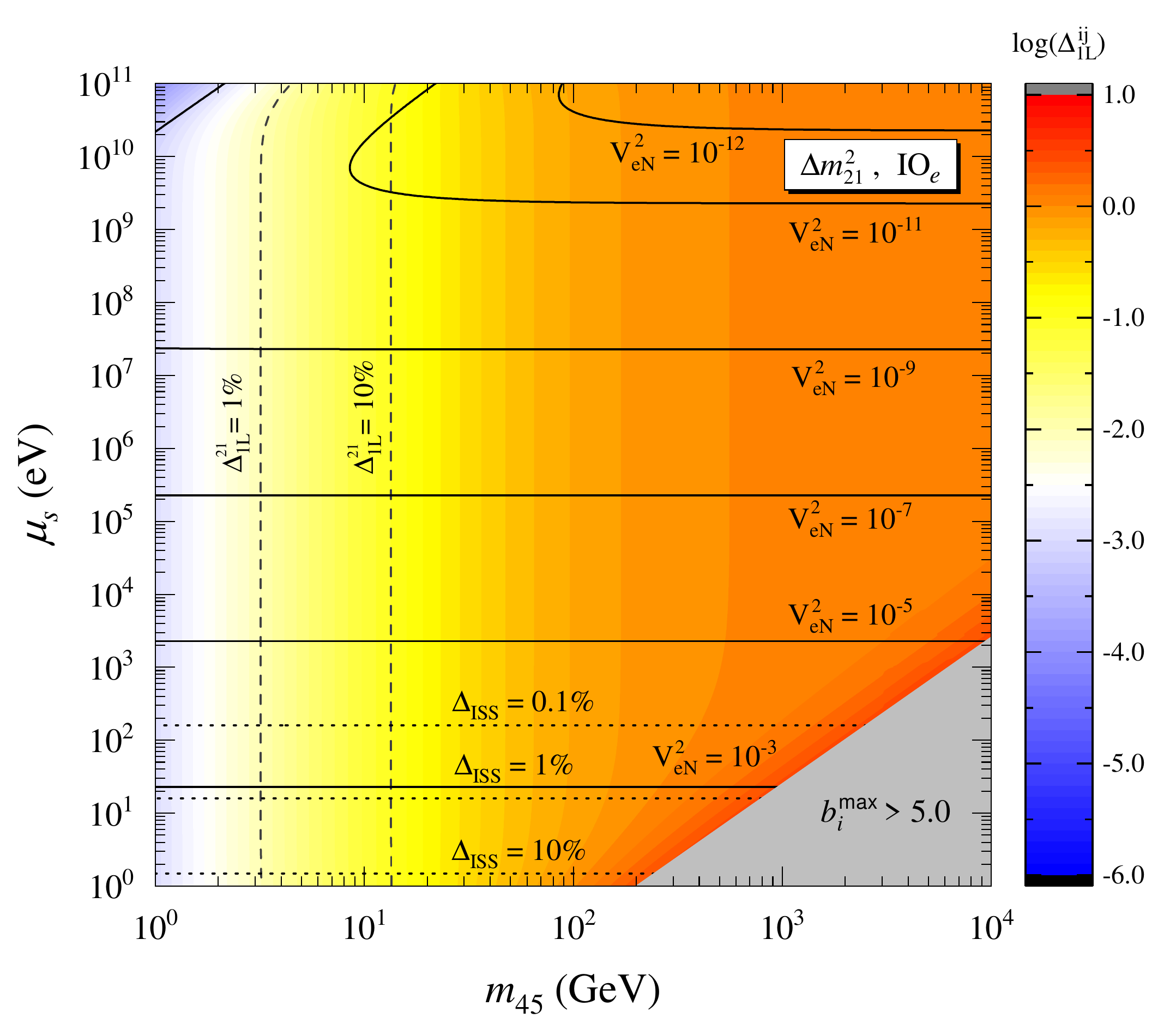}
	&\includegraphics[width=0.4\textwidth]{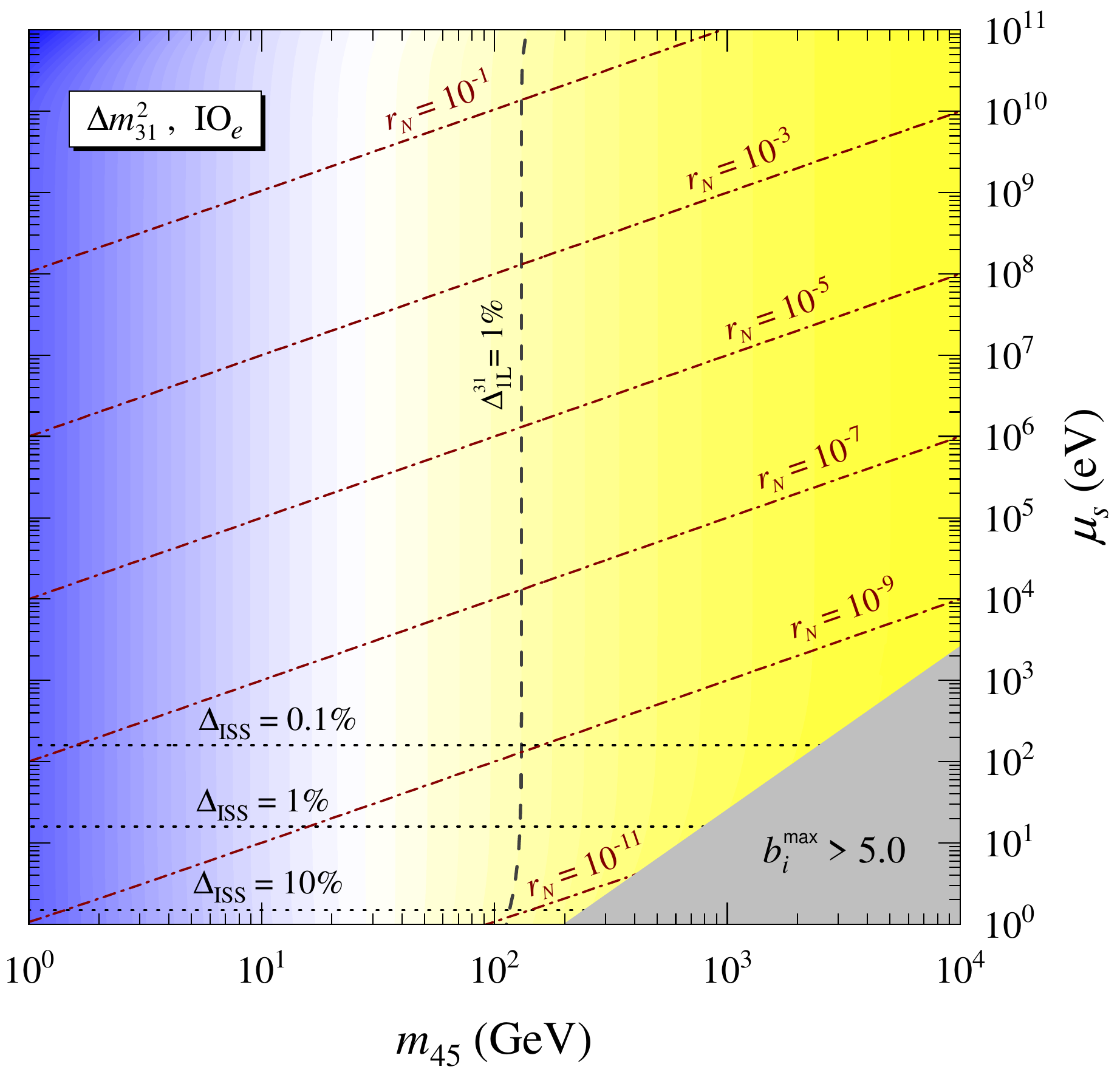}
	\end{tabular}
	\caption{Impact of one-loop corrections on $\dmsol$ (left panels) and $\dmatm$ (right panels) for the NO$_e$ (upper panels) and IO$_e$ (lower panels) cases. The coloured contour levels are for the $\Delta_{\rm 1L}^{ij}$ parameter defined in Eq.~\eqref{eq:D1L}. For reference, we show the dashed contours with $\Delta_{\rm 1L}^{ij}=1\%,10\%$. The solid, dash-dotted and dotted lines correspond to $V_{eN}$, $r_N$ and $\Delta_{\rm ISS}$ contours, respectively. Within the grey shaded region on the lower-right corner of each panel the largest Yukawa coupling of $\Y_D^{1,2}$ [see Eq.~\eqref{eq:YYY}] obeys $b_i^{\rm max}>5$.}
\label{Fig3}
\end{figure} 

In Fig.~\ref{Fig3} we show the contour-level plots for $\Delta_{\rm 1L}^{21}$ (left panels) and $\Delta_{\rm 1L}^{32}$ (right panels) defined in Eq.~\eqref{eq:D1L} for the NO$_e$ (upper panels) and IO$_e$ (lower panels) cases. The contours of $V_{eN}^2$, $r_N$ and $\Delta_{\rm ISS}$ are also shown by solid, dash-dotted and dotted lines, respectively. Within the grey shaded region $b_i^{\rm max}>5$. From the inspection of these plots we conclude the following:
\begin{itemize}
\item The validity of the (tree-level) inverse-seesaw approximation is verified at less than the percent level  for $\mu_s \gtrsim 10-20$~eV (region above the $\Delta_{\rm ISS}=1\%$ horizontal line). The reason why $\Delta_{\rm ISS}$ does not depend on $m_{45}$ for a given $\mu_s$, can be understood taking into account that the next-to-leading-order approximation of the inverse seesaw scales as $\mu_s m_{D}^4/M^4$ and, consequently, $\Delta_{\rm ISS}\sim m_{D}^2/M^2 \simeq m_{D}^2/m_{45}^2$. Thus, if $\mu_s$ is kept constant ($b=1$ in Eq.~\eqref{eq:resc}), $m_{D_i}$ scales as $m_{45}$ leaving $\Delta_{\rm ISS}$ invariant. Moreover, for a given $m_{45}$, a rescaling of $\mu_s \rightarrow b\mu_s$ leads to a rescaling of $\Delta_{\rm ISS} \rightarrow \Delta_{\rm ISS}/b$, as can be seen in Fig.~\ref{Fig3}.
\item The chosen intervals for $\mu_s$ and $M$ allows us to swipe a wide range for the heavy-light mixing parameter, namely $V_{eN}^2$ lies between $10^{-13}$ (or even less) and $10^{-3}$ in the parameter-space region where the inverse seesaw approximation holds up to 1\%. The $V_{eN}^2$ contours are approximate horizontal lines since, as shown in Eq.~\eqref{eq:Baj}, heavy-light mixing scales as $m_{D_i}/m_{45}$ and, consequently, $V_{eN}^2$ remains invariant for constant $\mu_s$ due to the same rescaling of $m_{D_i}$ and $m_{45}$. Notice that this feature fails for $\mu_s \sim m_{45}$ (upper-left corner of the left panels in Fig.~\ref{Fig3}) since in this case the leading-order ISS approximation for $V_{eN}$ is no longer valid. For the degeneracy parameter $r_N$, the contours follow the linear relation $r_N \simeq \mu_s/m_{45}$ already shown in Eq.~\eqref{eq:m45VeN} (dash-dotted lines in the right panels of Fig.~\ref{Fig3}).
\item The nearly-vertical levels of $\Delta_{\rm 1L}^{ij}$ indicate that radiative corrections to neutrino masses depend mainly on $m_{45}$. For NO$_e$, the one-loop corrected $\dmsol$ deviates from the tree-level result by less than 1\% for $m_{45} \lesssim 700$~GeV, while for $\dmatm$ that threshold is at $m_{45} \lesssim 20$~GeV. Instead, the corresponding upper limits on $m_{45}$ for IO$_e$ are at 3~GeV and 100~GeV, respectively. Notice that, being the one-loop corrections typically larger than the experimental uncertainty (see Table~\ref{tab:dataref}) for $m_{45}$ above those values, this does not mean that the symmetry realisation presented in Section~\ref{sec:symmetries} is no longer valid. It just signals the fact that the tree-level relations~\eqref{eq:xyzwdf} start failing. The take-home message to learn from this analysis is that the scale invariance of the tree-level inverse-seesaw approximation holds up to a certain level depending on the sterile neutrino mass. 
\end{itemize}

\begin{table}[!t]
\renewcommand{\arraystretch}{1.4}
\centering
\begin{tabular}{lll}   
\hline
LFV process & Present limit ($90 \%$ CL) & Future sensitivity \\ \hline
$\BR(\mu \rightarrow e \gamma)$ & $ 4.2 \times 10^{-13}$ (MEG \cite{TheMEG:2016wtm}) &  $ 6 \times 10^{-14}$ (MEG II \cite{Baldini:2018nnn}) \\
$\BR(\tau \rightarrow e \gamma)$ & $ 3.3 \times 10^{-8}$ (BaBar \cite{Aubert:2009ag}) & $3 \times 10^{-9}$ (Belle II \cite{Kou:2018nap})\\
$\BR(\tau \rightarrow \mu \gamma)$ & $ 4.4 \times 10^{-8}$ (BaBar \cite{Aubert:2009ag}) & $10^{-9}$ (Belle II \cite{Kou:2018nap})\\ \hline
$\BR(\mu^{-}  \rightarrow e^{-} e^{+} e^{-})\quad\quad$ & $ 1.0 \times 10^{-12}$ (SINDRUM \cite{Bellgardt:1987du}) & $10^{-16}$ (Mu3e \cite{Blondel:2013ia})\\
$\BR(\tau^{-} \rightarrow e^{-} e^{+} e^{-})$ & $ 2.7 \times 10^{-8}$  (Belle \cite{Hayasaka:2010np})& $5 \times 10^{-10}$ (Belle II \cite{Kou:2018nap})\\
$\BR(\tau^{-} \rightarrow e^{-} \mu^{+} \mu^{-})$ & $ 2.7 \times 10^{-8}$  (Belle \cite{Hayasaka:2010np})& $5 \times 10^{-10}$ (Belle II \cite{Kou:2018nap})\\
$\BR(\tau^{-} \rightarrow e^{+} \mu^{-} \mu^{-})$ & $ 1.7 \times 10^{-8}$  (Belle \cite{Hayasaka:2010np})& $3 \times 10^{-10}$ (Belle II \cite{Kou:2018nap})\\
$\BR(\tau^{-} \rightarrow \mu^{-} e^{+} e^{-})$ & $ 1.8 \times 10^{-8}$  (Belle \cite{Hayasaka:2010np})& $3 \times 10^{-10}$ (Belle II \cite{Kou:2018nap})\\
$\BR(\tau^{-} \rightarrow \mu^{+} e^{-} e^{-})$ & $ 1.5 \times 10^{-8}$  (Belle \cite{Hayasaka:2010np})& $3 \times 10^{-10}$ (Belle II \cite{Kou:2018nap})\\
$\BR(\tau^{-} \rightarrow \mu^{-} \mu^{+} \mu^{-})$ & $ 2.1 \times 10^{-8}$  (Belle \cite{Hayasaka:2010np})& $4 \times 10^{-10}$ (Belle II \cite{Kou:2018nap})\\ \hline
$\CR(\mu - e, \text{Al})$ & $-$ & $3 \times 10^{-17}$ (Mu2e \cite{Bartoszek:2014mya}) \\
&  & $10^{-15}-10^{-17}$ (COMET I-II~\cite{Adamov:2018vin}) \\
$\CR(\mu - e, \text{Ti})$ & $ 4.3 \times 10^{-12}$ (SINDRUM II \cite{Dohmen:1993mp})$\quad\quad$& $10^{-18}$ (PRISM/PRIME \cite{Alekou:2013eta}) \\
$\CR(\mu - e, \text{Au})$ & $ 7 \times 10^{-13}$ (SINDRUM II \cite{Bertl:2006up})& $-$ \\
$\CR(\mu - e, \text{Pb})$ & $ 4.6 \times 10^{-11}$ (SINDRUM II \cite{Honecker:1996zf}) & $-$ \\ 
\hline\hline
LFV process & Present limit ($95 \%$ CL) & Future sensitivity \\ \hline
$\BR(Z \rightarrow e^{\mp} \mu^{\pm})$ & $7.3 \times 10^{-7}$ (CMS \cite{Nehrkorn:2017fyt}) & $10^{-8}-10^{-10}$ (FCC-ee \cite{Dam:2018rfz}) \\
$\BR(Z \rightarrow e^{\mp} \tau^{\pm})$ & $9.8 \times 10^{-6}$   (OPAL \cite{Akers:1995gz})& $10^{-9}$ (FCC-ee \cite{Dam:2018rfz})\\
$\BR(Z \rightarrow \mu^{\mp} \tau^{\pm})$ & $1.2 \times 10^{-5}$  (DELPHI \cite{Abreu:1996mj})& $10^{-9}$ (FCC-ee \cite{Dam:2018rfz})\\ \hline
\end{tabular}
\caption{Current experimental bounds and future sensitivities for the branching ratios (BRs) and capture rates (CRs) of cLFV processes.}
\label{tab:boundsCLFV}
\end{table}

\section{Charged lepton flavour violation}
\label{sec:pheno}
Lepton flavour violating processes are in the front line of experimental searches for physics beyond the SM~\cite{Raidal:2008jk}. The ISS model, being a paradigm for low-scale neutrino mass generation, provides a natural scenario for the observation of flavour transitions beyond neutrino oscillations. The charged and neutral current interactions with heavy sterile neutrinos induce new phenomena which are strongly suppressed in the SM extended with light neutrinos only. Nevertheless, the predictive power of the general ISS is limited by the arbitrariness of its parameter values. In the present framework, the symmetries discussed in Section~\ref{sec:symmetries} provide the ground for a testable scenario in the light of present and future experimental probes on LFV processes. Thus, we now study LFV in the context of the scenarios set in the previous sections. Our attention will be focused on the cLFV processes listed in Table~\ref{tab:boundsCLFV}, where the present experimental upper limits and future sensitivities for the BRs and CRs are shown. We have revisited the computation of the rates taking into account the field content in our framework. The details are shown in Appendices~\ref{sec:Allint}-\ref{sec:LFFF}.

As stated in Section~\ref{sec:numass}, we focus our study on the case of the charged lepton matrix texture $5^{\ell}_{1}$. To justify this choice, let us look at the cLFV decay $\ell_{\alpha}^{-} \rightarrow \ell_{\beta}^{-} \ell_{\gamma}^{+} \ell_{\delta}^{-}$, which is mediated at tree level by the neutral scalars $R$ and $I$ coming from the two Higgs doublets (in the alignment limit). The corresponding BR for this process reads~\cite{Correia:2019vbn}
\begin{equation}
\frac{\BR(\ell_{\alpha}^{-} \rightarrow \ell_{\beta}^{-} \ell_{\gamma}^{+} \ell_{\delta}^{-} )}{\BR\left(\ell_{\alpha} \rightarrow \ell_{\beta} \nu_{\alpha} \overline{\nu_{\beta}} \right)} = \frac{1}{16 (1+ \delta_{\beta \delta})} \left\{ \left[\left|g_{LL}^{\alpha \beta, \gamma \delta}\right|^2 + \left|g_{LR}^{\alpha \beta, \gamma \delta}\right|^2 + \left( \beta \leftrightarrow \delta \right) \right]-\text{Re}\left[g_{LL}^{\alpha \beta, \gamma \delta} g_{LL}^{\alpha \delta, \gamma \beta *} \right] + \left(L \leftrightarrow R \right) \right\},
\end{equation}
where
\begin{equation}
\begin{aligned}
&g_{LL}^{\alpha \beta, \gamma \delta} = (\mathbf{N}_e^{\dagger})_{\beta \alpha} (\mathbf{N}_e^{\dagger})_{\delta \gamma} \Big(\frac{1}{m_{R}^2}- \frac{1}{m_{I}^2}\Big), \quad g_{RR}^{\alpha \beta, \gamma \delta} = (\mathbf{N}_e)_{\beta \alpha} (\mathbf{N}_e)_{\delta \gamma} \Big(\frac{1}{m_{R}^2} - \frac{1}{m_{I}^2}\Big), \\
&g_{LR}^{\alpha \beta, \gamma \delta} = (\mathbf{N}_e^{\dagger})_{\beta \alpha} (\mathbf{N}_e)_{\delta \gamma} \Big(\frac{1}{m_{R}^2} + \frac{1}{m_{I}^2}\Big), \quad
g_{RL}^{\alpha \beta, \gamma \delta} = (\mathbf{N}_e)_{\beta \alpha} (\mathbf{N}_e^{\dagger})_{\delta \gamma} \Big(\frac{1}{m_{R}^2}  + \frac{1}{m_{I}^2}\Big)\,,
\label{eq:Gfacts}
\end{aligned}
\end{equation}
being $m_{R,I}$ the neutral scalar masses. The matrix $\mathbf{N}_e$ dictates the interactions between charged leptons and the neutral scalars as shown in Eq.~\eqref{eq:deltanuR}. For the $5^{\ell}_{1}$ case, the structure of $\mathbf{N}_e$ is
\begin{equation}
5_1^{e}: \mathbf{N}_e \sim \begin{pmatrix} 
\times & 0 & 0 \\ 
0 & \times & \times \\
0 & \times & \times
\end{pmatrix}, \ 5_1^{\mu}: \mathbf{N}_e \sim \begin{pmatrix} 
\times & 0 & \times \\ 
0 & \times & 0 \\
\times & 0 & \times
\end{pmatrix},\ 5_1^{\tau}: \mathbf{N}_e \sim \begin{pmatrix} 
\times & \times  & 0 \\ 
\times & \times & 0 \\
0 & 0 & \times
\end{pmatrix}\,.
\end{equation}
The presence of zeros in $\mathbf{N}_e$ imposed by the flavour symmetry leads to a natural suppression of the above BRs. In fact, considering the most constraining three-body decay $\mu \rightarrow 3e$ (see Table~\ref{tab:boundsCLFV}), the tree-level contribution vanishes for the textures $5_1^{e,\mu}$ (decoupled electron or muon) irrespective of $m_{R,I}$. Although this does not hold for $5_1^\tau$, in this case the BRs are strongly suppressed by the tiny couplings in the $\mu-e$ sector. The same conclusion cannot be drawn for the texture $4^{\ell}_{3}$ since the flavour symmetry does not yield the charged-lepton decoupling feature. To suppress the decay rates in this case, not only very large $m_{R,I}$ masses are required but also they must be extremely fine tuned~\cite{Correia:2019vbn}, as can be readily seen from Eq.~\eqref{eq:Gfacts}. Similarly, the analysis of the one-loop contribution of the neutral scalars $R$ and $I$ to the decay $\ell_{\alpha} \rightarrow \ell_{\beta} \gamma$ reveals that in the $4^{\ell}_{3}$ case requires fine-tuned scalar masses. On the other hand, for $5_1^{e,\mu}$ the scalar contribution to the $\mu \rightarrow e \gamma$ amplitude vanishes. \\
\begin{figure}[ht!]
	\centering
	\begin{tabular}{ll}
	\includegraphics[width=0.37\textwidth]{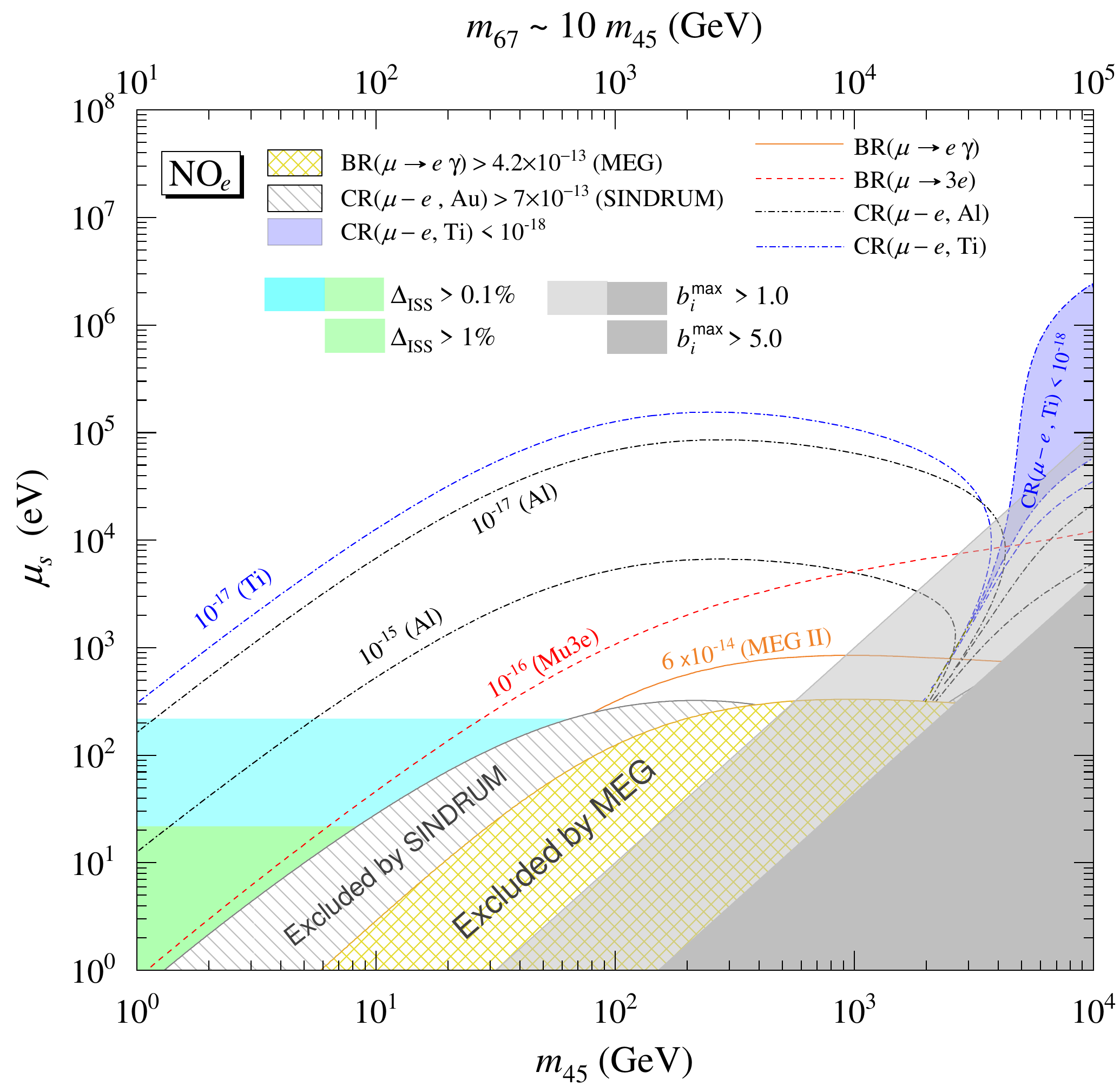}
	&\includegraphics[width=0.37\textwidth]{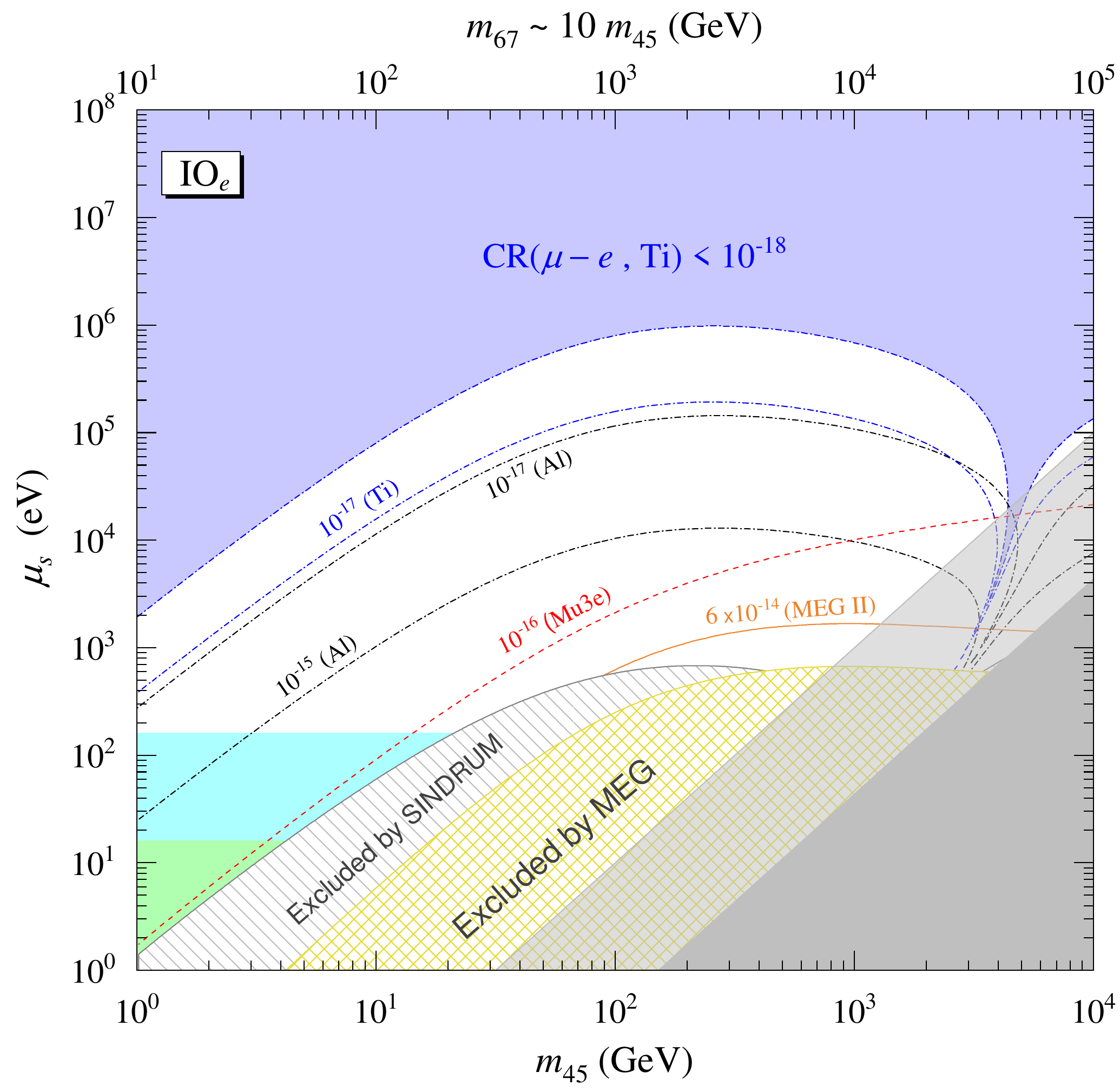}\\
		\includegraphics[width=0.37\textwidth]{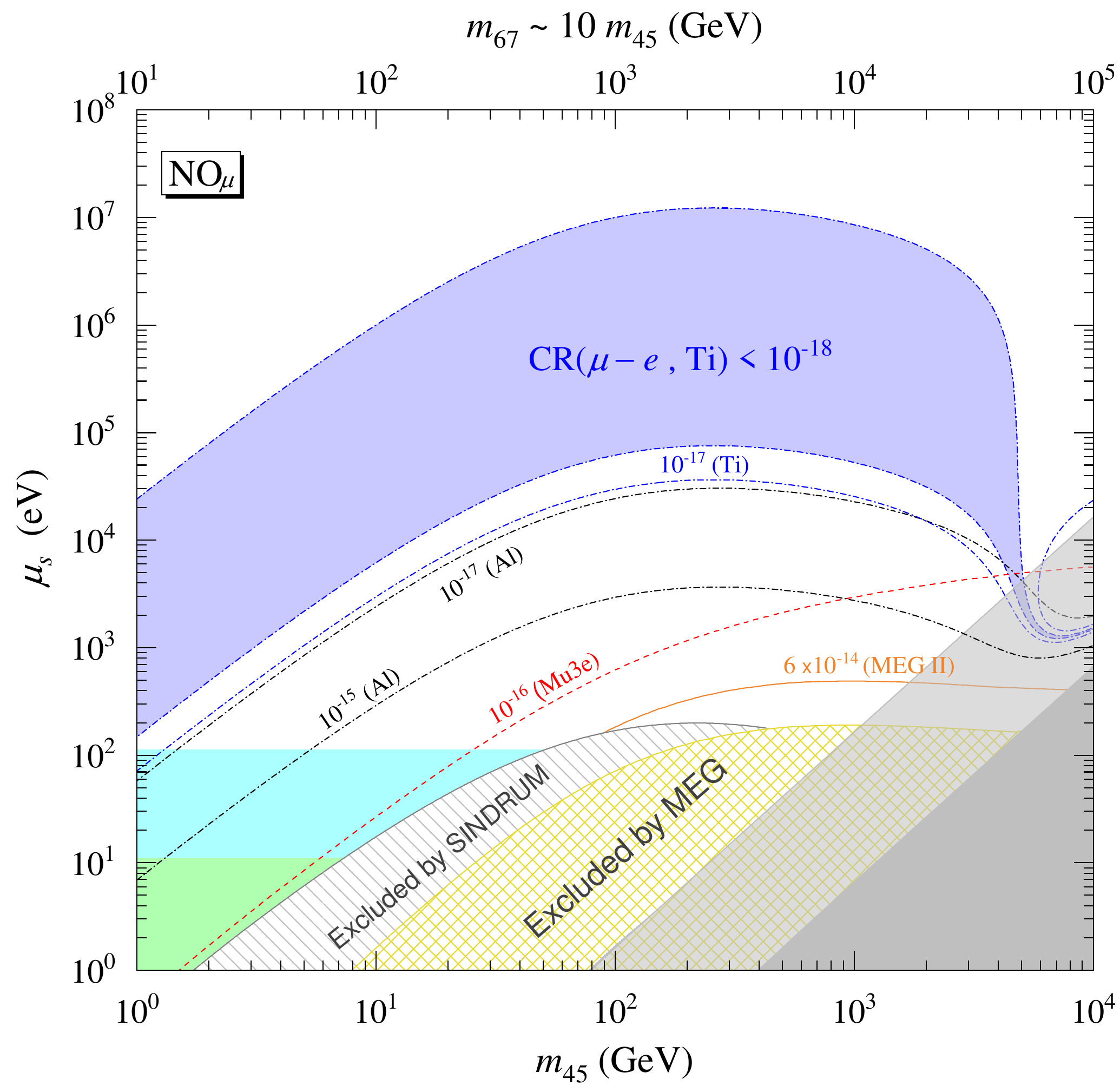}
	&\includegraphics[width=0.37\textwidth]{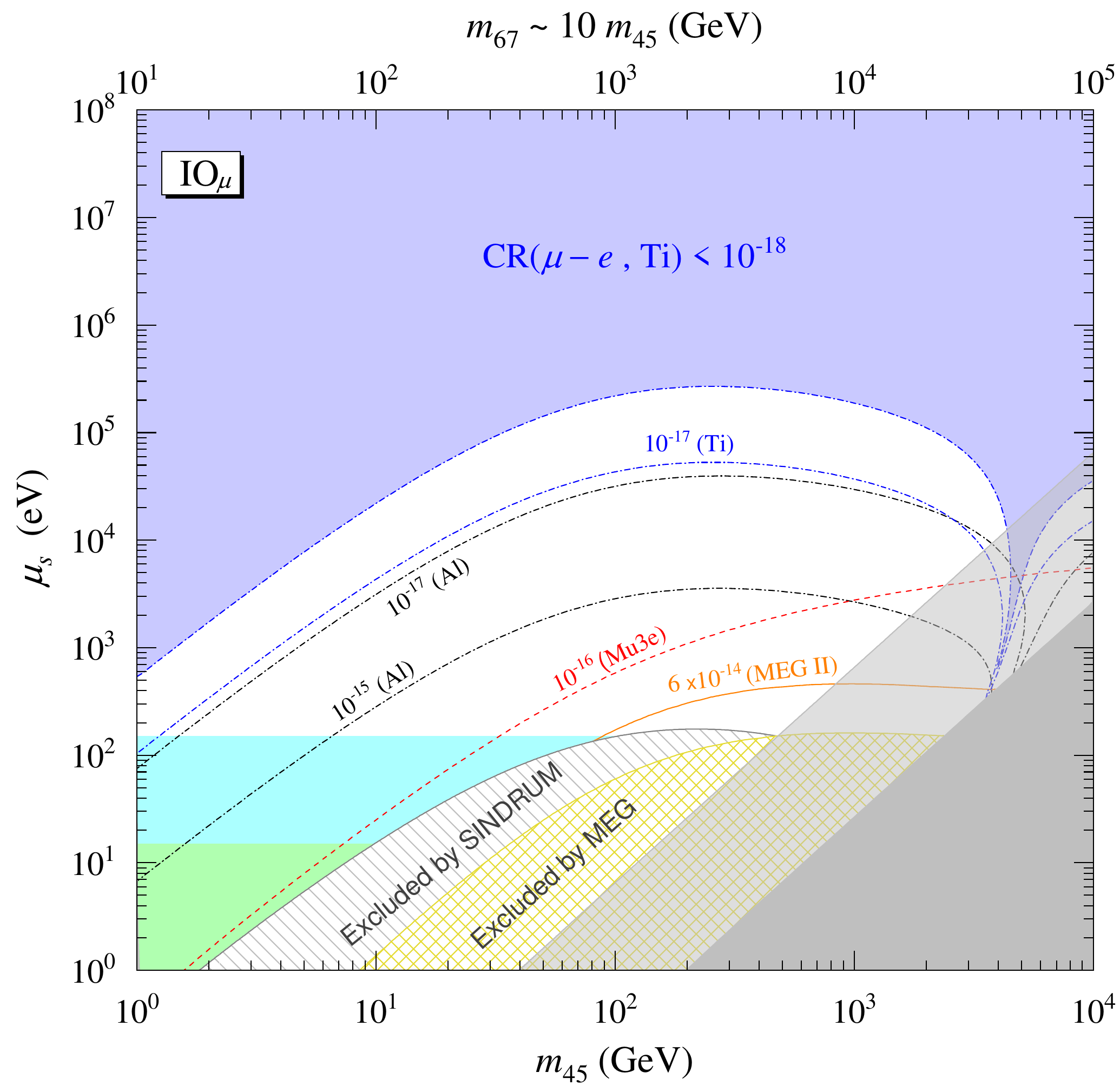}\\
		\includegraphics[width=0.37\textwidth]{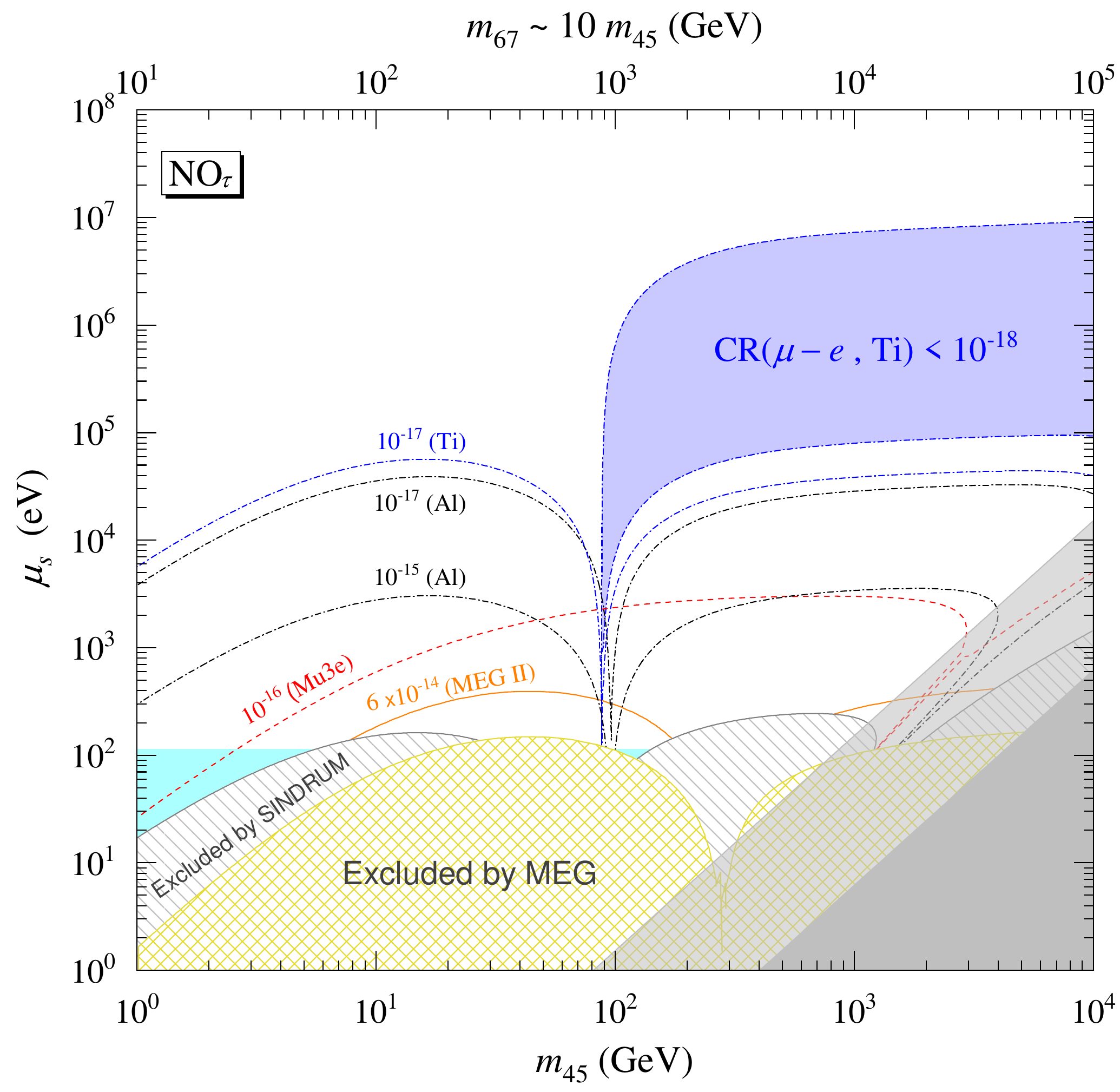}
	&\includegraphics[width=0.37\textwidth]{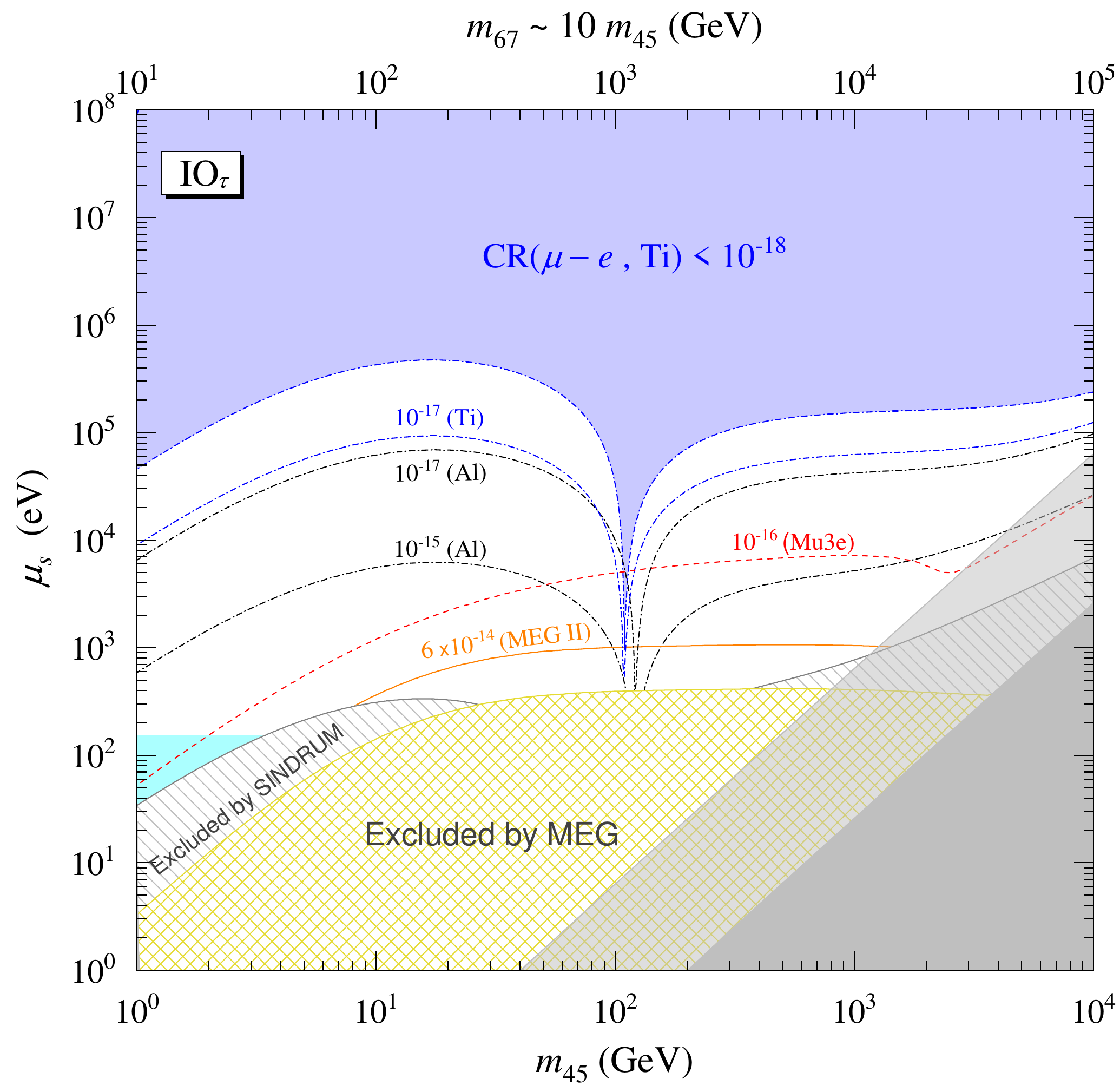}
	\end{tabular}
\caption{Constraints on the $(m_{45},\mu_s)$ parameter space imposed by the MEG bound on $\BR(\mu\rightarrow e \gamma)$ (yellow crosshatched region) and the SINDRUM II limit on $\CR(\mu-e,{\rm Au})$ (grey hatched region). The contours corresponding to the future sensitivities of the MEG II (solid orange) and Mu3e (red dashed) experiments are also given. The black and blue dash-dotted lines show the contours of $\CR(\mu-e,{\rm Al})$ and $\CR(\mu-e,{\rm Ti})$, respectively, for values within the sensitivity of future experiments (see Table~\ref{tab:boundsCLFV}). In the blue shaded region $\CR(\mu-e,{\rm Ti})<10^{-18}$. Limits on  $b_i^{\rm max}$ and $\Delta_{\rm ISS}$ are also shown (grey, green and cyan shaded regions). The results are shown for all cases found to be realisable through Abelian symmetries, i.e. for the $5_1^{e,\mu,\tau}$ cases with NO (left panels) and IO (right panels) neutrino mass spectra.}
\label{Fig4}
\end{figure} 

We now analyse the constraints imposed on the parameter space taking into account the present limits and future sensitivities for the processes indicated in Table~\ref{tab:boundsCLFV}. We focus on $\mu \rightarrow e\gamma$, $\mu \rightarrow 3e$ and $\mu-e$ conversion in Au, Ti and Al, since these are either the most constraining at present or the ones for which the projected sensitivity is higher. Later on, we will comment on other LFV three-body and $Z$ decays. The $(m_{45},\mu_s)$ parameter space is covered by means of a full numerical analysis as described in the previous section, using the results given in Appendices~\ref{sec:Allint}-\ref{sec:LFFF}. For simplicity, we work in the alignment limit for the two Higgs doublets and in the decoupling limit for the two neutral scalar singlets, as explained in Appendix~\ref{sec:ScalarSector}. The interaction terms between fermions and scalar eigenstates are those given in Section~\ref{sec:Scalarint}. Since the analysis of textures and symmetries for the quark sector is out of the scope of this work, the up-quark and down-quark mass matrices are taken to be diagonal, i.e., $\mathbf{M}_{u}=\text{diag}(m_{u},m_{c},m_{t})$ and $\mathbf{M}_{d}=\text{diag}(m_{d},m_{s},m_{b})$~\cite{Zyla:2020zbs}. Consequently, the CKM matrix $\mathbf{V}$ is the identity matrix, and the flavour-changing matrices $\mathbf{N}_{d}$ and $\mathbf{N}_{u}$, defined in Eqs.~\eqref{eq:Nd} and \eqref{eq:Nu}, vanish.

The results of our numerical analysis are shown in Fig.~\ref{Fig4} for the six cases NO$_{e,\mu,\tau}$ (left panels) and IO$_{e,\mu,\tau}$ (right panels). The colour codes in the legend of the upper-left panel apply to the whole figure. By inspecting these results we conclude the following:
\begin{itemize}
    \item The validity of the inverse-seesaw approximation up to 1\% level, i.e. $\Delta_{\rm ISS } < 1\%$, imposes lower bounds on the LNV parameter $\mu_s > 10-20$~eV (cyan shaded regions), which correspond to upper bounds on the mixing $V_{eN}^2 \lesssim 10^{-4}-10^{-3}$ (see Fig.~\ref{Fig3}). The light (dark) grey regions show that a considerable fraction of the parameter space is excluded if one takes into account $b_i^{\rm max}<1\,(5)$ as a perturbativity requirement.
    \item The MEG and SINDRUM II limits on $\BR(\mu\rightarrow e\gamma)$ and $\CR(\mu-e,{\rm Au})$ exclude $m_{45}\gtrsim 1-10$~GeV for $\Delta_{\rm ISS}\gtrsim 1\%$. Moreover, the improvement on $\BR(\mu\rightarrow e\gamma)$ foreseen by MEG II (solid orange contour) would have a marginal impact in covering the parameter space in our framework. On the other hand, reaching a sensitivity of $\BR(\mu\rightarrow 3e)$ at the $10^{-16}$ level would be more relevant in constraining the parameter space, especially for heavier sterile neutrinos, i.e., for larger $m_{45}$.
    \item The COMET and PRISM/PRIME projected sensitivities for $\CR(\mu-e,{\rm Al})$ and $\CR(\mu-e,{\rm Ti})$, represented by black and blue dash-dotted contours, respectively, cover a considerable part of the parameter space, leaving unprobed the regions in shaded blue where $\CR(\mu-e,{\rm Ti}) < 10^{-18}$. In the best-case scenario (NO$_e$), probing $\CR(\mu-e,{\rm Ti})$ down to $10^{-18}$ would cover the whole parameter space, as can be seen in the upper left panel. 
\end{itemize}

\begin{figure}[t!]
	\centering
	\begin{tabular}{l}
	\includegraphics[width=1.02\textwidth]{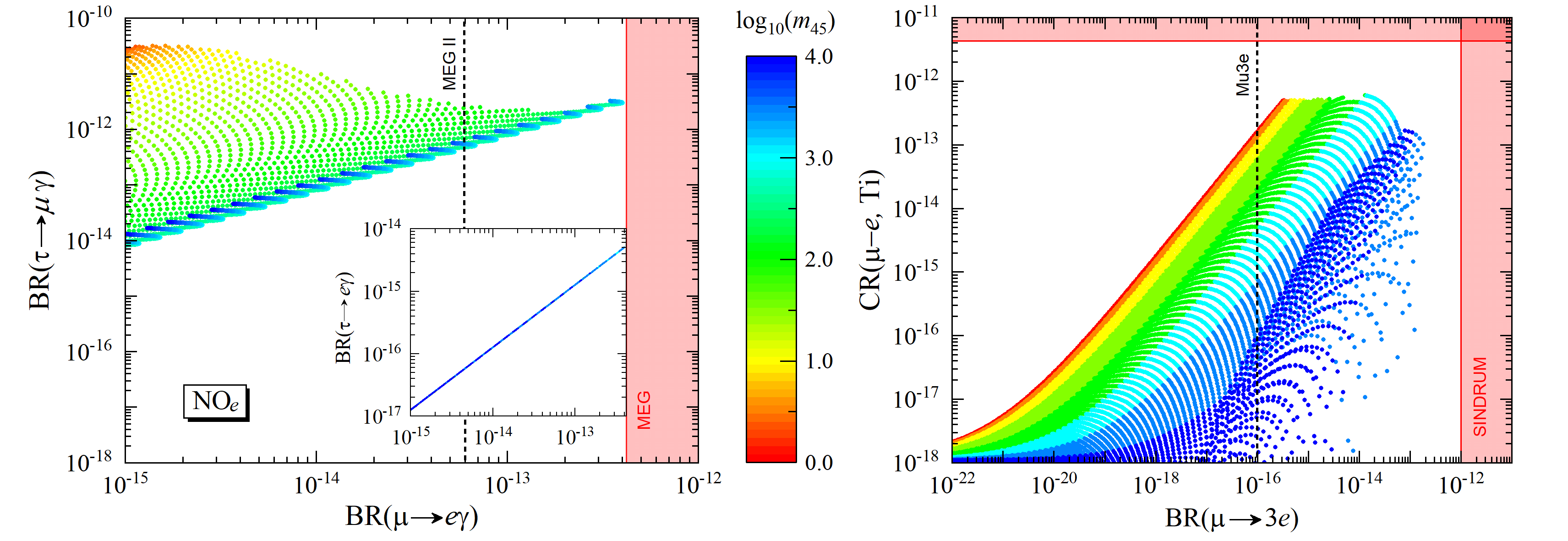}\\
	\includegraphics[width=1.02\textwidth]{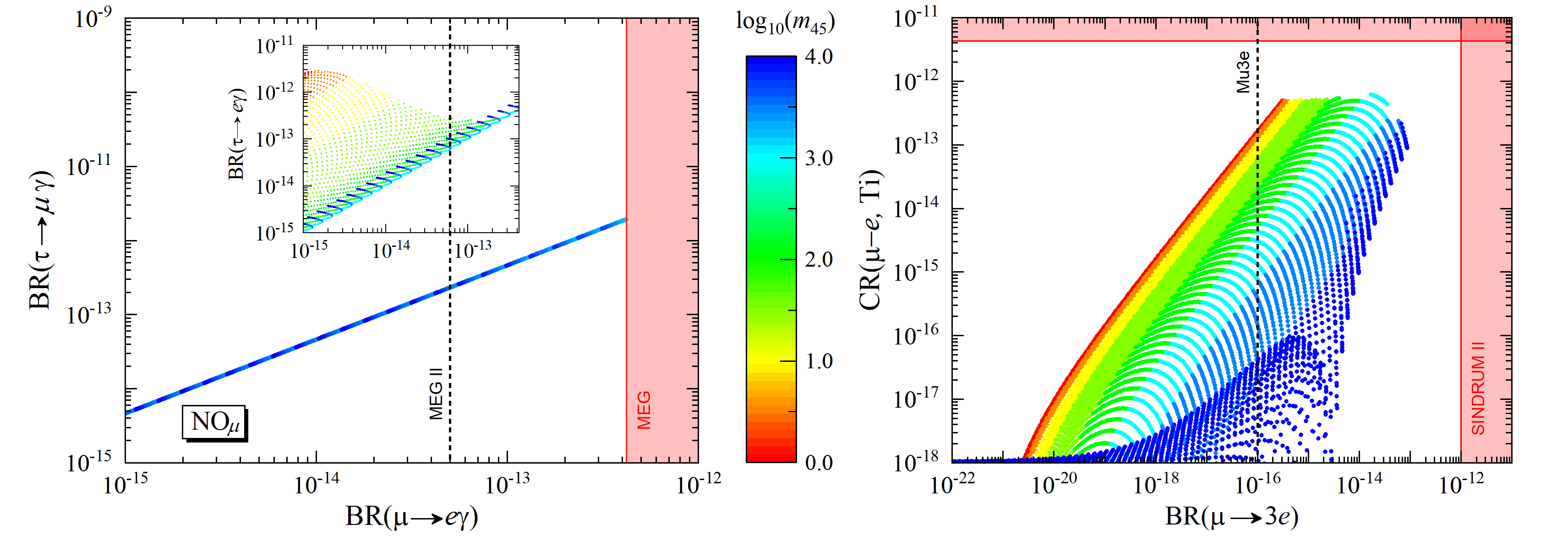}\\
		\includegraphics[width=1.02\textwidth]{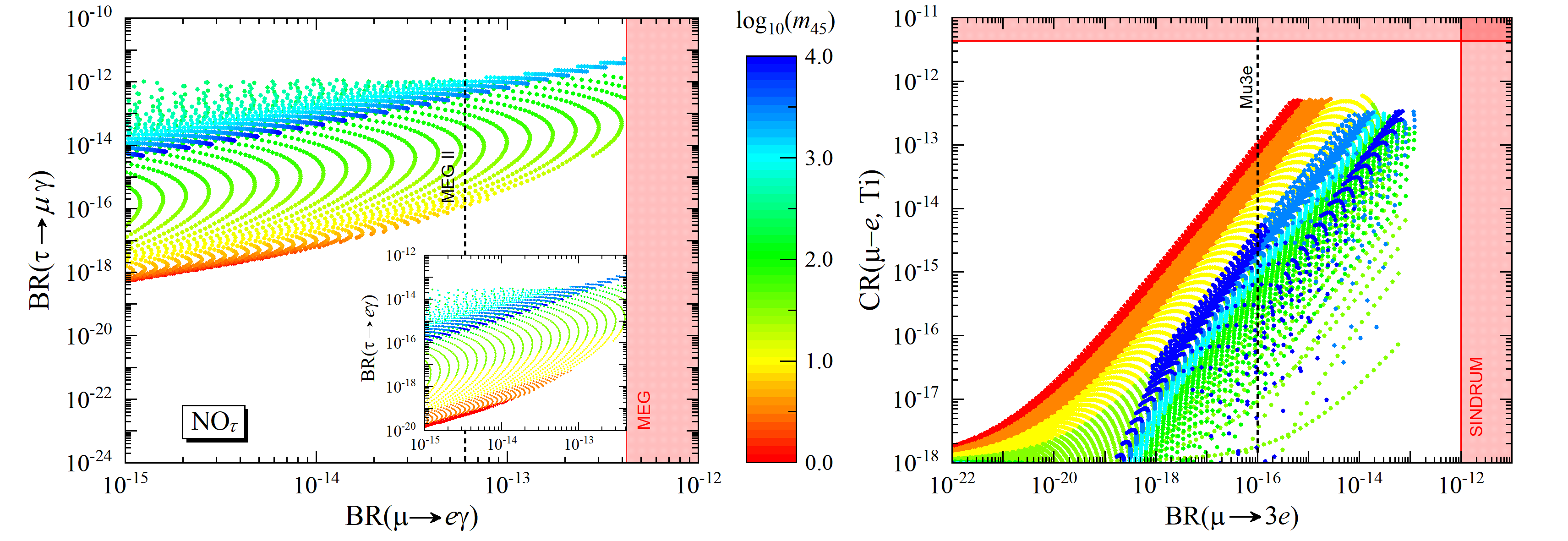}
	\end{tabular}
\caption{[Left] $\BR(\tau \rightarrow \mu \gamma)$ and $\BR(\tau \rightarrow e\gamma)$ (inner plots) vs. $\BR(\mu \rightarrow e \gamma)$ for NO$_{e,\mu,\tau}$ from top to bottom. The MEG bound $\BR(\tau \rightarrow e\gamma) < 4.2\times 10^{-13}$ sets the vertical red exclusion band, while the vertical dashed line corresponds to the MEG II projected sensitivity. [Right] $\CR(\mu-e,{\rm Ti})$ vs. $\BR(\mu \rightarrow 3e)$ for the same NO case as in the left panel. The vertical and horizontal red exclusion bands result from the SINDRUM and SINDRUM II limits on $\BR(\mu \rightarrow 3e)$ and $\CR(\mu-e,{\rm Ti})$, respectively (see Table~\ref{tab:boundsCLFV}). The vertical dashed line corresponds to the Mu3e projected sensitivity for $\mu \rightarrow 3e$. In all panels, the scatter points obey all constraints shown in Table~\ref{tab:boundsCLFV}, being their colour linked to the value of $m_{45}$ according to the colourmap shown in the middle.}
\label{Fig5}
\end{figure}

\begin{figure}[ht!]
	\centering
	\begin{tabular}{c}
	\includegraphics[width=1\textwidth]{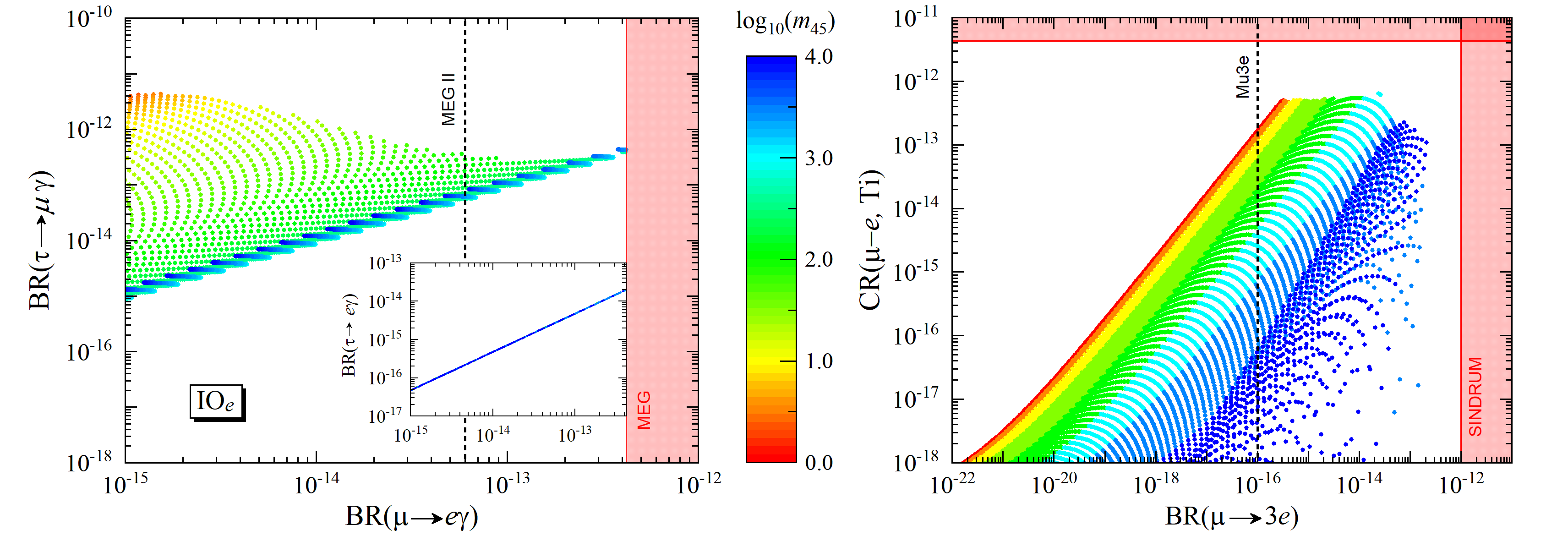}\\
		\includegraphics[width=1\textwidth]{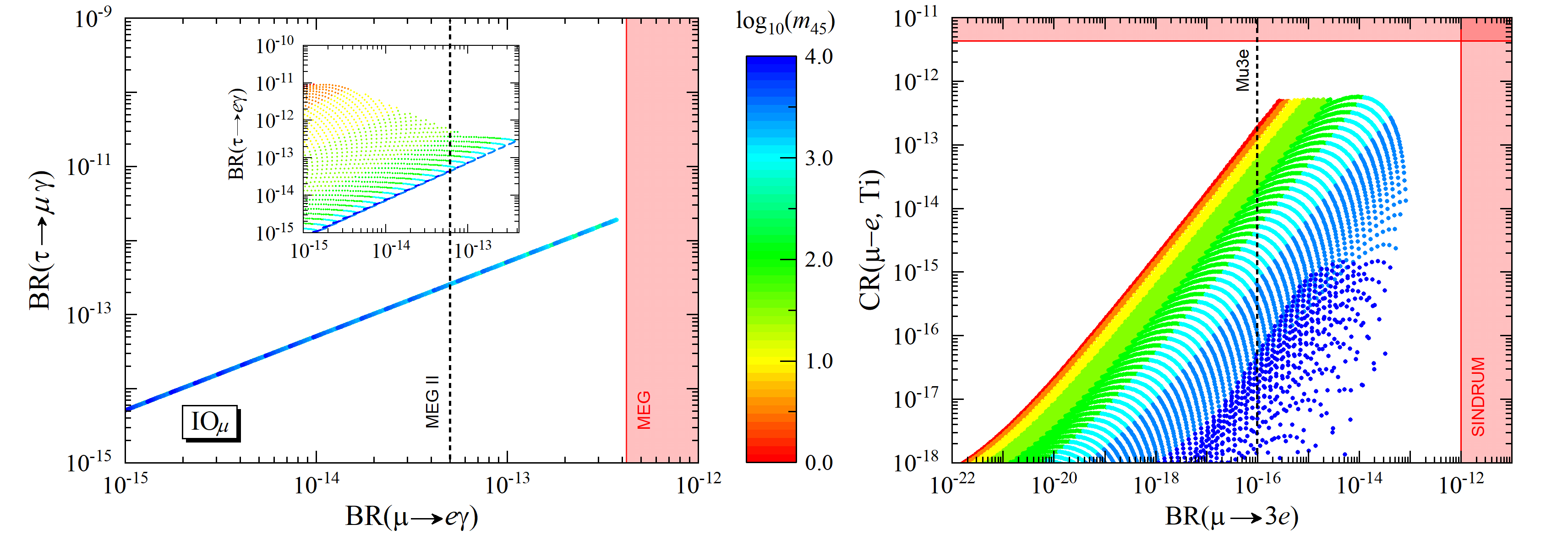}\\
	\includegraphics[width=1\textwidth]{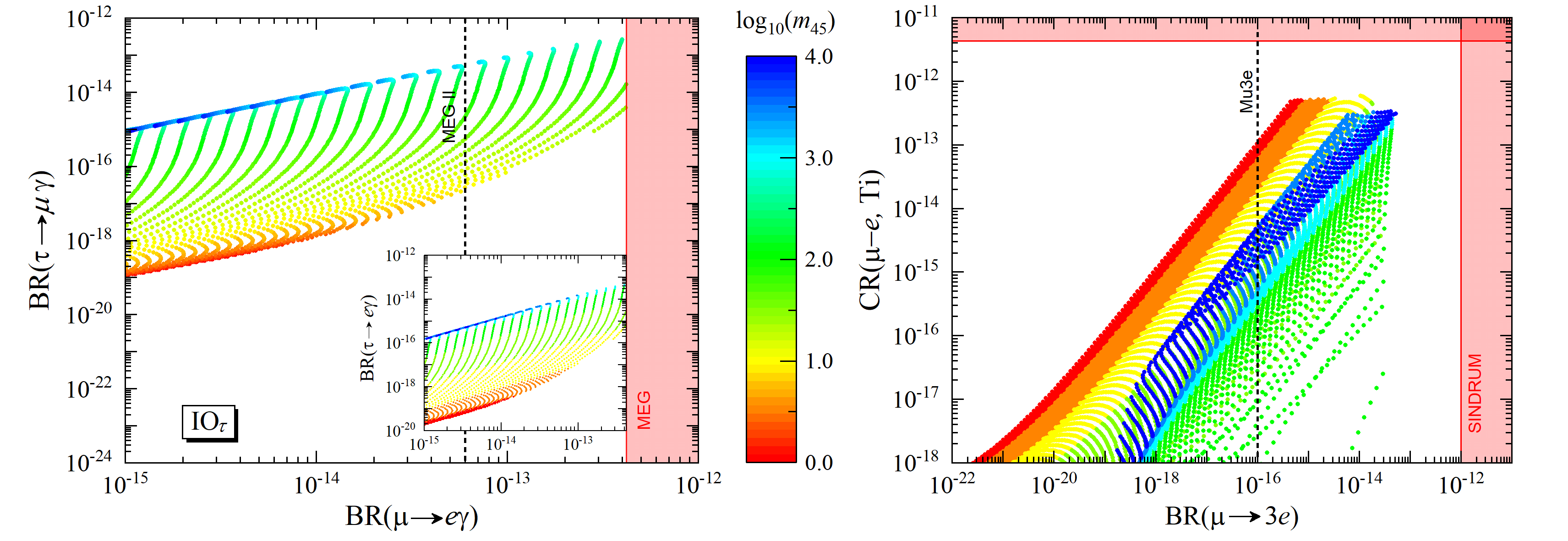}
	\end{tabular}
\caption{The same as in Fig.~\ref{Fig5} for IO$_{e}$, IO$_{\mu}$ and IO$_{\tau}$.}
\label{Fig6}
\end{figure}

The above results provide a general idea regarding how present experimental data constrain the minimal ISS with Abelian symmetries, and how future experiments would further probe its parameter space. However, it is interesting to investigate possible relations among different processes and, in particular, to ask whether the observation of a particular cLFV decay would allow us to draw conclusions regarding others. Notice that, in general, this is only possible when there is some relation among the LFV parameters and/or masses, as it is our case (see Table~\ref{tab:BREL}). With this purpose, we compare $\BR(\mu\rightarrow e \gamma)$ with $\BR(\tau\rightarrow \ell_{\alpha} \gamma)$, and $\CR(\mu-e,{\rm Ti})$ with $\BR(\mu \rightarrow 3e)$. The results are shown in Figs.~\ref{Fig5} and \ref{Fig6} for NO and IO, respectively, where all points respect the present limits shown in Table~\ref{tab:boundsCLFV}. The colour of each point is linked to the corresponding value of $m_{45}$ following the colour map shown in the middle of the figure. Some important conclusions stand out from the observation of these results. Namely, it is clear that any future observation of a radiative (or three-body) $\tau$ decay with a BR down to $10^{-9}$ would exclude all scenarios presented in this work. In fact, for both $\tau \rightarrow \mu \gamma$ and $\tau \rightarrow e \gamma$ the BRs are $10^{-11}$ at most, well below the Belle II sensitivity (left panels). In general, the spreading of the scatter points is due to variations in heavy-light mixing and heavy neutrino masses. However, in some cases, we observe a linear relation between $\BR(\tau\rightarrow \mu \gamma)$ or $\BR(\tau\rightarrow e\gamma)$ and $\BR(\mu\rightarrow e\gamma)$. For instance, using Eqs.~\eqref{eq:Gamma_abg} and~\eqref{eq:Brmutau} together with the approximate results of Table~\ref{tab:BREL}, it can be shown that for NO$_e$ the relation
\begin{align}
    \frac{\BR(\tau\rightarrow e\gamma)}{\BR(\mu\rightarrow e\gamma)} \simeq \left|\dfrac{\B_{\tau 4}}{\B_{\mu 4}}\right|^2 \BR(\tau\rightarrow \nu_\tau e \bar{\nu}_e) \simeq 0.013
\end{align}
holds for the whole range of $m_{45}$, in perfect agreement with the numerical results shown in the inner plot of the upper-left panel in Fig.~\ref{Fig5}. From the comparison of $\CR(\mu-e,{\rm Ti})$ with $\BR(\mu \rightarrow 3e)$ (right panels of Figs.~\ref{Fig5} and \ref{Fig6}) we see that in all cases $\BR(\mu \rightarrow 3e)$ is at most $10^{-13}$. Notice also that, for $m_{45}\lesssim 100$~GeV, there is an approximate linear relation between $\CR(\mu-e,{\rm Ti})$ with $\BR(\mu \rightarrow 3e)$, which is no longer valid for higher masses due to cancellations among the various contributions to $\mu-e$ conversion amplitudes. We have also investigated whether our framework could lead to observable signals in LVF $Z$ decays by computing $\BR(Z \rightarrow \ell_{\alpha} \ell_\beta)$ with $\alpha\neq \beta =e,\mu,\tau$. We have concluded that, after imposing the present constraints on $\mu\rightarrow e \gamma$ and~$\mu-e$ conversion in nuclei, the $Z \rightarrow \ell_{\alpha} \ell_\beta$ rates are well below the future sensitivities given in Table~\ref{tab:boundsCLFV}.
%
%*******************************************************************************************************
%*******************************************************************************************************
\section{Constraints on heavy sterile neutrinos and future prospects}
\label{sec:CONST}

In the previous section we have analysed the constraints imposed by LFV experimental searches on the minimal inverse seesaw model with Abelian flavour symmetries, adopting as reference parameters the LNV parameter $\mu_s$ and the average sterile neutrino mass $m_{45}$. Although this provides a clear understanding on how constrained the original scales in the Lagrangian are, it is convenient to look at the problem from a perspective where $\mu_s$ is replaced by the active-sterile mixing parameters $\B_{\alpha j}$. Notice that, in our framework, we only need to consider one of these quantities since, as seen in Section~\ref{sec:numass} and summarised in Table~\ref{tab:BREL}, they are all correlated. From now on, we will take as constrained parameters $m_{45}$ and $V_{eN}^2$ defined in Eq.~\eqref{eq:m45VeN}. For each of the scenarios analysed in this work, the correspondence between the $(m_{45},\mu_s)$ and $(m_{45},V_{eN}^2)$ parameter spaces is established by figures like Fig.~\ref{Fig3} (left panels). Following this approach, we will be able to compare the constraining power of the cLFV processes discussed in the previous section with other experimental searches which are usually translated into constraints on mass and mixing parameters~(see e.g. Refs.~\cite{,Atre:2009rg,Deppisch:2015qwa,deGouvea:2015euy,Abada:2015oba,Fernandez-Martinez:2016lgt,Caputo:2017pit,Strategy:2019vxc}). We are not interested in carrying out an exhaustive analysis of all sensitivity studies performed so far. Instead, we will consider the following searches:\\

\begin{itemize}
    \item {\bf Beam-dump experiments:} In a beam-dump experiment a primary beam strikes a high-density target and produces a large number of secondary heavy mesons which, in the presence of active-sterile mixing, can decay to final states with sterile neutrinos. Part of them fly towards a detector, decaying inside its volume. The NA3 experiment used a 300~GeV $\pi^-$ beam incident on 2 meter long beam dump to, among other purposes, look for the decays of heavy neutrinos $N$ into $\ell\ell\nu$ and $\pi^+\ell^-$ ($\ell=e,\mu$) final states. In such experiment, heavy neutrinos may originate from the rare $\pi$ and $K$ meson decays, or in semi-leptonic decays of charm $D$ and $F$ or beauty $B$ mesons. At NA3 the limit $V_{eN}^2 \lesssim 10^{-4}$ has been set within a heavy neutrino mass region $1-2$~GeV. The CHARM experiment~\cite{Bergsma:1985is} has conducted similar $N\rightarrow \ell\ell\nu$ searches using a prompt neutrino beam produced by dumping 400~GeV protons on a copper beam dump, setting $V_{eN}^2 \lesssim 10^{-7}$ for $1-2$~GeV heavy neutrinos. We will consider the NA3 and CHARM exclusion regions reported in Figs.~8a and 2 of Refs.~\cite{Badier:1986xz} and \cite{Bergsma:1985is}, respectively.\\
    
    {\bf Future:} The beam-dump experiment SHIP will use 400~GeV protons extracted from the CERN Super Proton Synchrotron (SPS) accelerator, dumped on a high-density target. The experiment will search for heavy neutrinos with mass up to $\sim 6$~GeV produced in the decays of $D$ and $B$ mesons. As for active-sterile mixing, SHIP will be able to probe it down to $V_{eN}^2\sim 10^{-10}$ for 1.6~GeV heavy neutrinos. In this work, we will consider the SHIP exclusion region given in Fig.~3 of Ref.~\cite{SHiP:2018xqw}. Although designed to measure active neutrino oscillation parameters with high precision~\cite{Acciarri:2016crz}, the DUNE experiment will also be able to search for heavy sterile neutrinos~\cite{Krasnov:2019kdc}. This will be achieved by striking a target with a very high-energy proton beam (up to 120 GeV), leading to the production of mainly pions and kaons which may produce sterile neutrinos in their decays. For illustration, we will consider the results presented in Fig.~2 of Ref.~\cite{Krasnov:2019kdc}.
    
    \item {\bf High-energy colliders:} At the Large Electron-Positron (LEP) collider, the L3 and DELPHI collaborations have looked for heavy neutrinos $N$ produced via on-shell $Z$ boson decays $e^+ e^- \rightarrow Z \rightarrow N \nu$. Several $N$ decay modes were considered, namely $N\rightarrow Z^\ast \nu \,(Z^\ast \rightarrow \ell\ell,\nu\nu,jj)$ and $N\rightarrow W^\ast \ell \,(W^\ast \rightarrow \ell\nu_\ell,jj')$. The L3 results were able to probe $V_{eN}^2$ down to $10^{-4}$ for $\sim 20$~GeV heavy neutrinos, while the DELPHI collaboration conducted similar searches excluding $V_{eN}^2\gtrsim 3\times 10^{-5}$ for masses in the range $3-50$~GeV. We will consider the L3 and DELPHI exclusion regions given in Figs. 6 and 10 of Refs.~\cite{Adriani:1992pq} and \cite{Abreu:1996pa}, respectively.
    
    At the LHC, the ATLAS and CMS collaborations are also looking for heavy neutrino signals. Both collaborations search for $N$ production in $W^\pm\rightarrow \ell^\pm N$ followed by subsequent decays $N\rightarrow W^{\pm\ast} \ell^\mp\,(W^{\pm\ast}  \rightarrow \ell^\pm\nu_\ell)$ with $\ell=e,\mu$. ATLAS has explored event signatures consisting of three charged leptons (electrons and muons) with same-sign dileptons of the same flavour (LNV mode). CMS has extended the search to include events with lepton number conservation, thus being sensitive to displaced decays. Overall, the ATLAS and CMS analyses on trilepton signatures excluded $V_{eN}^2 \gtrsim 10^{-4}-10^{-5}$ for heavy neutrino masses in the $5-50$~GeV range. Both collaborations have also searched for the decays of heavy neutrinos produced in $pp\rightarrow W^{\pm\ast} \rightarrow \ell^\pm N $ into same-sign dileptons and jets $N\rightarrow W^\pm \ell^\pm\, (W^\pm\rightarrow jj')$. For the ATLAS exclusion regions we will consider the results given in Figs.~6 and 8 of Refs.~\cite{Aad:2019kiz} and \cite{Aad:2015xaa}, while for the CMS ones we take the results of Figs.~2 and 4 of Refs.~\cite{Sirunyan:2018mtv} and \cite{Sirunyan:2018xiv}, respectively. In the presence of active-sterile neutrino mixing, new interactions of the SM Higgs boson may arise, opening the $H^0 \rightarrow N\nu$ decay channel (if kinematically allowed). The subsequent decays $N\rightarrow \ell W^\ast\,(W^\ast \rightarrow \ell \nu)$ and $N\rightarrow \ell Z^\ast\,(Z^\ast \rightarrow \ell^+ \ell^-)$ at the LHC have been studied in Ref.~\cite{Das:2017zjc} to constrain the mixing-mass parameter space as shown in Fig.~3 of that reference.

    {\bf Future:} Future high-energy colliders will play a crucial role in searching for heavy sterile neutrinos. In particular, during the high-luminosity LHC phase (HL-LHC), ATLAS and CMS will be able to cover masses up to $2-3$~TeV. Sensitivity studies have also been performed for a Future Circular Hadron Collider (FCC-hh) at a 100~TeV centre-of-mass energy~\cite{Antusch:2016ejd,Pascoli:2018heg}. For the HL-LHC and FCC-hh cases we will consider the exclusion regions given in Fig.~25 of Ref.~\cite{Pascoli:2018heg} corresponding to LHC14 and LHC100 with integrated luminosities $\mathcal{L}=3~{\rm ab}^{-1}$ and 15~ab$^{-1}$, respectively. Heavy-neutrino searches performed at a future high-luminosity $e^+e^-$ storage ring collider (FCC-ee) can drastically improve the limits on active-sterile mixing down to $10^{-11}$ for $\sim 60$~GeV neutrinos (Fig.~8 of Ref.~\cite{Blondel:2014bra} ). In a future $e^+e^-$ linear collider, as the International Linear Collider (ILC), the sensitivity on heavy-light neutrino mixing can reach values down to $10^{-4}$ for a 500~GeV centre-of-mass energy and an integrated luminosity of $100\,{\rm fb}^{-1}$ (Fig.~15 of Ref.~\cite{Banerjee:2015gca}). For a Compact Linear Collider (CLIC) operating at 3~TeV and with $\mathcal{L}=1\,{\rm ab}^{-1}$, values of $V_{eN}\sim 10^{-5}-10^{-4}$ can be probed for a $600\,\GeV-2.5\,\TeV$ mass range (Fig.~24 of Ref.~\cite{Das:2018usr}).
    
    Detectors placed near LHC interaction points would allow for searches of heavy-sterile neutrinos produced in $pp$ collisions through the reconstruction of displaced vertices in a low-background environment. Several proposals have been been put forward to conduct this kind of analyses, namely the AL3X~\cite{Dercks:2018wum}, CODEX-b~\cite{Gligorov:2017nwh}, FASER2~\cite{Feng:2017uoz}, MATHUSLA~\cite{Curtin:2018mvb} and MoEDAL~\cite{Frank:2019pgk} detectors. The sensitivity improvement with respect to that achieved by the main detectors (ATLAS, CMS and LHCb) could be of several orders of magnitude in the low-mass regime. In this work we will consider the FASER2 and MATHUSLA exclusion regions given in Figs.~5 and 37 of Refs.~\cite{Kling:2018wct} and \cite{Curtin:2018mvb}, respectively.\\
    
    Given that we are dealing with nearly-degenerate sterile neutrinos due to the smallness of $\mu_s$, LNV decay modes are expected to be suppressed as a result of the quasi-Dirac nature of the heavy sterile neutrinos. This is, however, not the case if the average decay width is of the order of the mass splitting, i.e. $\Gamma_N = (\Gamma_{N_1}+\Gamma_{N_2})/2 \simeq \Delta m_N$. This issue is especially relevant when looking for the same-sign dilepton signatures discussed above. In order to provide an insight regarding whether LNV decays are suppressed or not, we will use the same-sign to opposite-sign ratio~\cite{Deppisch:2015qwa,Anamiati:2016uxp}
    \begin{align}
    R_{ll}=\frac{\Delta m_N^2}{2\Gamma_N^2+\Delta m_N^2}\,,
    \end{align}
    such that $R_{ll}\ge 1/3$ can be adopted as a criterion to identify the regions of the parameter space where LNV decays are unsuppressed~\cite{Drewes:2019byd}. To compute $\Gamma_N = (\Gamma_{4}+\Gamma_{5})/2$ in terms of the sterile neutrino masses and mixings we use the results of Ref.~\cite{Atre:2009rg}.\\
    
   \item {\bf Electroweak precision data (EWPD):} As already mentioned, in the presence of sterile neutrinos, the active neutrino mixing matrix~$\mathbf{U}$ relevant for neutrino oscillations [cf. Eq.~\eqref{eq:nonunitpmns}] is no longer unitary. Deviations from unitarity are constrained by neutrino oscillation data, electroweak precision tests and lepton flavour violating decays~\cite{delAguila:2008pw,Antusch:2006vwa,delAguila:2008pw,Akhmedov:2013hec,Antusch:2014woa,Antusch:2016brq,Fernandez-Martinez:2016lgt,Blennow:2016jkn,Coutinho:2019aiy,Coutinho:2020xhc}. In fact, the off-diagonal elements of $\boldsymbol{\eta}_{\alpha\beta}$ defined in Eq.~\eqref{eq:epsB} are mainly restricted by the LFV decays studied in the previous section. On the other hand, $\boldsymbol{\eta}_{\alpha\alpha}$ are restricted by SM gauge boson decays, namely $W \rightarrow \ell_{\alpha} \nu_{\alpha}$ and $Z \rightarrow \nu \nu$, and universality tests in $W$ and $\pi$ decays. We will use here the limits for $|\boldsymbol{\eta}_{\alpha\alpha}|$ obtained in Ref.~\cite{Fernandez-Martinez:2016lgt}, namely: 
\begin{equation}
\left|\boldsymbol{\eta}_{ee}\right|< 1.25 \times 10^{-3}, \quad \left|\boldsymbol{\eta}_{\mu \mu}\right|<2.2 \times 10^{-4}, \quad \left|\boldsymbol{\eta}_{\tau \tau}\right|<2.8\times 10^{-3}\,.
\label{eq:epslim}
\end{equation}
Notice that, in our framework, the $\boldsymbol{\eta}_{\alpha\beta}$ are not independent since the $\B_{\alpha i}$ are related to each other as a result of the Abelian flavour symmetries. These relations are written in terms of low-energy neutrino observables as shown in Eq.~\eqref{eq:Baj}, implying that $\boldsymbol{\eta}_{\alpha\beta}$ can be expressed by a single mixing parameter which, for convenience of our analysis, we choose to be $V_{eN}^2=|\B_{e 4}|^2$. In this case, we have
\begin{equation}
|\boldsymbol{\eta}_{\alpha\alpha}|=\frac{1}{2} \sum_{i=4}^7 |\B_{\alpha i}|^2 
\simeq  V_{eN}^2 (x_{\alpha 4} + x_{\alpha 6})\;,\; x_{\alpha j}=\left|\frac{\B_{\alpha j}}{\B_{e 4}}\right|^2\,,
\end{equation}
where we have used Eq.~\eqref{eq:epsB} and the fact that $|\B_{\alpha 4}|\simeq |\B_{\alpha 5}|$ and $|\B_{\alpha 6}|\simeq |\B_{\alpha 7}|$. It is then possible to use the above equations together with Table~\ref{tab:BREL} to compute the $x_{\alpha i}$ factors and extract upper bounds on $V_{eN}^2$ from the limits given in Eq.~\eqref{eq:epslim} for $|\boldsymbol{\eta}_{\alpha\alpha}|$ (see Table~\ref{tab:VeNlimits}). \\

\begin{table}[!t]
\renewcommand{\arraystretch}{1.5}
\setlength{\tabcolsep}{8pt}
\centering
\begin{tabular}{l|ccc}   
\hline
 & $\left|\boldsymbol{\eta}_{ee}\right|$%< 1.25 \times 10^{-3}$ 
 & $\left|\boldsymbol{\eta}_{\mu \mu}\right|$%<2.2 \times 10^{-4}$ 
 & $\left|\boldsymbol{\eta}_{\tau \tau}\right|$%<2.8\times 10^{-3}$
 \\ \hline
NO$_e$ [IO$_e$]  & $1.25\times 10^{-3}\,[1.25\times 10^{-3}]$
& $3.90\times 10^{-6}\,[3.42\times 10^{-4}]$
& $1.97\times 10^{-4}\,[4.01\times 10^{-3}]$ \\ \hline
NO$_\mu$ [IO$_\mu$] 
& $2.27\times 10^{-4}\,[2.44\times 10^{-5}]$
& $6.58\times 10^{-6}\,[9.29\times 10^{-6}]$
& $7.43\times 10^{-5}\,[9.05\times 10^{-5}]$\\ \hline
NO$_\tau$ [IO$_\tau$]  
& $3.86\times 10^{-4}\,[7.64\times 10^{-5}]$
& $4.17\times 10^{-6}\,[3.35\times 10^{-5}]$
& $1.02\times 10^{-4}\,[3.09\times 10^{-4}]$ \\ \hline
\end{tabular}
\caption{Upper bounds on $V_{eN}^2$ imposed by EWPD (see text for details) for NO$_{e,\mu,\tau}$ and IO$_{e,\mu,\tau}$.}
\label{tab:VeNlimits}
\end{table}

\item {\bf Neutrinoless double beta decay:} In the presence of sterile neutrinos, the effective neutrino mass parameter $m_{\beta\beta}$, relevant for neutrinoless double beta decay, is~\cite{Blennow:2010th}
\begin{equation}
m_{\beta\beta} \simeq \sum_{i=1}^{n_f} \mathbf{B}_{e i}^2 \ p^2 \frac{m_i}{p^2-m_i^2} \simeq  \sum_{i=1}^{3} \mathbf{B}_{e i}^2 m_i + \sum_{i=4}^{7} p^2 \mathbf{B}_{e i}^2 \ \frac{m_i}{p^2-m_i^2}\,,
\label{eq:ndbd}
\end{equation}
where $p^2 \simeq - (100\ \text{MeV})^2$ is the virtual momentum of the neutrinos and $m_i$ are the physical neutrino masses. The first and second sums run over the number of light and heavy neutrinos which, in the present case, is three and four, respectively. For the 1~GeV--10~TeV mass range studied in this work, the contributions of the second term in Eq.~\eqref{eq:ndbd} are negligible and, thus, the results in Fig.~\ref{Fig2} remain valid (for neutrinoless double beta decay studies in the presence of sterile neutrinos see e.g. Refs.~\cite{Mitra:2011qr,LopezPavon:2012zg,Girardi:2013zra,Abada:2014nwa,Lopez-Pavon:2015cga,Abada:2018qok,Bolton:2019pcu}).
\end{itemize}
\begin{figure}[ht!]
	\centering
	\begin{tabular}{c}
	\includegraphics[width=0.76\textwidth]{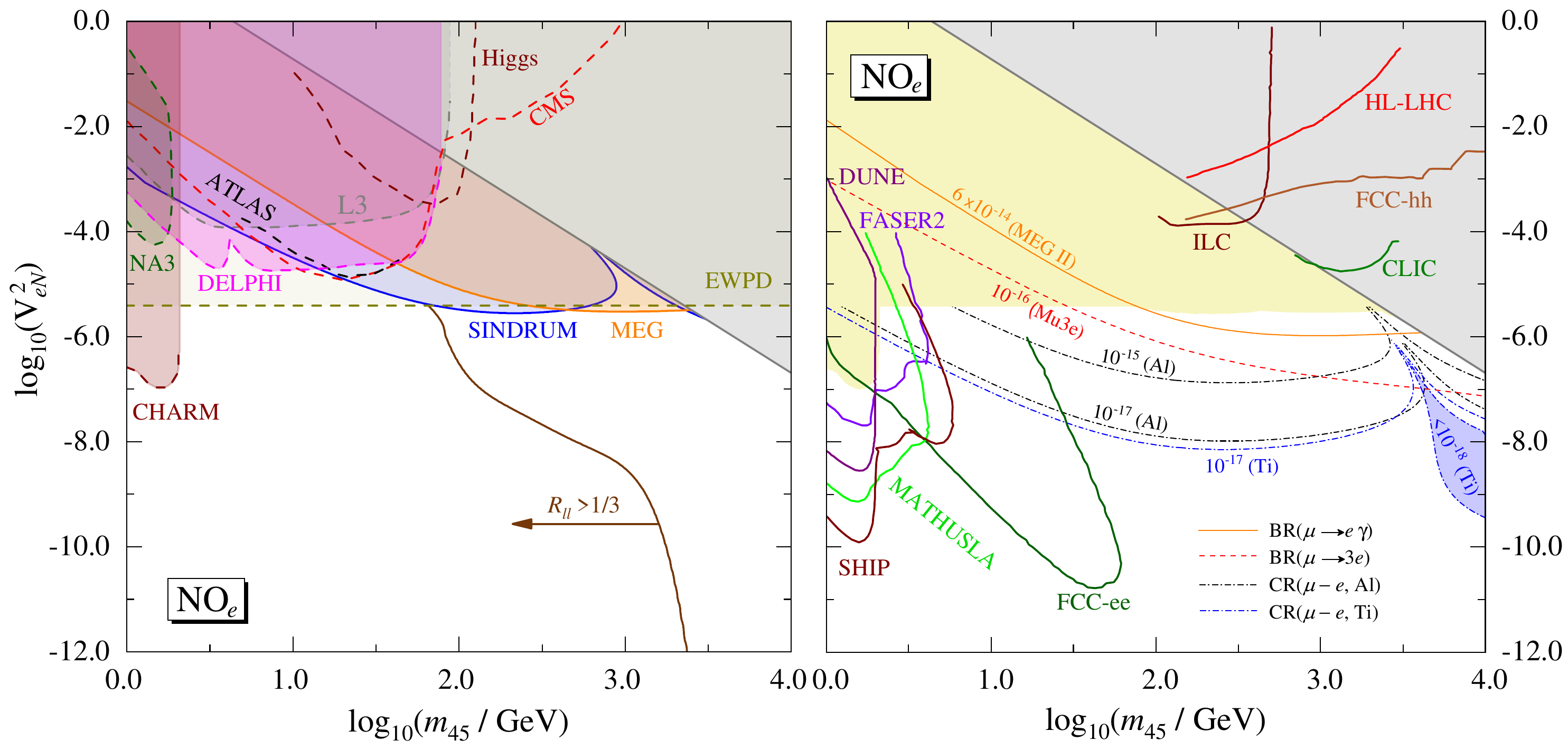}\\
	\includegraphics[width=0.76\textwidth]{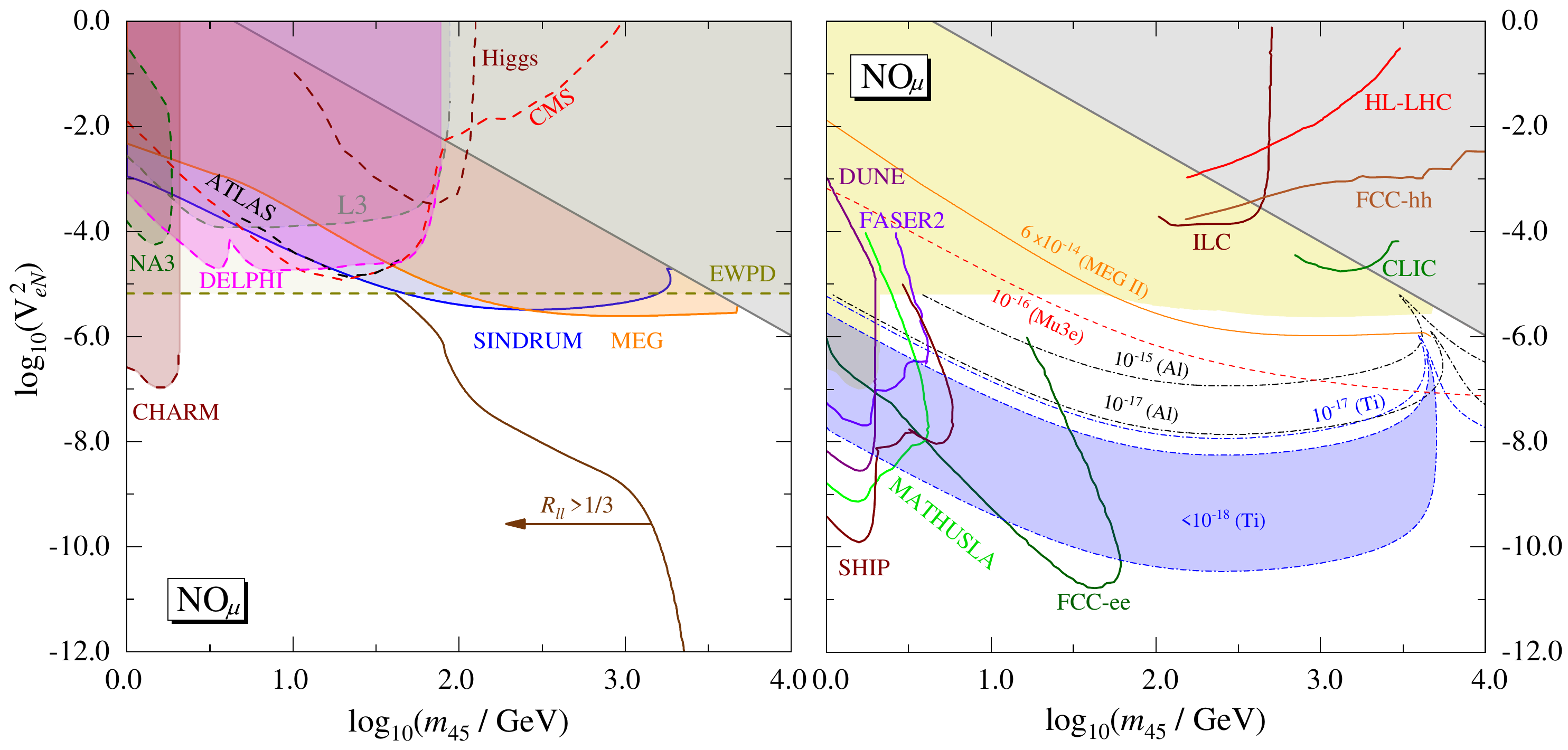}\\
	\includegraphics[width=0.76\textwidth]{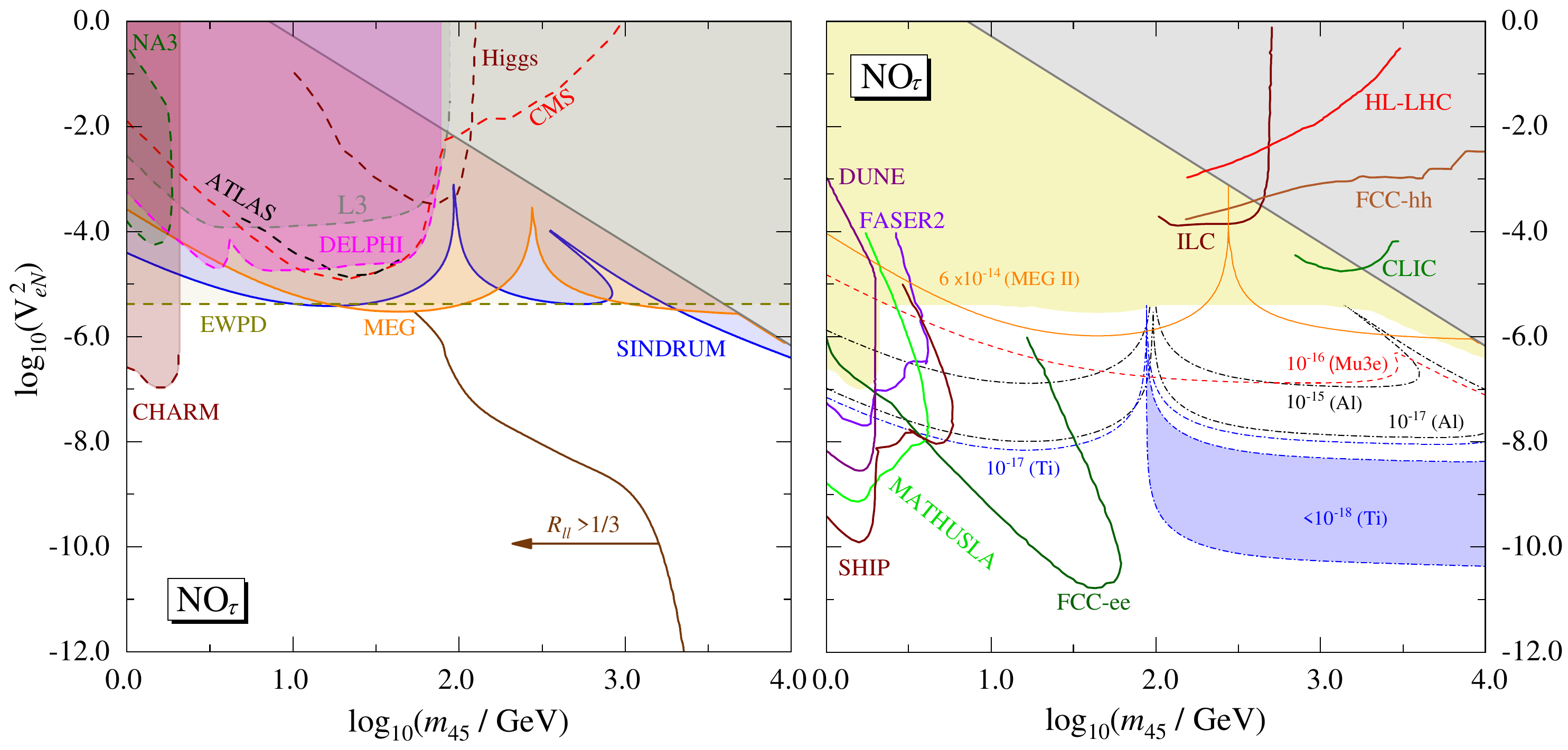}
	\end{tabular}
\caption{[Left] Constraints imposed on the $(m_{45},V_{eN}^2)$ parameter space by the MEG and SINDRUM limits on $\BR(\mu\rightarrow e \gamma)$ and $\CR(\mu-e,{\rm Au})$ (see Section~\ref{sec:pheno}), by the current searches conducted at colliders and beam-dump experiments and by EWPD (see discussion in the main text where the sources of the several exclusion regions are indicated). As in Figs.~\ref{Fig5} and~\ref{Fig6}, $b_i^{\rm max} >5$ within the grey-shaded region. To the left of the solid brown line $R_{ll} > 1/3$. [Right] Projected sensitivities for cLFV searches and other experiments discussed in the main text. The yellow-shaded regions correspond to overlapping the current constraints shown on the left panels. Inside the blue shaded region $\CR(\mu-e,{\rm Ti})<10^{-18}$. The top (middle) [bottom] panels correspond to the NO$_{e}$ (NO$_{\mu}$) [NO$_{\tau}$] case.}
\label{Fig7}
\end{figure} 
\begin{figure}[ht!]
	\centering
	\begin{tabular}{c}
	\includegraphics[width=0.77\textwidth]{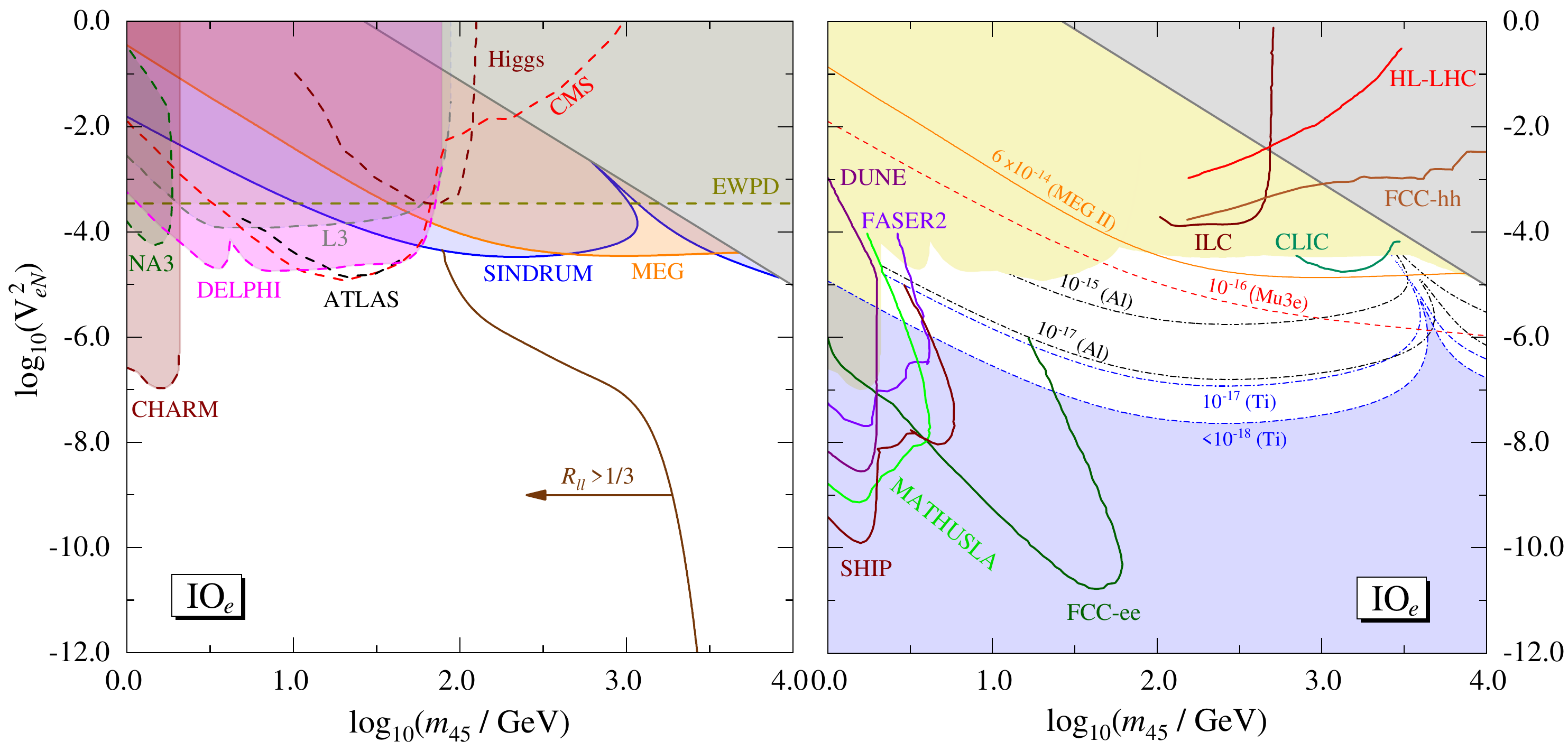}\\
	\includegraphics[width=0.77\textwidth]{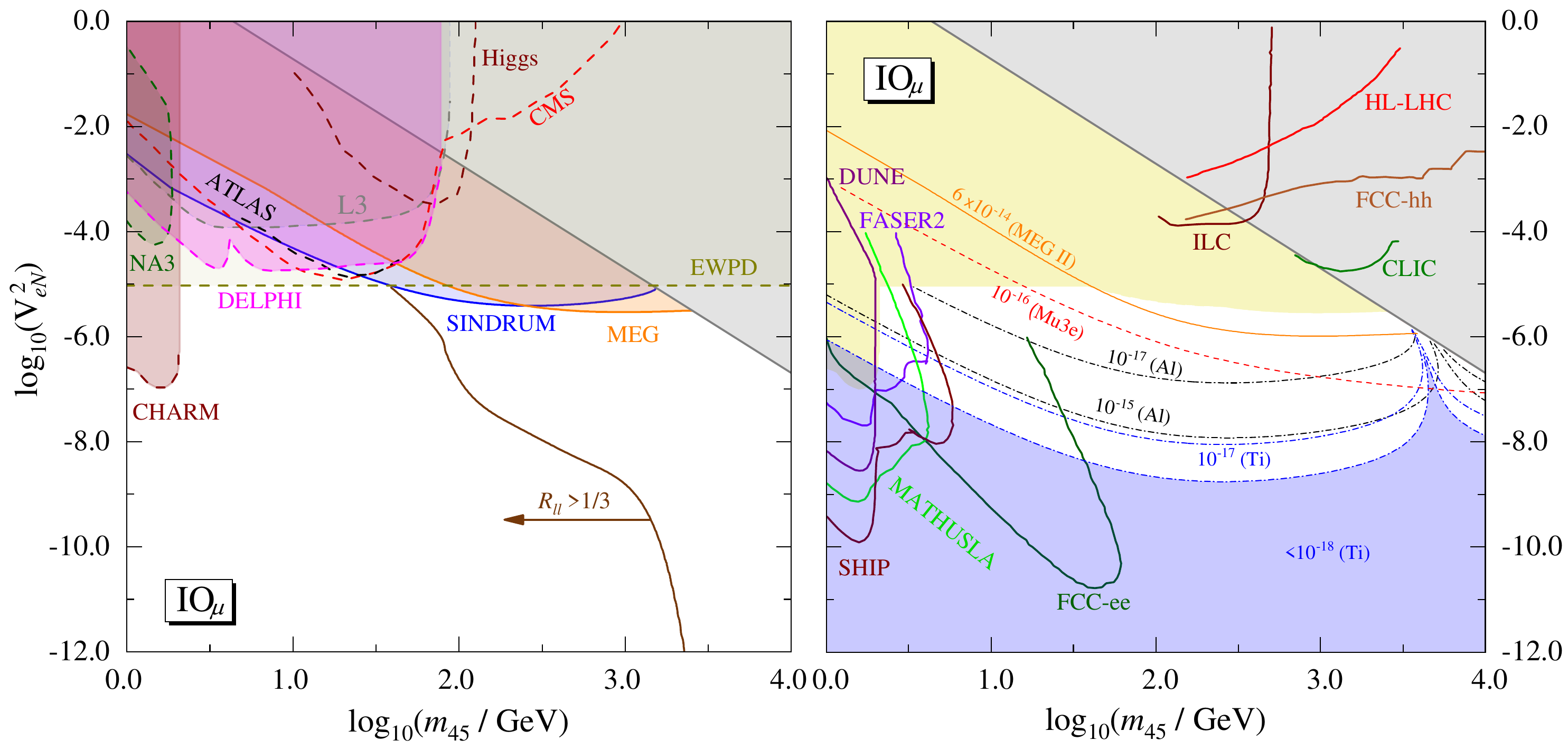}\\
	\includegraphics[width=0.77\textwidth]{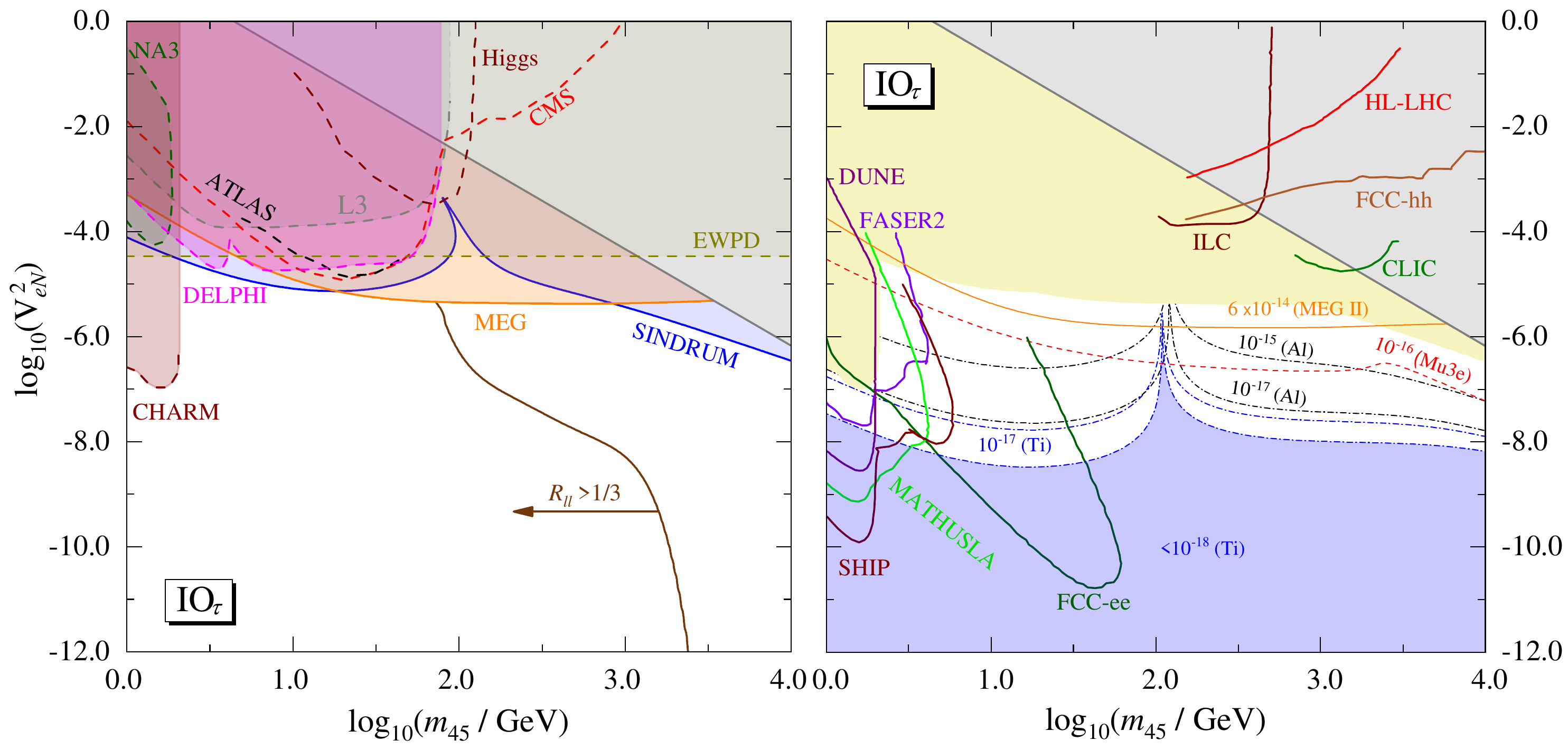}
	\end{tabular}
\caption{The same as in Fig.~\ref{Fig7} for IO$_{e}$, IO$_{\mu}$ and IO$_{\tau}$.}
\label{Fig8}
\end{figure} 

In the left panels of Figs.~\ref{Fig7} and \ref{Fig8} we present a summary of all the current constraints discussed above, together with those stemming from $\mu\rightarrow e \gamma$ (MEG) and $\mu-e$ conversion in Au (SINDRUM) searches (see Fig.~\ref{Fig5}), now shown in the $(m_{45},V_{eN}^2)$ plane. For the EWPD exclusion regions we consider the most restrictive $V_{eN}^2$ limits given in Table~\ref{tab:VeNlimits}, i.e. those extracted from $\left|\boldsymbol{\eta}_{\mu \mu}\right|$ (third column). On the right of the same figures, the projected sensitivities of the several experiments enumerated above are shown, including the cLFV ones already presented in Fig.~\ref{Fig5} in the~$(m_{45},\mu_s)$ plane. For all cases, the overlap of the current exclusion regions (left panels) is shown in light yellow. By looking at these two figures one can conclude that:

\begin{itemize}
    \item For $m_{45}\gtrsim 2~\GeV$, the strongest constraints are typically those imposed by the SINDRUM and MEG limits on $\BR(\mu\rightarrow e \gamma)$ and $\CR(\mu-e,{\rm Au})$, respectively, and by EWPD (left panels). One exception is the IO$_e$ case where, for $2\,\GeV \lesssim m_{45}\lesssim 50~\GeV$, the DELPHI, ATLAS and CMS limits are stronger. In all situations, the CHARM exclusion region is more constraining when $m_{45}=1-2$~GeV. Notice that the EWPD exclusion regions are not the same for the different NO and IO scenarios since the U(1) flavour symmetries, together with present neutrino data, impose different relations among the $\B_{\alpha j}$. Thus, the limits on  $\left|\boldsymbol{\eta}_{\alpha\alpha}\right|$ cannot be directly translated into limits of a single $\B_{\alpha j}$ by neglecting the remaining $\B_{\alpha k}$ with different $k\neq j$. 
    
    \item Any signal of sterile neutrinos with $V_{eN}^2 \gtrsim 10^{-4}$ at future hadron or linear colliders (HL-LHC, FCC-hh, CLIC and ILC regions) would not be compatible with the limits already imposed by current constraints from LFV searches and EWPD (see right panels). Therefore, in the context of the present work, high-energy collider probes conducted at the FCC-ee and at experiments like SHIP, MATHUSLA, DUNE and FASER2 turn out to be of utmost importance (obviously, taking into account the considered sensitivity studies).
    
    \item While for NO$_e$ cLFV indirect searches are fully complementary to the aforementioned direct ones, this is not the case for the remaining NO and IO scenarios. In particular, for inverted neutrino masses, the region with $V_{eN}^2 \lesssim 10^{-9}-10^{-8}$ cannot be probed by future $\mu-e$ conversion experiments. In this case, such mixing regimes can be covered by displaced-vertex experiments and by a high-luminosity $Z$ factory like the FCC-ee. Notice that $R_{ll} \ge 1/3$ within the sensitivity regions of those searches (see the brown solid lines in the left panels), indicating that LNV sterile neutrino decays are not suppressed by their quasi-Dirac nature. It should also be mentioned that in the absence of a positive $\mu\rightarrow e \gamma$ signal, the impact of MEG II data would be mild (compare the yellow regions with the solid orange line in the right panels). Instead, if that decay is observed, ranges for $m_{45}$ and $V_{eN}^2$ can be set in most of the cases, being the latter relatively narrow. As for $\mu \rightarrow 3e$, future probes conducted by the Mu3e collaboration will be able to cover $V_{eN}^2$ down to $10^{-6}-10^{-7}$ for wide ranges of sterile neutrino masses.
\end{itemize}

To conclude this section, we remark that although we have shown the results of our analysis in the $(m_{45},V_{eN}^2)$ plane, it is relatively easy to infer how the obtained exclusion regions and sensitivity contours would appear choosing a different mixing parameter by taking into account the relations in Table~\ref{tab:BREL}.

%
%*******************************************************************************************************
%*******************************************************************************************************
\section{Concluding remarks}
\label{sec:conclusion}

In this paper we have thoroughly investigated the minimal inverse-seesaw mechanism with couplings constrained by U(1) flavour symmetries, and with all fermion masses generated via spontaneous symmetry breaking through VEVs of doublet and singlet scalar fields. After finding the maximally-restrictive mass matrices compatible with present neutrino data, we have identified all possible U(1) symmetry realisations and concluded that at least two Higgs doublets and two complex scalar singlets are required to successfully implement those symmetries. The presence of such singlets opens up the possibility for SCPV, which turns out to be successfully communicated to the lepton sector via their couplings to the new sterile fermions. As a result of SCPV and the Abelian symmetries, the low-energy Majorana and Dirac CP phases turn out to be related to each other. We have also shown that, including one-loop corrections to neutrino masses and requiring them to be at the one percent level, sterile-neutrino mass ranges are established, within which the tree-level results are still valid in light of the present experimental precision in the determination of the oscillation parameters. Due to the flavour symmetries, the heavy-light mixings are not independent, being their ratios entirely determined by the lepton masses, mixing angles and CP phases. This provides a very constrained setup for phenomenological studies in the framework of current and future probes sensitive to the presence of sterile neutrino states. We have studied several cLFV decays and obtained the exclusion regions set by the experimental limits on $\BR(\mu\rightarrow e \gamma)$ and $\CR(\mu-e,{\rm Au})$. These results establish upper bounds on $V_{eN}^2$ of about $10^{-4}-10^{-5}$. The prospects to further explore the parameter space were discussed in view of the projected sensitivities of future LFV searches, especially those dedicated to $\mu\rightarrow e \gamma$, $\mu \rightarrow 3e$ and $\mu-e$ conversion in nuclei.

After analysing the constraining power of cLFV processes, we focused our analysis on alternative probes, namely collider and beam-dump experimental searches that are sensitive to the presence of sterile neutrinos. With minor exceptions, we concluded that the HL-LHC, FCC-hh, ILC and CLIC sensitivity regions are already excluded by current LFV and EWPD constraints for all possible U(1) symmetry realisations. On the other hand, searches at a high-luminosity $Z$ factory as the FCC-ee and at experiments like SHIP, MATHUSLA and FASER2 would be highly complementary to the Mu3e, Mu2e, COMET and PRISM/PRIME projects. Although we have not explored all possible future scenarios which could arise from independent results of different searches, it is clear that a single positive signal in any of those experiments would definitely put at test the scenarios studied in this paper. In this sense, further symmetry-motivated studies performed in the context of sterile neutrino searches are most welcome.

\vspace{1cm}
{\bf Acknowledgements:} This work has been supported by Funda\c{c}{\~a}o para a Ci{\^e}ncia e a Tecnologia (FCT, Portugal) through the projects UIDB/00777/2020, UIDP/00777/2020, CERN/FIS-PAR/0004/2019, and PTDC/FIS-PAR/29436/2017.

\appendix

\section{The scalar sector and spontaneous CP violation}
\label{sec:ScalarSector}

The Abelian symmetries that realise some of the maximally-restrictive textures of Section~\ref{sec:compdata} were presented in Section~\ref{sec:symmetries}, being the minimal scalar-field content required to implement those symmetry realisations given in Eq.~\eqref{eq:scalarfield}, i.e., two Higgs doublets $\Phi_{1,2}$ and two complex singlets $S_{1,2}$:
\begin{equation}
\Phi_a =\begin{pmatrix}
\phi_a^{+} \\
\phi_a^0
\end{pmatrix}= \frac{1}{\sqrt{2}}  \begin{pmatrix}
 \sqrt{2} \phi_a^{+} \\
 v_a e^{i \theta_a} + \rho_a + i \eta_a
\end{pmatrix} \; , \; S_a = \frac{1}{\sqrt{2}}\left( u_a e^{i \xi_a} + \rho_{a+2} + i \eta_{a+2}\right)\;,\; a=1,2\,,
\label{eq:scalarfield2}
\end{equation}
where $v_a$ and $u_a$ are the VEVs of the $\Phi_a$ neutral component and of $S_a$, respectively. In the present case, only the phase difference $\theta=\theta_2-\theta_1$ is relevant. The above fields transform under the Abelian symmetries as shown in Table~\ref{tab:symcharges}. The most general potential invariant under those symmetries can be written as,
\begin{equation}
    V_{\rm U(1)}= V_{\Phi\Phi} + V_{\Phi S} + V_{S S},
\end{equation}
with
\begin{equation}
\begin{aligned}
   V_{\Phi\Phi} &= \mu^2_{11} \Phi_1^{\dagger} \Phi_1 + \mu^2_{22} \Phi_2^{\dagger} \Phi_2 + \frac{\lambda_{1}}{2} \left(\Phi_1^{\dagger} \Phi_1\right)^2 + \frac{\lambda_{2}}{2} \left(\Phi_2^{\dagger} \Phi_2\right)^2+ \lambda_{3} \left(\Phi_1^{\dagger} \Phi_1\right) \left(\Phi_2^{\dagger}\Phi_2\right) + \lambda_{4} \left(\Phi_1^{\dagger} \Phi_2\right) \left(\Phi_2^{\dagger} \Phi_1\right), \\
   V_{\Phi S} &= \left(\lambda_{5} \Phi_1^{\dagger} \Phi_1 + \lambda_{6} \Phi_2^{\dagger} \Phi_2\right)\left|S_1\right|^2 + \left(\lambda_{9} \Phi_1^{\dagger} \Phi_1 + \lambda_{10} \Phi_2^{\dagger} \Phi_2\right)\left|S_2\right|^2,\\
    V_{S S} &= \mu^2_{1} \left|S_1\right|^2 + \mu^2_{2} \left|S_2\right|^2 + \frac{\lambda_{7}}{2} \left|S_1\right|^4 + \frac{\lambda_{8}}{2} \left|S_2\right|^4
   + \lambda_{11} \left|S_1\right|^2\left|S_2\right|^2\,.
\end{aligned}
\label{eq:U(1)pot}
\end{equation}
For reasons that will become clear later, we consider all parameters to be real. To avoid massless Goldstone bosons, we add terms to the scalar potential, which break softly the Abelian flavour symmetries. Such terms could, for instance, originate from the spontaneous breaking of a larger symmetry at very high-energies. Possible soft-breaking terms are
\begin{equation}
   V_{\text{soft}}(\Phi_a,S_a) = \mu_{12}^2 \ \Phi_1^{\dagger} \Phi_2 + \mu_{3}^2 \ S_1^2 + \mu_{4} \left|S_1\right|^2 S_1 + \mu_{5} \ \left|S_2\right|^2 S_2 + \text{H.c.}\,,
\label{soft}
\end{equation}
with all parameters real. This specific form of $V_{\text{soft}}$ is chosen not only to avoid unwanted massless scalars, but also to open up the possibility for SCPV coming from the complex VEV of the scalar singlet $S_1$, as will be discussed later. The full scalar potential is then given by $V(\Phi_a,S_a)=V_{\text{U}(1)}+V_{\text{soft}}$.

As argued in Section~\ref{sec:symmetries}, one of the motivations for adding two complex singlet scalars is to account for the mass hierarchy in the inverse-seesaw approximation through ${u}_{1},{v}_{1},{v}_{2} \ll {u}_{2}$. In order to simplify the analysis, we will use that VEV hierarchy to consider the case in which $S_2$ is decoupled from the remaining scalars and, thus, the quartic term $\sim \left|S_2\right|^4$ will dominate over the terms $\sim \Phi_a^{\dagger} \Phi_a\left|S_2\right|^2$ and $\sim \left|S_1\right|^2\left|S_2\right|^2$. The analysis is then simplified by taking
\begin{equation}
 V_{\Phi S} \simeq \left(\lambda_{5} \Phi_1^{\dagger} \Phi_1 + \lambda_{6} \Phi_2^{\dagger} \Phi_2\right)\left|S_1\right|^2, \
 V_{S S} \simeq \mu^2_{1} \left|S_1\right|^2 + \mu^2_{2} \left|S_2\right|^2 + \frac{\lambda_{7}}{2} \left|S_1\right|^4 + \frac{\lambda_{8}}{2} \left|S_2\right|^4.
\end{equation}
In order to ensure vacuum stability, the scalar potential has to be bounded from below in any direction of the field space as the fields become large. This can be guaranteed by requiring the Hessian matrix of the quartic couplings in the potential to be copositive~\cite{Kannike:2012pe}. In the case under analysis, this is translated into the following conditions among $\lambda_{1-8}$~\cite{Arhrib:2018qmw}:
\begin{equation}
\begin{aligned}
&\lambda_{1},\lambda_{2},\lambda_{7},\lambda_{8} > 0\,,\,\lambda_{3} + \sqrt{\lambda_1 \lambda_2} > 0\;,\;\lambda_{3} + \lambda_{4} + \sqrt{\lambda_1 \lambda_2} > 0\;,\;\lambda_{5} + \sqrt{\lambda_1 \lambda_{7}} > 0\;,\;\lambda_{6} + \sqrt{\lambda_2 \lambda_{7}} > 0, \\
    &\lambda_{3}\lambda_{7} - \lambda_{5} \lambda_{6} + \sqrt{\left(\lambda_1 \lambda_{7} - \lambda_{5}^2\right) \left(\lambda_2 \lambda_{7} - \lambda_{6}^2\right) } > 0\;,\; \left(\lambda_{3} + \lambda_{4}\right) \lambda_{7} - \lambda_{5} \lambda_{6} + \sqrt{\left(\lambda_1 \lambda_{7} - \lambda_{5}^2\right) \left(\lambda_2 \lambda_{7} - \lambda_{6}^2\right) } > 0\,.
\end{aligned}
\end{equation}

Since we considered all Yukawa and scalar parameters to be real, the full Lagrangian is CP-conserving. Yet, CP can be spontaneously broken if some scalar field acquires a complex VEV. The main motivation for studying this possibility is to provide a dynamical explanation of CPV in the lepton sector manifested through non-trivial Dirac and Majorana phases $\delta$ and $\alpha$, respectively (see Section~\ref{sec:numass}). To show that this is indeed possible, we start by minimising 
\begin{equation}
\begin{aligned}
V_0 &= \ \frac{\lambda_1}{8} v_1^4  + \frac{\lambda_2}{8} v_2^4  + \frac{\lambda_{3}}{4} v_1^2 v_{2}^2 + \frac{\lambda_{4}}{4} v_1^2 v_{2}^2 + \frac{\lambda_{5}}{4} u_{1}^2 v_1^2  + \frac{\lambda_{6}}{4} u_{1}^2 v_2^2 + \frac{\lambda_{7}}{8} u_{1}^4 + \frac{\lambda_{8}}{8} u_{2}^4
+ \frac{\mu_{11}^2}{2} v_1^2 + \frac{\mu_{22}^2}{2} v_2^2 \\ &+ \frac{\mu_{1}^2}{2} u_{1}^2 + \frac{\mu_{2}^2}{2} u_{2}^2 + \mu_{12}^2 v_1 v_2 \cos \theta + \frac{\mu_{5}}{\sqrt{2}} u_{2}^3 \cos \xi_2 +\frac{\mu_{4}}{\sqrt{2}} u_{1}^3 \cos \xi_1 + \mu_{3}^2 u_{1}^2 \cos\left(2\xi_1\right),
\end{aligned}
\end{equation}
with respect to $v_a, u_a, \theta, \xi_1$ and $\xi_2$. In particular, the extrema conditions for $\theta$ and $\xi_2$  lead to $\theta,\xi_2=0,\pi$. Two possible solutions for $\mu_{11}^2$, $\mu_{22}^2$ and $\mu_2^2$ can be then obtained from the minimisation equations of $v_{1,2}$ and $u_2$, respectively. Namely,
\begin{equation}
\begin{aligned}
\theta=0,\pi:\;\mu_{11}^2 &= - \frac{1}{2} \left[\lambda_1 v_1^2  + \left( \lambda_3 +  \lambda_4\right) v_2^2  + \lambda_{5} u_{1}^2 \right] \mp \frac{v_2}{v_1} \mu_{12}^2, \\
\mu_{22}^2 &= - \frac{1}{2} \left[\lambda_1 v_2^2 + \left( \lambda_3 +  \lambda_4\right) v_1^2 + \lambda_{6} u_{1}^2 \right] \mp  \frac{v_1}{v_2} \mu_{12}^2\,,\\
\xi_2=0,\pi:\;\mu_{2}^2 &= - \frac{u_2}{2} \Big(\lambda_{8} u_2  \pm 3 \sqrt{2} \mu_5 \Big)\,.
\end{aligned}
\label{eq:min1}
\end{equation}
Instead, minimising with respect to $\xi_1$ leads to three solutions:
\begin{equation}
\begin{aligned}
\xi_1 =0,\pi: \;\quad &\mu_{1}^2 = - \frac{1}{2} \left(\lambda_{5} v_1^2 + \lambda_{6} v_2^2 + \lambda_{7} u_{1}^2 + 4 \mu_{3}^2 \pm 3\sqrt{2} \mu_{4} u_{1}\right),\\
\xi_1 = \arctan \left( \frac{\sqrt{32 \mu_{3}^4 - \mu_{4}^2 u_{1}^2}}{\mu_{4} u_{1}} \right):\; \quad &\mu_{1}^2 = \frac{\mu_{4}^2 u_{1}^2}{4\mu_{3}^2} - \frac{1}{2} \left( \lambda_{5} v_1^2 + \lambda_{6} v_2^2 + \lambda_{7} u_{1}^2 - 4 \mu_{3}^2 \right)\,.
\end{aligned}
\label{eq:min3}
\end{equation}
Notice that the last solution in \eqref{eq:min3} is the only one which provides a non-trivial phase $\xi_1$, leading to the possibility for SCPV. Setting $\theta,\xi_2= 0,\pi$ in $V_0$, the value of the potential at each $\xi_1$ minimum $(V_{\rm min})$ is
\begin{equation}
\begin{aligned}
\xi_1 = 0,\pi: \quad &V_{\rm min} = \mp \frac{ \mu_{4} u_{1}^3}{2 \sqrt{2}} - V^\prime, \\
\xi_1 = \arctan \left( \frac{\sqrt{32 \mu_{3}^4 - \mu_{4}^2 u_{1}^2}}{\mu_{4} u_{1}} \right): \quad &V_{\rm min} = \frac{ \mu_{4}^2 u_{1}^4}{16 \mu_{3}^2} - V^\prime,
\end{aligned}
\label{gmin1}
\end{equation}
with
\begin{equation}
V^\prime = \frac{\lambda_1}{8} v_1^4 + \frac{\lambda_2}{8} v_2^4 + \frac{\lambda_{3}}{4} v_1^2 v_{2}^2 + \frac{\lambda_{4}}{4} v_1^2 v_{2}^2+\frac{\lambda_{5}}{4} u_1^2 v_{1}^2 + \frac{\lambda_{6}}{4} u_1^2 v_{2}^2 + \frac{\lambda_{7}}{8} u_{1}^4 + \frac{\lambda_{8}}{8} u_{2}^4 \pm \frac{\mu_5}{2\sqrt{2}} u_2^3.
\end{equation}
Therefore, the CPV solution with $\xi_1 \neq 0,\pi$ corresponds to the deepest minimum if $\mu_{4} u_1 < \mp 4\sqrt{2}\, \mu_{3}^2$.

%
%*******************************************************************************************************
\subsection{Scalar mass spectrum}
\label{sec:Scalarmassspectrum}

We now briefly discuss the main features of the scalar mass spectrum of our model. In total we have two charged complex scalar fields $\phi_{1,2}^+$ and four neutral ones, $\phi^0_{1,2}$ and $S_{1,2}$. The mass matrix for the charged scalars is 
\begin{equation}
\mathbf{M}^2_{+} = - \frac{v_1 v_2 \lambda_4 \pm 2 \mu_{12}^2}{2} \begin{pmatrix}
 \dfrac{v_2}{v_1} & \mp 1 \\ \mp 1 & \dfrac{v_1}{v_2}
\end{pmatrix},
\label{eq:charged}
\end{equation}
where the upper (lower) sign corresponds to $\theta=0\,(\pi)$. This matrix is diagonalised via the basis transformation
\begin{equation}
\mathbf{R}=\begin{pmatrix}
c_{\beta} & \pm s_{\beta} \\
- s_{\beta} & \pm c_{\beta} 
\end{pmatrix} \rightarrow 
\begin{pmatrix}
H_1 \\
H_2
\end{pmatrix} 
= \mathbf{R}
\begin{pmatrix}
\Phi_1 \\
 \Phi_2
\end{pmatrix},
\label{higgsbasis}
\end{equation}
with 
\begin{equation}
\begin{aligned}
c_{\beta} = \cos\beta = \frac{v_1}{v},\quad s_{\beta} = \sin\beta = \frac{v_2}{v}, 
\end{aligned}
\label{eq:betadef}
\end{equation}
where $v = \sqrt{v_1^2 + v_2^2} \simeq 246 \ \text{GeV}$. This leads to the SM massless Goldstone boson $G^{\pm}$ and massive charged states $H^{\pm}$ with mass
\begin{equation}
m^2_{H^{\pm}} = - v^2  \frac{\lambda_4}{2} \mp \frac{2\mu_{12}^2}{\sin(2\beta)}.
\end{equation}
The above rotation brings $\Phi_{1,2}$ into the Higgs basis with $H_{1,2}$~\cite{Branco:2011iw} given by
\begin{equation}
    H_1 = \frac{1}{\sqrt{2}} \begin{pmatrix}
    \sqrt{2} G^{+} \\
    v + H^0 + i G^{0}
    \end{pmatrix}, \quad H_2 = \frac{1}{\sqrt{2}} \begin{pmatrix}
    \sqrt{2} H^{+} \\
  R + i I
    \end{pmatrix},
\label{inhiggsbasis}
\end{equation}
where $\left< H_1^0 \right> = v/\sqrt{2}$, $\left< H_2^0 \right> = 0$. Here, $H^0$ coincides with the 125~GeV SM Higgs in the alignment limit, $G^{\pm}$ and $G^0$ are the charged and neutral Goldstone bosons, and $H^{\pm}$ is the physical charged Higgs field. 
The neutral scalar mass matrix $\mathbf{M}^2_{0}$ is diagonalised through a unitary transformation $\mathbf{T}$ which relates weak and mass eigenstates through
\begin{equation}
\left(G^0,S^0_1, ..., S^0_7\right)^{T} = \mathbf{T}
\left(\rho_1, \rho_2, \rho_3, \rho_4,\eta_1, \eta_2,  \eta_3, \eta_4\right)^{T}.
\label{diagneutral}
\end{equation}
As mentioned before, we assume that the mixing of the complex singlet $S_2$ with the remaining fields is negligible and, thus, $\eta_{4}$ and $\rho_{4}$ are decoupled. Furthermore, the CP-odd scalars $\eta_{1,2}$ from $\phi^0_{1,2}$ also do not mix with the other scalars. However, $\rho_{1,2,3}$ and $\eta_3$ do mix among themselves. 
For the SCPV solution in Eq.~\eqref{eq:min3}, the mixing among $\eta_3$, $\rho_3$ and $\rho_{1,2}$ will depend on the soft-breaking parameters $\mu_{3,4}$ and on VEV products $u_1^2$ and $u_1 v_a$ ($a=1,2$). Moreover, since the naturally small soft-breaking parameters must fulfil the SCPV condition $\mu_{4} u_1 < \mp 4\sqrt{2}\, \mu_{3}^2$, and given that $u_1 \lesssim v_1, v_2$, it is reasonable to consider that the mixing among the $\phi_{1,2}^0$ and $S_1$ will be small. In this limit, the neutral mass matrix can be recast into a block-diagonal form composed of four $2 \times 2$ matrices: the CP-even $\mathbf{M}^2_{\text{CP-even}}$ for $\rho_{1,2}$, the CP-odd $\mathbf{M}^2_{\text{CP-odd}}$ for $\eta_{1,2}$, and the $S_{1,2}$ mass matrices $\mathbf{M}^2_{S_1}$ and $\mathbf{M}^2_{S_2}$. The former is given by
\begin{equation}
\mathbf{M}^2_{\text{CP-even}} = \begin{pmatrix}
v_1^2 \lambda_1 \mp \dfrac{v_2 \mu_{ 1 2}^2}{v_1} \   &\pm v_1 v_2 \left(\lambda_3 + \lambda_4 \right) + \mu_{12}^2   \\
\pm v_1 v_2 \left(\lambda_3 + \lambda_4 \right) + \mu_{12}^2  & v_2^2 \lambda_2 \mp \dfrac{v_1 \mu_{ 1 2}^2}{v_2} 
\end{pmatrix},
\label{eq:cpeven}
\end{equation}
being diagonalised by
\begin{equation}
\begin{pmatrix}
S_1^0 \\
S_2^0
\end{pmatrix} 
= \begin{pmatrix}
c_{\alpha_1-\beta} & s_{\alpha_1-\beta} \\
- s_{\alpha_1-\beta} & c_{\alpha_1-\beta} 
\end{pmatrix}
\begin{pmatrix}
H^0 \\
R
\end{pmatrix}\;,\; \tan \left(2 \alpha_1 \right) =  \frac{\pm v^2 \left(\lambda_3 + \lambda_4 \right)\sin(2\beta) + 2\mu_{12}^2 }{v^2( \lambda_1c_\beta^2 -  \lambda_2s_\beta^2) \pm 2\mu_{1 2}^2 \tan(2\beta) } \; ,
\label{alpha}
\end{equation}
where, as before, the upper (lower) sign corresponds to $\theta=0\,(\pi)$. The angle $\alpha_1$ parameterises the mixing in the~$(\rho_1,\rho_2)$ sector. Throughout this work we set $\beta = \alpha_1 + \pi/2$, which is known as the alignment limit~\cite{Gunion:2002zf}. In this case, there is no mixing between $H^0$ and $R$ and, thus, $S_1^0 = H^{0}$ and $S_2^0 = R$. As already mentioned, this allows us to identify $H^{0}$ with the 125~GeV Higgs boson discovered at the LHC~\cite{Aad:2015zhl}. For the CP-odd scalars we have
\begin{equation}
\mathbf{M}^2_{\text{CP-odd}} = \mu_{12}^2 \begin{pmatrix}
 \mp \dfrac{v_2}{v_1} &  1 \\
 1 & \mp\dfrac{v_1}{v_2}
\end{pmatrix}\,,
\label{eq:cpodd}
\end{equation}
which is diagonalised through the rotation matrix defined in Eq.~\eqref{higgsbasis}, leading to the massless Goldstone boson $G^{0}$ and to the massive scalar $S_5^0=I$ with mass
\begin{equation}
m^2_{I} = m^2_{H^{\pm}} + v^2  \frac{\lambda_4}{2} = \mp \frac{2\mu_{12}^2}{\sin(2\beta)}.
\end{equation}
The $S_1$ mass matrix is
\begin{align}
&\mathbf{M}^2_{S_1} = \begin{pmatrix}
  \dfrac{u_1^2\mu_4^2}{32 \mu_3^4} \left(u_1^2 \lambda_7 - 16 \mu_3^2 \right)+ 4 \mu_3^2 &  \dfrac{u_1\mu_4}{32 \mu_3^4} \sqrt{32 \mu_{3}^4 - \mu_{4}^2 u_{1}^2}  \left(u_1^2 \lambda_7 - 8 \mu_3^2 \right)  \\
  \dfrac{u_1\mu_4}{32 \mu_3^4} \sqrt{32 \mu_{3}^4 - \mu_{4}^2 u_{1}^2}  \left(u_1^2 \lambda_7 - 8 \mu_3^2 \right) &  u_1^2 \lambda_7 \left(1 -  \dfrac{u_1^2\mu_4^2}{32 \mu_3^4}\right)
\end{pmatrix}\,,
\end{align}
which is diagonalised by a rotation with an angle $\alpha_2$ given by,
\begin{align}
&\tan \left(2 \alpha_2 \right) = \frac{ u_1 \mu_4 \sqrt{32 \mu_{3}^4 - \mu_{4}^2 u_{1}^2}  \left(u_1^2 \lambda_7 - 8 \mu_3^2 \right)}{16 \mu_3^4 \left(4 \mu_3^2 - u_1^2 \lambda_7\right) + u_1^2 \mu_4^2 \left(u_1^2 \lambda_7 - 8 \mu_3^2\right) }.
\label{eq:MS1}
\end{align}
After diagonalizing the mass matrix $\mathbf{M}^2_{S_1}$ there remains a very small mixing in the $(\rho_3,\eta_3)$ sector, which depends on the soft-breaking couplings. The mass terms for the resulting physical fields, $S_3^0 \simeq \rho_3$ and $S_6^0 \simeq \eta_3$, are given by
\begin{equation}
\xi_1 = \arctan \left(\frac{\sqrt{32 \mu_{3}^4 - \mu_{4}^2 u_{1}^2}}{\mu_{4} u_{1}} \right), \ m_{S_3^0}^2 \simeq  m_{S_6^0}^2 \simeq 2 \mu_3^2-\frac{u_1^2 \left(\mu_4^2-2 \lambda_7 \mu_3^2 \right)}{4 \mu_3^2},
\label{eq:S1masses}
\end{equation}
which are approximately degenerate for the SCPV solution. Notice that in the absence of soft-breaking terms proportional to $\mu_3$ and $\mu_4$, unwanted massless Goldstone bosons appear. SCPV originated from the complex VEV of the singlet $S_1$ is possible if the masses above are positive and the condition for the global minimum is satisfied. Lastly, the matrix $\mathbf{M}^2_{S_2}$ is diagonal and the corresponding scalar masses for $S_4^0 = \rho_4$ and $S_7^0 = \eta_4$ are
\begin{equation}
m_{\eta_4}^2 = \mp  \frac{\mu_5 u_2}{\sqrt{2}}, \
m_{\rho_4}^2 =  u_2^2 \left(\lambda_8 \pm \frac{3 \mu_5}{\sqrt{2} u_2}\right),
\end{equation}
for $\xi_2$ equal $0$ or $\pi$, respectively. Once again, if the soft-breaking term proportional to $\mu_5$ vanishes, $\eta_4$ would be a massless Goldstone boson. Since $u_2$ can be naturally large, the scalar fields $\eta_4$ and $\rho_4$ can have a large mass, which further justifies the decoupling behaviour of the singlet $S_2$.

%*******************************************************************************************************
\section{Interactions in the mass-eigenstate basis}
\label{sec:Allint}

In this appendix we collect the interactions relevant for our work in the mass-eigenstate basis of fermions, scalars and gauge bosons. We consider an arbitrary number of Majorana neutrinos $\nu_i$ ($i=1, \dots, n_f$) with masses $m_i$, so that the results can be applied for scenarios with any number of sterile neutrinos. In the ISS(2,2) considered in this work~$n_f=7$, being $\nu_{1,2,3}$ the three light active neutrinos, and $\nu_{4-7}$ the heavy sterile ones.

%*******************************************************************************************************
\subsection{Charged-current and neutral-current interactions}
\label{sec:NCCCint}

In the Feynman-’t Hooft gauge and mass eigenstate basis, the charged-current, weak neutral-current and Goldstone boson interactions read:
\begin{align}
&\mathcal{L}_{W^{\pm}} =\frac{g}{\sqrt{2}} W_{\mu}^{-} \sum_{\alpha=1}^{3} \sum_{j=1}^{n_f} \mathbf{B}_{\alpha j} \ \overline{e_{\alpha}} \gamma^{\mu} P_{L} \nu_j + \text{H.c.},
\label{eq:Wint} \\
&\mathcal{L}_{Z} =\frac{g}{2 c_{W}} Z_{\mu}  \sum_{i,j=1}^{n_f} \bm{\mathcal{C}}_{ij} \ \overline{\nu_i} \gamma^{\mu} P_{L} \nu_j = \frac{g}{4 c_{W}} Z_{\mu}  \sum_{i,j=1}^{n_f} \ \overline{\nu_i} \gamma^{\mu} \left(\bm{\mathcal{C}}_{ij} P_{L} -\bm{\mathcal{C}}_{ij}^{*} P_{R} \right) \nu_j,
\label{eq:Zint} \\
&\mathcal{L}_{G^{\pm}} = -\frac{g}{\sqrt{2} M_W} G^{-} \sum_{\alpha=1}^{3} \sum_{j=1}^{n_f} \mathbf{B}_{\alpha j} \ \overline{e_{\alpha}} \left( m_\alpha  P_{L} - m_j P_{R} \right) \nu_j +\text{H.c.},
\label{eq:Gpint} \\
&\begin{aligned}
\mathcal{L}_{G^0} &= - \frac{i g}{2 M_W} G^0 \sum_{i,j=1}^{n_f} \ \bm{\mathcal{C}}_{ij} \overline{\nu_i} \left(P_L m_i - P_{R} m_j \right) \nu_j \\
                  &= - \frac{i g}{4 M_W} G^0 \sum_{i,j=1}^{n_f} \  \overline{\nu_i} \left[\bm{\mathcal{C}}_{ij} \left(P_L m_i - P_{R} m_j \right) - \bm{\mathcal{C}}_{ij}^{*} \left(P_R m_i - P_{L} m_j \right) \right]\nu_j,
\end{aligned}
\label{eq:G0int}
\end{align}
where we have followed closely the notation of Refs.~\cite{Ilakovac:1994kj,Alonso:2012ji}. $P_{L,R} = ( 1\mp \gamma_{5})/2$ are the chirality projectors, $g$ is the weak coupling constant, and $c_W \equiv \cos\theta_W$, with $\theta_W$ being the Weinberg angle.  The $\mathbf{B}$ and $\bm{\mathcal{C}}$ matrices have been defined in~Eq.~\eqref{eq:BC}. The last equalities in Eqs.~\eqref{eq:Zint} and \eqref{eq:G0int}  result from the Majorana character of neutrinos~($\nu = \nu^c$). Therefore, for Majorana neutrinos, the coupling $Z\, \overline{\nu_i}\, \nu_j$ is non-diagonal and involves both chiralities.

\subsection{Scalar-fermion interactions}
\label{sec:Scalarint}

In this section we present the scalar-fermion interactions extracted from the Yukawa Lagrangian in Eq.~\eqref{eq:Yuk}, using the notation for the scalar fields introduced in Section~\ref{sec:Scalarmassspectrum}. For charged and neutral scalars $S_{a}^{\pm}$ and $S_{a}^{0}$ we have: 
\begin{align}
&\begin{aligned}
\mathcal{L}_{S^{\pm}_a} &= S^{-}_{a} \Bigg\{\sum_{\alpha=1}^{3} \sum_{j=1}^{n_f} \overline{e_{\alpha}} \left[\left(\mathbf{\Gamma}_{e}^{a} \right)_{\alpha j} P_{L}-\left(\mathbf{\Gamma}_{\nu}^{a} \right)_{\alpha j} P_{R} \right] \nu_j +\mathop{\sum_{u_{i}=u,c,t}}_{d_{i}=d,s,b} \ \overline{d_{i}} \left[\left(\mathbf{\Gamma}_{d}^{a}\right)_{d_{i} u_{j}} P_{L} - \left(\mathbf{\Gamma}_{u}^{a}\right)_{d_{i} u_{j}} P_{R} \right] u_{j} \Bigg\}  
+\text{H.c.},
\end{aligned} \\
&\mathcal{L}_{S^{0}_a}= S^{0}_{a}\left[ \sum_{\alpha,\beta=1}^{3}   \overline{e_{\alpha}}\,(\mathbf{\Delta}_{e}^{a})_{\alpha \beta} P_{R} e_{\beta} + \sum_{i,j=1}^{n_f}   \overline{\nu_i}\, (\mathbf{\Delta}_{\nu}^{a})_{i j} P_{R} \nu_j \right]  +\text{H.c.},
\label{eq:genericS0}
\end{align}
respectively, where $\mathbf{\Gamma}$ and $\mathbf{\Delta}$ are general Yukawa matrices. In this work, the scalar sector contains two Higgs doublets and two neutral scalar singlets, which we will consider to obey the alignment and decoupling limits discussed in Appendix~\ref{sec:ScalarSector}. In such case, the Yukawa coupling matrices between the fermions and the charged Higgs $S_{1}^{\pm}=H^{\pm}$ are given by
\begin{equation}
\begin{aligned}
&\mathbf{\Gamma}_{e}^{1} = \frac{g}{\sqrt{2} M_{W}} \mathbf{N}_{e}^{\dagger} \mathbf{B}\;,\;
\mathbf{\Gamma}_{\nu}= \frac{g}{\sqrt{2} M_{W}} \mathbf{B}\mathbf{N}_{\nu}\;,\;\mathbf{\Gamma}_{d}^{1} = \frac{g}{\sqrt{2} M_{W}} \mathbf{N}_{d}^{\dagger} \mathbf{V}^{\dagger}\;,\;\mathbf{\Gamma}_{u}^{1}= \frac{g}{\sqrt{2} M_{W}} \mathbf{V}^{\dagger} \mathbf{N}_{u}\,,
\end{aligned}
\end{equation}
where $\mathbf{B}$ is defined in Eq.~\eqref{eq:BC} and
\begin{align}
&\mathbf{N}_e = \mathbf{V}_{L}^{\dagger} \mathbf{N}_e^{0}  \mathbf{V}_{R}, \ \mathbf{N}_e^{0} = \frac{v}{\sqrt{2}} \left( s_{\beta} \mathbf{Y}^1_{\ell} - c_{\beta} \mathbf{Y}^2_{\ell} e^{i \theta} \right),
\label{eq:Ne} \\
&(\mathbf{N}_{\nu})_{i j} = \sum_{\alpha=1}^{3} \sum_{k=4}^{3+n_R} \bm{\mathcal{U}}^{*}_{\alpha i} \left(\mathbf{N}_{\nu}^{0}\right)_{\alpha k} \bm{\mathcal{U}}_{k j}^*, \ \mathbf{N}_{\nu}^{0} = \frac{v}{\sqrt{2}} \left( s_{\beta} \mathbf{Y}^1_{D} - c_{\beta} \mathbf{Y}^2_{D} e^{-i \theta} \right),
\label{eq:Nnu} \\
&\mathbf{N}_d = \mathbf{V}_{L}^{d \dagger} \mathbf{N}_d^{0}  \mathbf{V}_{R}^{d}, \ \mathbf{N}_d^{0} = \frac{v}{\sqrt{2}} \left( s_{\beta} \mathbf{Y}^1_{d} - c_{\beta} \mathbf{Y}^2_{d} e^{i \theta} \right) ,
\label{eq:Nd} \\
&\mathbf{N}_{u} = \mathbf{V}_{L}^{u \dagger} \mathbf{N}_u^{0}  \mathbf{V}_{R}^{u}, \ \mathbf{N}_{u}^{0} = \frac{v}{\sqrt{2}} \left( s_{\beta} \mathbf{Y}^1_{u} - c_{\beta} \mathbf{Y}^2_{u} e^{-i \theta} \right)\,.
\label{eq:Nu}
\end{align}
The unitary matrices $\mathbf{V}_{L,R}$ and $\bm{\mathcal{U}}$ are given in Eqs.~\eqref{diagcharg} and \eqref{diagnutr}, respectively. Additionally, $\mathbf{Y}^{1,2}_{d,u}$ are the Yukawa matrices appearing in the interaction terms among the Higgs doublets and the quarks fields, and $\mathbf{V}_{L,R}^{d,u}$ are the unitary transformations that diagonalise the quark mass matrices. The Cabbibo-Kobayashi-Maskawa (CKM) quark mixing matrix is denoted by $\mathbf{V}$. We explicitly present the interaction Lagrangian involving the $H^{\pm}$ charged scalar since it has been extensively used, for instance, in the computation of the cLFV amplitudes (see Appendices~\ref{sec:clfv} and~\ref{sec:LFFF}). Namely,
\begin{equation}
\begin{aligned}
\mathcal{L}_{H^{\pm}} &= - \frac{g}{\sqrt{2} M_{W}} \Bigg\{ \sum_{\alpha,\beta=1}^{3} \sum_{i,j=1}^{n_f} \overline{e_{\alpha}} \left[\left(\mathbf{B} \mathbf{N}_{\nu}\right)_{\alpha j}  P_{R} - \left(\mathbf{N}_{e}^{\dagger} \mathbf{B}\right)_{\alpha j} P_{L} \right] \nu_j \\
& +\mathop{\sum_{u_{i}=u,c,t}}_{d_{i}=d,s,b} \overline{d_{i}} \left[\left(\mathbf{V}^{\dagger} \mathbf{N}_{u}\right)_{d_{i} u_{j}} P_{R} - \left(\mathbf{N}_{d}^{\dagger} \mathbf{V}^{\dagger}\right)_{d_{i} u_{j}} P_{L} \right] u_{j} \Bigg\}  H^{-}
+\text{H.c.}.
\end{aligned}
\label{eq:Hpint}
\end{equation}

The Yukawa matrices entering the interaction terms \eqref{eq:genericS0} among leptons and the the neutral scalars $S_{1}^{0}=H^{0}$, $S_{2}^{0}=R$ and $S_{5}^{0}=I$ (those stemming from the Higgs doublets) are
\begin{equation}
\begin{aligned}
&\left(\mathbf{\Delta}_{e}^{1}\right)_{\alpha \beta}= - \frac{g}{2 M_{W}} \delta_{\alpha \beta} m_{\alpha} \; , \; \left(\mathbf{\Delta}_{\nu}^{1}\right)_{i j}= -\frac{g}{4 M_W} \left(\bm{\mathcal{C}}_{ij} m_j +  \bm{\mathcal{C}}_{ij}^{*} m_i\right),
 \\
&\left(\mathbf{\Delta}_{e}^{2}\right)_{\alpha \beta}= \frac{g}{2 M_{W}} (\mathbf{N}_e)_{\alpha \beta} = -i \left(\mathbf{\Delta}_{e}^{5}\right)_{\alpha \beta}\; , \; \left(\mathbf{\Delta}_{\nu}^{2}\right)_{i j}= \frac{g}{4 M_W} \left[\left(\mathbf{N}_{\nu}\right)_{i j} + \left(\mathbf{N}_{\nu}^T\right)_{i j} \right]=i\left(\mathbf{\Delta}_{\nu}^{5}\right)_{i j},
\label{eq:deltanuR} 
\end{aligned}
\end{equation}
where the matrix $\bm{\mathcal{C}}$ is defined in Eq.~\eqref{eq:BC}. We explicitly present the Lagrangians involving the above neutral scalars and $\nu_i$ since they were useful in our calculations of the radiative corrections to light neutrino masses performed in Section~\ref{sec:radcorr}. We have,
\begin{align}
\mathcal{L}_{H^0} &= -\frac{g}{2 M_W} H^0 \sum_{i,j=1}^{n_f} \ \bm{\mathcal{C}}_{ij} \overline{\nu_i} \left( P_{L} m_i + P_R m_j \right) \nu_j \nonumber\\
                  &= -\frac{g}{4 M_W} H^0 \sum_{i,j=1}^{n_f} \  \overline{\nu_i} \left[\bm{\mathcal{C}}_{ij} \left( P_{L} m_i + P_R m_j \right) + \bm{\mathcal{C}}_{ij}^{*} \left( P_{R} m_i + P_L m_j \right) \right] \nu_j,
\label{eq:H0int}\\
\mathcal{L}_{R}&= \frac{g}{4 M_{W}} R \sum_{i,j=1}^{n_f} \ \overline{\nu_i} \left\{\left[\left(\mathbf{N}_{\nu}\right)_{i j} + \left(\mathbf{N}_{\nu}^T\right)_{i j} \right] P_{R} +  \left[\left(\mathbf{N}_{\nu}^\dagger\right)_{i j} + \left(\mathbf{N}_{\nu}^*\right)_{i j} \right] P_{L} \right\}\nu_j,
\label{eq:Rint} \\
\mathcal{L}_{I} &= -\frac{i g}{4 M_{W}} I \sum_{i,j=1}^{n_f} \ \overline{\nu_i} \left\{ \left[\left(\mathbf{N}_{\nu}\right)_{i j} + \left(\mathbf{N}_{\nu}^T\right)_{i j} \right] P_{R} -  \left[\left(\mathbf{N}_{\nu}^\dagger\right)_{i j} + \left(\mathbf{N}_{\nu}^*\right)_{i j} \right] P_{L} \right\}\nu_j.
\label{eq:Iint}
\end{align}

Lastly, the coupling matrices appearing in interaction terms between the neutrinos and the neutral scalars $S_{3}^{0} \simeq \rho_3$, $S_{6}^{0} \simeq \eta_3$,  $S_{4}^{0}=\rho_4$ and $S_{7}^{0}=\eta_4$ (coming from the complex scalar singlets $S_{1,2}$) are,
\begin{equation}
\begin{aligned}
 \left(\mathbf{\Delta}_{\nu}^{3}\right)_{i j}&= -\frac{1}{4 u_1} \left[\left(\tilde{\mathbf{N}}_{s}^\dagger\right)_{i j} + \left(\tilde{\mathbf{N}}_{s}^*\right)_{i j} \right], \ \left(\mathbf{\Delta}_{\nu}^{6}\right)_{i j}= \frac{i}{4 u_1} \left[\left(\tilde{\mathbf{N}}_{s}^\dagger\right)_{i j} + \left(\tilde{\mathbf{N}}_{s}^*\right)_{i j} \right], \\
(\mathbf{\Delta}_{\nu}^{4})_{i j}&= -\frac{1}{2 u_2} \left[ \left(\mathbf{N}_{R}^\dagger\right)_{i j} + \left(\mathbf{N}_{R}^*\right)_{i j}\right]\,\,, \ \left(\mathbf{\Delta}_{\nu}^{7}\right)_{i j}= \frac{i}{2 u_2} \left[ \left(\mathbf{N}_{R}^\dagger\right)_{i j} + \left(\mathbf{N}_{R}^*\right)_{i j}\right],
\label{eq:deltanusinglets}
\end{aligned}
\end{equation}
where
\begin{align}
(\tilde{\mathbf{N}}_{s})_{i j} &= \sum_{k,l=4+n_R}^{n_f} \bm{\mathcal{U}}_{k i} \left(\tilde{\mathbf{N}}_{s}^{0}\right)_{k l} \bm{\mathcal{U}}_{l j}\; , \; \tilde{\mathbf{N}}_{s}^{0} = \frac{u_1}{\sqrt{2}} \left(\mathbf{Y}^1_{s}+\mathbf{Y}^2_{s}\right) \, ,
\label{eq:NSt} \\
\left(\mathbf{N}_{s}\right)_{i j} &= \sum_{k,l=4+n_R}^{n_f} \bm{\mathcal{U}}_{k i} \left(\mathbf{N}_{s}^{0}\right)_{k l} \bm{\mathcal{U}}_{l j}\;\, , \; \mathbf{N}_{s}^{0} = \frac{u_1}{\sqrt{2}} \left(\mathbf{Y}^1_{s}-\mathbf{Y}^2_{s}\right) ,
\label{eq:NS} \\
\left(\mathbf{N}_{R}\right)_{i j} &= \sum_{k=4}^{3+n_R} \ \sum_{l=4+n_R}^{n_f} \bm{\mathcal{U}}_{k i} \left(\mathbf{N}_{R}^{0}\right)_{k l} \bm{\mathcal{U}}_{l j}\;\,, \; \mathbf{N}_{R}^{0} = \frac{u_2}{\sqrt{2}} \mathbf{Y}_{R}.
\label{eq:NR}
\end{align}
Note that, in order to obtain the Feynman rules using the interactions between $\nu_i$ and the $Z$-boson or $S^0$ neutral scalars, one must multiply by a factor of 2 since Majorana neutrinos are self-conjugate fields.

\section{Charged-lepton flavour violation}
\label{sec:clfv}

\begin{figure}[t!]
	\centering
	\begin{tabular}{ll}
	\includegraphics[width=0.60\textwidth]{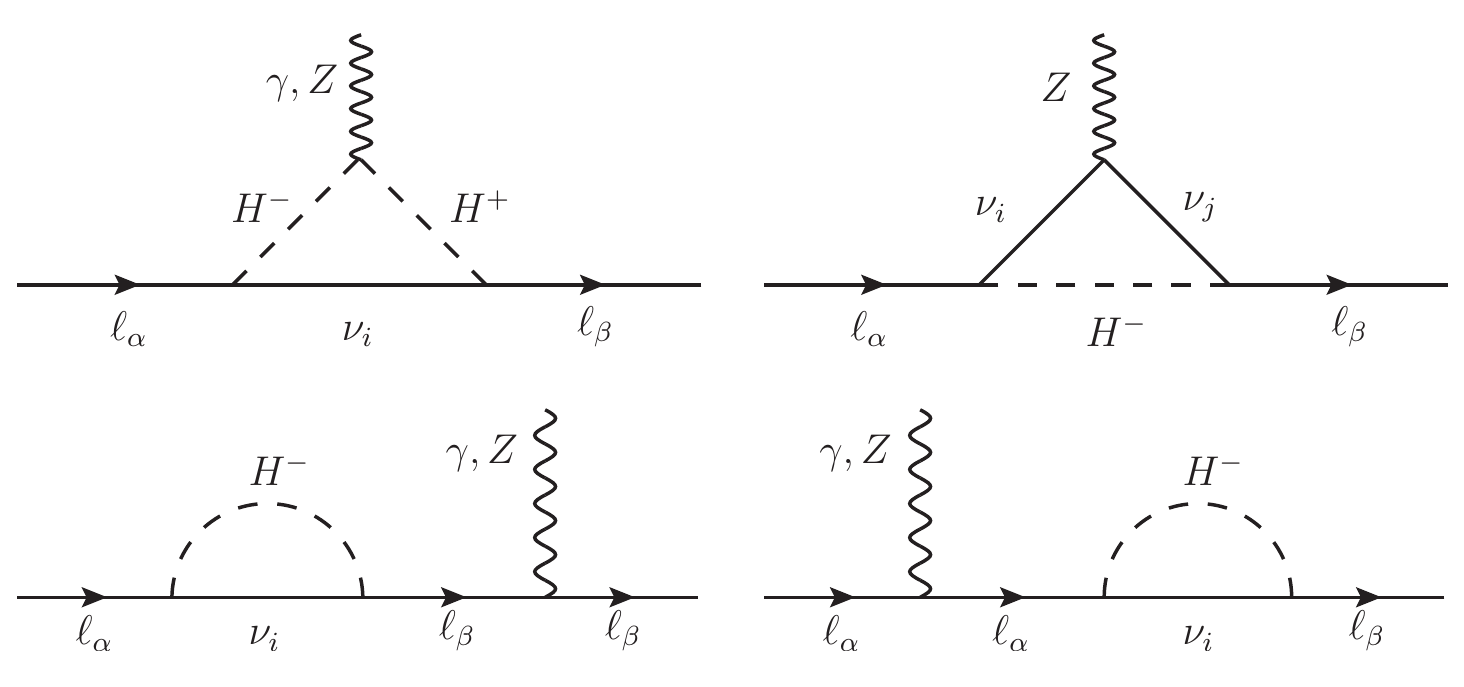}
	&\includegraphics[width=0.295\textwidth]{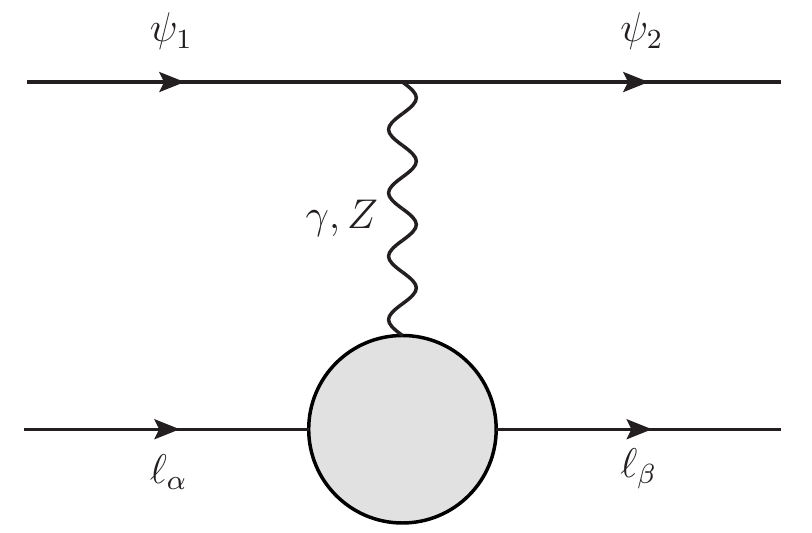}
	\end{tabular}
\caption{$\gamma$ and $Z$-penguin diagrams contributing to radiative decays $\ell_{\alpha} \rightarrow \ell_{\beta} \gamma$ and $Z \rightarrow \ell_{\alpha}^{\pm} \ell_{\beta}^{\mp}$, respectively. Although only the $H^{\pm}$ loop contributions are presented, similar diagrams with $W^{\pm}$/$G^{\pm}$ were also considered in the calculations. On the right, we show the penguin-diagram topology for $\ell_{\alpha}^{-} \rightarrow \ell_{\beta}^{-} \ell_{\gamma}^{+} \ell_{\delta}^{-}$ ($\psi_1 = \psi_2 = \ell_{\gamma}^{+} = \ell_{\delta}^{-}$) and $\mu - e$ conversion in nuclei ($\psi_1=\psi_2 =u,d$).}
\label{fig:photonZ}
\end{figure}

\begin{figure}[t!]
\centering
\includegraphics[width=0.85\textwidth]{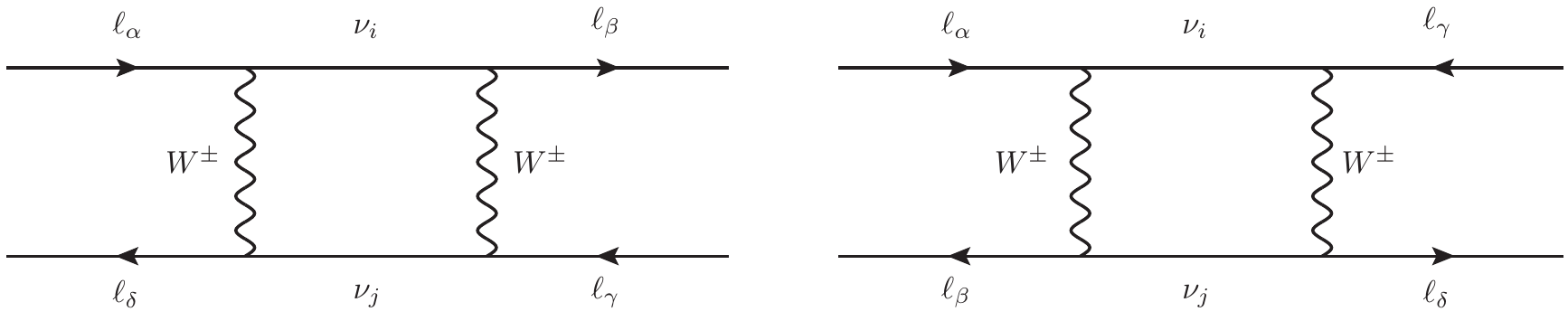}
\caption{Generic box diagrams contributing to $\ell_{\alpha}^{-} \rightarrow \ell_{\beta}^{-} \ell_{\gamma}^{+} \ell_{\delta}^{-}$. Although we only display the $W^{\pm}$ contributions, similar diagrams with $G^{\pm}$ and $H^{\pm}$ were also included in the calculations. We have also taken into account LNV diagrams (on the right) due to the Majorana character of neutrinos, and crossed diagrams corresponding to the flavour index interchange $\beta \leftrightarrow \delta$.}
\label{fig:boxleptonic}
\end{figure}

\begin{figure}[t!]
\centering
\includegraphics[width=0.85\textwidth]{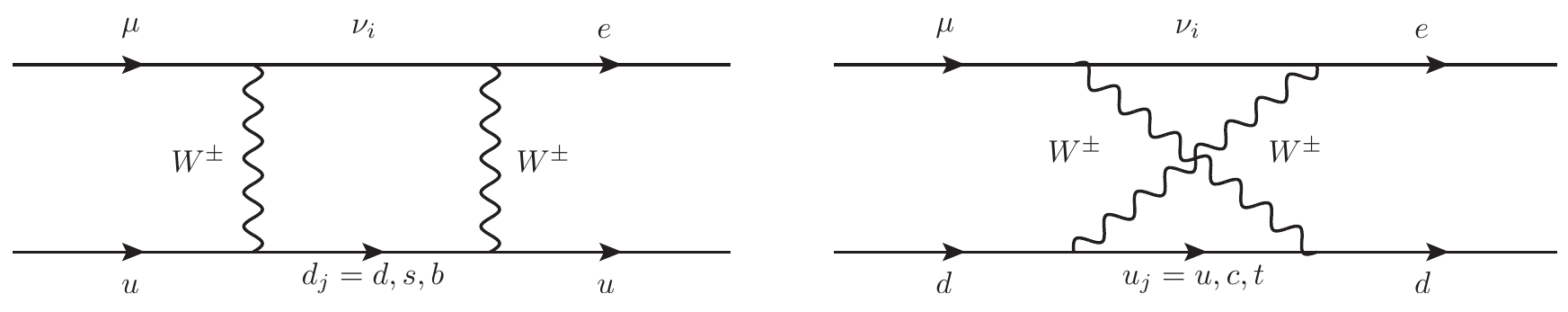}
\caption{Generic $u$-type (on the left) and $d$-type (on the right) semi-leptonic box diagrams contributing to the $\mu - e$ conversion process. Although for illustrative purposes we only display the $W^{\pm}$-boson diagrams, the $G^{\pm}$ and $H^{\pm}$ contributions were also included in our calculations.}
\label{fig:boxsemi}
\end{figure}

In this appendix we present the amplitudes and decay rates for the cLFV processes $\ell_{\alpha} \rightarrow \ell_{\beta} \gamma$, $Z \rightarrow \ell_{\alpha}^{\pm} \ell_{\beta}^{\mp}$, $\ell_{\alpha}^{-} \rightarrow \ell_{\beta}^{-} \ell_{\gamma}^{+} \ell_{\delta}^{-}$ and coherent $\mu - e$ conversion in heavy nuclei. The corresponding current experimental bounds and future sensitivities are given in Table~\ref{tab:boundsCLFV}, while the various topologies of one-loop Feynman diagrams are summarised in Figs. \ref{fig:photonZ}, \ref{fig:boxleptonic} and \ref{fig:boxsemi}. We performed the computations in the Feynman-’t Hooft gauge following the interaction Lagrangians in the mass-eigenstate basis given in Appendix~\ref{sec:Allint}. We provide the results for the amplitudes and branching ratios in terms of the form factors collected in Appendix~\ref{sec:LFFF}.

In Fig.~\ref{fig:photonZ} we show the diagrams contributing to the effective vertex $\overline{\ell_{\beta}} \ell_{\alpha} \gamma$ ($\beta \neq \alpha$) at one-loop level. The transition amplitude can be written in the form
\begin{equation}
\mathcal{A}_{\gamma}^{\alpha \beta}=-\frac{e\, \alpha_W}{8 \pi M_W^2} \varepsilon^{\mu}_{\gamma} \ \overline{\ell_{\beta}} \left[\left(q^2 \gamma_{\mu} - \slashed{q} q_{\mu} \right) \left(F_{\gamma,L}^{\alpha \beta}  P_L + F_{\gamma,R}^{\alpha \beta}  P_R \right)- i \sigma_{\mu \nu} q^{\nu} \left(G_{\gamma,L}^{\alpha \beta} P_L + G_{\gamma,R}^{\alpha \beta} P_R \right)\right] \ell_{\alpha},
\label{eq:photonamp}
\end{equation}
where $\alpha_{W} = g^2/(4\pi)$, $\varepsilon^{\mu}_{\gamma}$ is the photon polarisation four-vector and $q=p_{\alpha}-p_{\beta}$ is the photon momentum. $F_{\gamma}$ is the local monopole contribution for an off-shell photon ($q^2 \neq 0$), while $G_{\gamma}$ stands for the non-local dipole contribution for an on-shell photon ($q^2 = 0$). 
The expressions for $F_{\gamma,L(R)}^{\alpha \beta}$ and $G_{\gamma,L(R)}^{\alpha \beta}$ are given in Section~\ref{sec:LFFFphoton}. 
The former contributes to $\ell_{\alpha}^{-} \rightarrow \ell_{\beta}^{-} \ell_{\gamma}^{+} \ell_{\delta}^{-}$ and $\mu - e$ conversion in nuclei, while the latter encodes the only contribution to $\ell_{\alpha} \rightarrow \ell_{\beta} \gamma$. In particular, for this process one has 
\begin{equation}
\label{eq:Gamma_abg}
\BR(\ell_{\alpha}^{-} \rightarrow \ell_{\beta}^{-} \gamma)= \frac{3 \alpha_{e}}{2 \pi m_{\alpha}^2} \left(\left|G_{\gamma,L}^{\alpha \beta}\right|^2 + \left|G_{\gamma,R}^{\alpha \beta}\right|^2 \right)\BR\left(\ell_{\alpha} \rightarrow \ell_{\beta} \nu_{\alpha} \overline{\nu_{\beta}} \right),
\end{equation}
where $\alpha_{e} = e^2/(4\pi)$ and the values of $\BR\left(\ell_{\alpha} \rightarrow \ell_{\beta} \nu_{\alpha} \overline{\nu_{\beta}}\right)$ are given by~\cite{Zyla:2020zbs},
\begin{equation}
\BR\left(\mu \rightarrow e \nu_{\mu} \overline{\nu_{e}}\right) \simeq 1.0 \; , \; \BR\left(\tau \rightarrow e \nu_{\tau} \overline{\nu_{e}}\right) \simeq 0.18 \; , \; \BR\left(\tau \rightarrow \mu \nu_{\tau} \overline{\nu_{\mu}}\right) \simeq 0.17.
\label{eq:Brmutau}
\end{equation}
In the limit where there is no charged scalar contribution, the branching ratio given in Eq.~\eqref{eq:Gamma_abg} is consistent with the results of Refs.~\cite{Minkowski:1977sc,Inami:1980fz,Beg:1982ex,Cheng:1980tp}.

The one-loop diagrams for the effective vertex $\overline{\ell_{\beta}} \ell_{\alpha} Z$ ($\beta \neq \alpha$) are also shown in Fig. \ref{fig:photonZ}. In this case, the transition amplitude is
\begin{equation}
\mathcal{A}_{Z}^{\alpha \beta}=\frac{g\, \alpha_W}{ 8 \pi c_W} \varepsilon^{\mu}_{Z} \ \overline{\ell_{\beta}} \left( F_{Z,L}^{\alpha \beta} \gamma_{\mu} P_L + F_{Z,R}^{\alpha \beta} \gamma_{\mu} P_R \right) \ell_{\alpha},
\end{equation}
where $\varepsilon^{\mu}_{Z}$ is the $Z$-boson polarisation four-vector. The branching ratio for the LFV decay $Z \rightarrow \ell_{\alpha}^{-} \ell_{\beta}^{+} + \ell_{\alpha}^{+} \ell_{\beta}^{-}$ is
\begin{equation}
\BR(Z \rightarrow \ell_{\alpha}^{-} \ell_{\beta}^{+} + \ell_{\alpha}^{+} \ell_{\beta}^{-}) = \frac{\alpha_W^3}{192 \ \pi^2 c_W^2} \frac{M_Z}{\Gamma_{Z}} \left(\left|F_{Z,L}^{\alpha \beta}\right|^2 + \left|F_{Z,R}^{\alpha \beta}\right|^2 \right),
\end{equation}
where $F_{Z,L}^{\alpha \beta}$ and $F_{Z,R}^{\alpha \beta}$, are given in Section~\ref{sec:LFFFZ} and the $Z$-boson total decay width is $\Gamma_{Z}=2.4952$~GeV~\cite{Zyla:2020zbs}. Note that, in the limit where the scalar content coincides with the SM one, the above branching ratio is consistent with the results of Refs.~\cite{Bernabeu:1993up,Illana:1999ww}.

The $\ell_{\alpha}^{-} \rightarrow \ell_{\beta}^{-} \ell_{\gamma}^{+} \ell_{\delta}^{-}$ amplitude receives contributions from the $\gamma$-penguin, $Z$-penguin and leptonic box diagrams shown in Figs. \ref{fig:photonZ} and \ref{fig:boxleptonic}, which we write as
\begin{align}
\mathcal{A}_{\gamma}^{\alpha \beta \gamma \delta}&= - \frac{\alpha_W^2 s_W^2}{2 M_W^2} \ \Big\{\delta_{\gamma \delta} \ \overline{\ell_{\beta}} \Big[\left(F_{\gamma,L}^{\alpha \beta} \gamma_{\mu} P_L + F_{\gamma,R}^{\alpha \beta} \gamma_{\mu} P_R \right) - i \sigma_{\mu \nu} \frac{q^{\nu}}{q^2} \left(G_{\gamma,L}^{\alpha \beta} P_L + G_{\gamma,R}^{\alpha \beta} P_R \right) \Big] \ell_{\alpha} \ \overline{\ell_{\delta}} \gamma^{\mu} \ell_{\gamma}^c-(\beta \leftrightarrow \delta) \Big\},
\label{eq:TAlepboxg} \\
\mathcal{A}_{Z}^{\alpha \beta \gamma \delta}&=\frac{\alpha_W^2}{2 M_W^2} \left[\delta_{\gamma \delta} \ \overline{\ell_{\beta}} \left( F_{Z,L}^{\alpha \beta} \gamma_{\mu} P_L + F_{Z,R}^{\alpha \beta} \gamma_{\mu} P_R \right) \ell_{\alpha} \ \overline{\ell_{\delta}} \left( g_L^{\ell} \gamma^{\mu} P_L + g_R^{\ell} \gamma^{\mu} P_R \right) \ell_{\gamma}^c - (\beta \leftrightarrow \delta)\right],
\label{eq:TAlepboxZ} \\
\mathcal{A}_{\text{Box}}^{\alpha \beta \gamma \delta}&= - \frac{\alpha_W^2}{4 M_W^2} \ \sum_{X,Y=L,R} \ \sum_{A=S,V,T}  B_{A,X Y}^{\alpha \beta \gamma \delta} \ \overline{\ell_{\beta}} \Lambda_{A}^{X} \ell_{\alpha} \ \overline{\ell_{\delta}} \Lambda_{A}^{Y} \ell_{\gamma}^c\,,
\label{eq:TAlepbox}
\end{align}
where $s_W=\sin \theta_W$, $g_L^{\ell} = s_W^2 - 1/2$, $g_R^{\ell} = s_W^2$ and $\Lambda_{A}^{X,Y}$ are given by the following combinations of Dirac matrices and chiral projectors:
\begin{equation}
\left(\Lambda_{S}^{L},\Lambda_{S}^{R},\Lambda_{V}^{L},\Lambda_{V}^{R},\Lambda_{T}^{L},\Lambda_{T}^{R}\right) \equiv \left(P_L,P_R,\gamma_{\mu}P_L,\gamma_{\mu}P_R,\sigma_{\mu \nu}P_L,\sigma_{\mu \nu}P_R\right),
\label{eq:chiralstruct}
\end{equation}
with $\sigma_{\mu \nu} = i \left[\gamma_{\mu}, \gamma_{\nu} \right] /2$. From the generic leptonic box diagrams presented in Fig.~\ref{fig:boxleptonic} we obtain the transition amplitude~\eqref{eq:TAlepbox} involving the form factor $B_{A,XY}^{\alpha \beta \gamma \delta}$, written in terms of the spinorial structure $\overline{\ell_{\beta}} \Lambda_{A}^{X} \ell_{\alpha} \ \overline{\ell_{\delta}} \Lambda_{A}^{Y} \ell_{\gamma}^c$. Since we are in the presence of Majorana neutrinos, LNV diagrams must be also considered, together with cross-diagrams with interchanged lepton indices $(\beta \leftrightarrow \delta)$. All these contributions can be written in the form \eqref{eq:TAlepbox} by using Fierz transformations~\cite{Nieves:2003in,Nishi:2004st}. In Section~\ref{sec:LFFFlep}, the form factors $B_{A,X Y}^{\alpha \beta \gamma \delta}$ are presented. In particular, $B_{T,L R}^{\alpha \beta \gamma \delta}=B_{T,R L}^{\alpha \beta \gamma \delta}=0$, which can be readily seen from the identity $\sigma_{\mu \nu} \gamma_{5} = - i \varepsilon_{\mu \nu \rho \lambda} \ \sigma^{\rho \lambda}/2$. 

The calculation of the decay rate for each process is done, whenever possible, assuming vanishing masses for the final lepton states, since $m_{\beta,\gamma,\delta} \ll m_{\alpha}$. However, an exception is made for terms involving the photon amplitude, where light lepton masses must be treated with care. As well known, the three-body phase space integrals for the photon contribution are singular in the limit $q^2 \rightarrow 0$. Therefore, one must first perform the phase space integration and only after take the limit $m_{\beta,\gamma,\delta} \rightarrow 0$ for non-divergent terms. This will lead to the logarithmic term $\ln (m_{\alpha}^2 / m_{\beta,\gamma,\delta}^2)$ appearing in the pure photonic dipole contributions. Also, a symmetry factor of $1/2$ has to be taken into account in the case of two identical charged leptons in the final state. The branching ratio for the 3-body LFV decays can be written in the general form:
\begin{align}
\frac{\BR(\ell_{\alpha}^{-} \rightarrow \ell_{\beta}^{-} \ell_{\gamma}^{+} \ell_{\delta}^{-})}{\BR\left(\ell_{\alpha} \rightarrow \ell_{\beta} \nu_{\alpha} \overline{\nu_{\beta}} \right)} &= \frac{k_1 \alpha_W^2}{64 \pi^2} \Big\{ k_2 \left|F_{V,LL}\right|^2 + \left|F_{V,LR}\right|^2 +\frac{1}{4} \Big( \left| F_{S,LL} \right|^2 + \left| F_{S,LR} \right|^2 \Big)+ 12 \left| F_{T,LL} \right|^2 \nonumber \\
&+ k_3 \frac{8 s_W^2}{m_{\alpha}} \text{Re} \left[ \left(k_2 F_{V,LL}+ F_{V,LR}\right) G_{\gamma,R}^{*} \right] + k_3 \frac{32 s_W^4}{m_{\alpha}^2} \left| G_{\gamma,L}\right|^2 \Big(\ln \frac{m_{\alpha}^2}{m_{\delta}^2} - k_4 \Big) + (L \leftrightarrow R) \Big\}\,,
\end{align}
where the values for the branching ratios $\BR\left(\ell_{\alpha} \rightarrow \ell_{\beta} \nu_{\alpha} \overline{\nu_{\beta}} \right)$ are given in Eq.~\eqref{eq:Brmutau}. The $k_i$ coefficients and the form factors $F$ depend on the charged-lepton final states, for which three combinations are possible. Namely,\\

(i) Two different flavours of leptons, where leptons with the same flavour have opposite charge $(\beta \neq \gamma \land \delta = \gamma)$
\begin{equation}
\begin{aligned}
&k_1=k_2=k_3= 1 \;,\; k_4= 3\,,\\
&F_{V,LL} = B_{V,LL} - 2 g_L^{\ell} F_{Z,L} + 2 s_W^2 F_{\gamma,L} \; , \; F_{V,LR} = B_{V,LR} - 2 g_R^{\ell} F_{Z,L} + 2 s_W^2 F_{\gamma,L}\\
&F_{S,LL} = B_{S,LL}\; , \; F_{S,LR} =  B_{S,LR} \; , \; F_{T,LL} = B_{T,LL};
\end{aligned}
\end{equation}

(ii) Three leptons with the same flavour $(\beta = \gamma = \delta)$
\begin{equation}
\begin{aligned}
&k_1= k_3= 1\;,\; k_2= 2\;,\; k_4= 11/4 \,,\\
&F_{V,LL} =  \frac{1}{2} B_{V,LL} - 2 g_L^{\ell} F_{Z,L} + 2 s_W^2 F_{\gamma,L}\; , \; F_{V,LR} = B_{V,LR} - 2 g_R^{\ell} F_{Z,L} + 2 s_W^2 F_{\gamma,L} ,\\
&F_{S,LL} = \frac{1}{\sqrt{2}} B_{S,LL}\; , \; F_{S,LR} = 0\; , \; F_{T,LL} = \frac{1}{\sqrt{2}} B_{T,LL};
\end{aligned}
\end{equation}

(iii) Two distinct flavours, where leptons with same flavour have the same charge $(\delta \neq \gamma \land \beta \neq \gamma)$
\begin{equation}
\begin{aligned}
&k_1= 1/2 \;,\; k_2= 1 \;,\; k_3=k_4= 0 \,,\\
&F_{V,LL} = B_{V,LL}  \; , \; F_{V,LR} = B_{V,LR} \; , \; F_{S,LL} = B_{S,LL}\; , \; F_{S,LR} =  B_{S,LR} \; , \; F_{T,LL} = B_{T,LL}.
\end{aligned}
\end{equation}
Our results agree with those obtained in Ref.~\cite{Ilakovac:2012sh,Crivellin:2013hpa} (see also Refs.~\cite{Hisano:1995cp,Arganda:2005ji,Abada:2014kba,Crivellin:2018mqz}). We have also checked that the BRs match the well-known results in the limit of a SM scalar content~\cite{Ilakovac:1994kj}.

Following Ref.~\cite{Kitano:2002mt}, the operators relevant for coherent $\mu - e$ conversion in nuclei have the general form
\begin{equation}
\mathcal{L}^{\mu- e}_{\text{int.}}=\frac{G_{\text{F}}}{\sqrt{2} \pi} \Bigg[ \frac{e}{8 \pi} G_{\gamma,L}^{\mu e} \ \overline{e} \sigma_{\lambda \rho} P_L \mu \ F^{\lambda \rho} + \alpha_W \sum_{q=u,d,s} \Bigg( F_{S,L}^{\mu e,q} \ \overline{e} P_L \mu \ \overline{q} q+  F_{V,L}^{\mu e,q} \ \overline{e} \gamma^{\rho} P_L \mu \ \overline{q} \gamma_{\rho} q \Bigg) \Bigg]+(L\leftrightarrow R)+\text{H.c.}\,,
\label{eq:CRLag}
\end{equation}
where $G_{\text{F}}$ is the Fermi constant \cite{Zyla:2020zbs}. The form factors receive contributions from $\gamma$-penguin and $Z$-penguin  diagrams as shown in Fig.~\ref{fig:photonZ}, the expression for their amplitudes are similar to the ones in Eqs.~\eqref{eq:TAlepboxg} and~\eqref{eq:TAlepboxZ}, respectively. The transition amplitude for the semi-leptonic box diagrams in Fig.~\ref{fig:boxsemi} is given by,
\begin{align}
\mathcal{A}_{\text{Box}}^{\mu e qq}&= \frac{\alpha_W^2}{4 M_W^2} \ \sum_{X,Y=L,R} \ \sum_{A=S,V,T}  B_{A,X Y}^{\mu e q q} \ \overline{\ell_{\beta}} \Lambda_{A}^{X} \ell_{\alpha} \ \overline{q} \Lambda_{A}^{Y} q\,,
\label{eq:TAlepboxqq}
\end{align}
leading to, 
\begin{equation}
\begin{aligned}
F_{V,X}^{\mu e,q} &=  Q_{q} s_{W}^2 F_{\gamma,X}^{\mu e} + \left(\frac{I^3_{q}}{2} - Q_q s_W^2\right) F_{Z,X}^{\mu e} + \frac{1}{4} \left(B_{V,XX}^{\mu e q q} + B_{V,XY}^{\mu e q q}\right), \\
F_{S,X}^{\mu e,q} &= \frac{1}{4} \left( B_{S,XX}^{\mu e q q} + B_{S,XY}^{\mu e q q} \right), \ \text{for} \ X,Y=L,R \ \text{and} \ X \neq Y\,,
\end{aligned}
\end{equation}
with $I_{u}^3 = - I_{d}^3 = 1/2$, $Q_u = 2/3$, $Q_d = -1/3$ and the semi-leptonic box form factors $B_{A,X Y}^{\alpha \beta q q}$~($\alpha=\mu$ and $\beta=e$) are given in Section~\ref{sec:LFFFsemilep}. From the Lagrangian \eqref{eq:CRLag} we obtain the coherent $\mu - e$ conversion rate
\begin{equation}
\begin{aligned}
\CR(\mu \rightarrow e, \text{N}) &= \frac{G_{\text{F}}^2 \alpha_W^2 m_\mu^5}{8 \pi^2 \ \Gamma_{\text{capt}}(Z)} \ \Bigg[ \bigg| 4 V^{(p)} \left(2 F_{V,L}^{\mu e,u} + F_{V,L}^{\mu e,d} \right) + 4 V^{(n)} \left(  F_{V,L}^{\mu e,u} + 2 F_{V,L}^{\mu e,d} \right) \\
&+ \frac{G_{\gamma,R}^{\mu e}}{m_{\mu}} s_w^2 \left( \frac{D}{2 e} \right) + 4\sum_{q=u,d,s} \left(S^{(p)}  G_{S}^{(q,p)} +  S^{(n)} G_{S}^{(q,n)} \right)F_{S,R}^{\mu e,q} \bigg|^2 + (L \leftrightarrow R) \Bigg],
\end{aligned}
\end{equation}
where the values for the nuclear form factors $D$, $S^{(p)}$, $S^{(n)}$, $V^{(p)}$ and $V^{(n)}$, as well as the muon capture rates $\Gamma_{\text{capt}}(Z)$ are reported in Table~\ref{tab:nuclearinfo} for the nuclei relevant to this work. The scalar-operator coefficients $G_{S}^{(q,n)}$ and $G_{S}^{(q,p)}$ are~\cite{Kosmas:2001mv}
\begin{equation}
G_{S}^{(u,p)} = G_{S}^{(d,n)} = 5.1, \  
G_{S}^{(d,p)} = G_{S}^{(u,n)} = 4.3, \
G_{S}^{(s,p)} = G_{S}^{(s,n)} = 2.5.
\end{equation}
The above expressions are general and thus can be applied to several models in which $\mu - e$ conversion is studied. We have also verified that with only the SM Higgs doublet the above CR coincides with that given in~\cite{Alonso:2012ji}.

\begin{table}[!t]
\renewcommand{\arraystretch}{1.4}
\centering
\setlength{\tabcolsep}{8pt}
\begin{tabular}{ccccccc}
\hline
Nucleus $^A_Z$N & $D m_{\mu}^{-5/2}$ & $S^{(p)} m_{\mu}^{-5/2}$ & $S^{(n)} m_{\mu}^{-5/2}$ & $V^{(p)} m_{\mu}^{-5/2}$ & $V^{(n)} m_{\mu}^{-5/2}$ & $\Gamma_{\text{capt}}$ ($10^6 s^{-1}$) \\ \hline
$^{27}_{13}$Al & 0.0362 & 0.0155 & 0.0167 & 0.0161 & 0.0173 & 0.7054\\
$^{48}_{22}$Ti & 0.0864 & 0.0368 & 0.0435 & 0.0396 & 0.0468 & 2.59 \\
$^{197}_{79}$Au &  0.189 & 0.0614 & 0.0918 & 0.0974 & 0.146 & 13.07 \\
$^{208}_{82}$Pb & 0.161 & 0.0488 & 0.0749 & 0.0834 & 0.128 & 13.45 \\
\hline
\end{tabular}
\caption{Nuclear form factors and muon capture rate for the $\mu - e$ conversion process~\cite{Kitano:2002mt}.}
\label{tab:nuclearinfo}
\end{table}

\section{Form factors and loop functions}
\label{sec:LFFF}

We now provide the form factors and loop functions entering the amplitudes for the cLFV observables analysed in Appendix~\ref{sec:clfv}. All contributions with $W^\pm$ and $H^\pm$ in the loops were included, being the form factors given at leading order in the momenta and masses of the external charged leptons.\footnote{We have checked that in the limit where there are no charged-scalar contributions, i.e., for a scalar sector matching the SM one, our results are in agreement with Refs.~\cite{Ilakovac:1994kj,Alonso:2012ji} (see also~\cite{Abada:2015oba,Abada:2018nio,Hernandez-Tome:2019lkb}).}

\subsection{Photon form factors}
\label{sec:LFFFphoton}

The photon form factors coming from the $\gamma$-penguin diagrams, generically presented in Fig.~\ref{fig:photonZ}, are given by
\begin{align}
F_{\gamma,L(R)}^{\alpha \beta} = F_{\gamma,L(R)}^{\alpha \beta (1)} + F_{\gamma,L(R)}^{\alpha \beta (2)}\quad , \quad
G_{\gamma,L(R)}^{\alpha \beta} = G_{\gamma,L(R)}^{\alpha \beta (1)} + G_{\gamma,L(R)}^{\alpha \beta (2)}\,,
  \label{eq:FG2}
\end{align}
where the superscript (1) and (2) correspond to the contributions stemming from $W^\pm$ and $H^\pm$, respectively. Looking at the general form of the amplitude $\mathcal{A}_{\gamma}^{\alpha \beta}$ in Eq.~\eqref{eq:photonamp}, we notice that due to electromagnetic gauge invariance these photon form factors vanish in the limit of zero external lepton momenta and masses. Therefore, to obtain a non-vanishing result we must expand the loop integrals up to next order in $q^2$. The $W^{\pm}$-boson contribution yields
\begin{align}
&F_{\gamma,L}^{\alpha \beta  (1)} = \sum_{i=1}^{n_f} \mathbf{B}_{\beta i} \mathbf{B}^{*}_{\alpha i} F^{(1)}_{\gamma}(\lambda_i)\;,\;
F_{\gamma,R}^{\alpha \beta  (1)} = 0\,, \\
&G_{\gamma,L}^{\alpha \beta  (1)} = m_{\beta} \sum_{i=1}^{n_f} \mathbf{B}_{\beta i} \mathbf{B}^{*}_{\alpha i} G^{(1)}_{\gamma}(\lambda_i) \;,\;
G_{\gamma,R}^{\alpha \beta  (1)} = m_{\alpha} \sum_{i=1}^{n_f} \mathbf{B}_{\beta i} \mathbf{B}^{*}_{\alpha i} G^{(1)}_{\gamma}(\lambda_i)\;,\; \lambda_i = \dfrac{m_i^2}{M_W^2}\,,
\end{align}
where $\mathbf{B}$ has been defined in Eq.~\eqref{eq:BC}. The $H^{\pm}$ scalar contribution yields
\begin{align}
&F_{\gamma,L}^{\alpha \beta (2)} = \frac{1}{m^2_{H^{\pm}}} \sum_{i=1}^{n_f} \left(\mathbf{B}\mathbf{N}_{\nu}\right)_{\beta i} \left(\mathbf{N}_{\nu}^{\dagger} \mathbf{B}^{\dagger}\right)_{i \alpha}  F^{(2)}_{\gamma}(\omega_i), \quad
F_{\gamma,R}^{\alpha \beta (2)} = \frac{1}{m^2_{H^{\pm}}} \sum_{i=1}^{n_f} \left(\mathbf{N}_{e}^{\dagger} \mathbf{B}\right)_{\beta i} \left(\mathbf{B}^{\dagger} \mathbf{N}_{e}\right)_{i \alpha}  F^{(2)}_{\gamma}(\omega_i), \\
&\begin{aligned}
G_{\gamma,L}^{\alpha \beta (2)} &= - \frac{1}{m^2_{H^{\pm}}} \sum_{i=1}^{n_f} \bigg\{ \left[m_i \left(\mathbf{N}_{e}^{\dagger} \mathbf{B}\right)_{\beta i} \left(\mathbf{N}_{\nu}^{\dagger} \mathbf{B}^{\dagger}\right)_{i \alpha} + m_i \left(\mathbf{B}\mathbf{N}_{\nu}\right)_{\beta i} \left(\mathbf{B}^{\dagger} \mathbf{N}_{e}\right)_{i \alpha} \right] G^{(2)}_{\gamma}(\omega_i) \\
&- \left[ m_{\alpha} \left(\mathbf{N}_{e}^{\dagger} \mathbf{B}\right)_{\beta i} \left(\mathbf{B}^{\dagger} \mathbf{N}_{e}\right)_{i \alpha} + m_{\beta} \left(\mathbf{B}\mathbf{N}_{\nu} \right)_{\beta i} \left(\mathbf{N}_{\nu}^{\dagger} \mathbf{B}^{\dagger}\right)_{i \alpha} \right] G^{(3)}_{\gamma}(\omega_i) \bigg\}, 
\end{aligned} \\
&\begin{aligned}
G_{\gamma,R}^{\alpha \beta (2)} &= - \frac{1}{m^2_{H^{\pm}}} \sum_{i=1}^{n_f} \bigg\{\left[m_i \left(\mathbf{N}_{e}^{\dagger} \mathbf{B}\right)_{\beta i}  \left(\mathbf{N}_{\nu}^{\dagger} \mathbf{B}^{\dagger}\right)_{i \alpha} +m_i \left(\mathbf{B}\mathbf{N}_{\nu}\right)_{\beta i} \left(\mathbf{B}^{\dagger} \mathbf{N}_{e}\right)_{i \alpha} \right] G^{(2)}_{\gamma}(\omega_i) \\
&- \left[ m_{\alpha} \left(\mathbf{B}\mathbf{N}_{\nu}\right)_{\beta i} \left(\mathbf{N}_{\nu}^{\dagger} \mathbf{B}^{\dagger}\right)_{i \alpha} + m_{\beta} \left(\mathbf{N}_{e}^{\dagger} \mathbf{B}\right)_{\beta i} \left(\mathbf{B}^{\dagger} \mathbf{N}_{e}\right)_{i \alpha} \right] G^{(3)}_{\gamma}(\omega_i) \bigg\}\;,\; \omega_i = \dfrac{m_i^2}{m^2_{H^{\pm}}}\,,
\end{aligned}
\end{align}
where scalar-lepton couplings $\mathbf{N}_{e}$ and $\mathbf{N}_{\nu}$ are given in Eqs.~\eqref{eq:Ne}~and~\eqref{eq:Nnu}. The loop functions $F^{(i)}_{\gamma}$ and $G^{(i)}_{\gamma}$ read 
\begin{equation}
\begin{aligned}
&F^{(1)}_{\gamma}(x) =  -\frac{x\left(12+x-7 x^2\right)}{12 (1-x)^3} - \frac{x^2 \left(12 - 10 x + x^2\right)}{6(1-x)^4} \ln x\;,\;F^{(2)}_{\gamma}(x) = \frac{-2 + 7 x - 11 x^2}{36 (1-x)^3} - \frac{x^3 \ln x}{6 (1-x)^4},
 \\
&G^{(1)}_{\gamma}(x) = \frac{x\left(1-5x-2x^2\right)}{4(1-x)^3} - \frac{3 x^3}{2(1-x)^4} \ln x\;,\;G^{(2)}_{\gamma}(x) = \frac{(1 + x)}{2(1-x)^2} + \frac{x \ln x}{(1-x)^3},  \\
&G^{(3)}_{\gamma}(x) = \frac{1 - 5 x - 2 x^2}{12 (1-x)^3} - \frac{x^2 \ln x}{2(1-x)^4}\,.
\end{aligned}
\end{equation}
The charged Higgs form factors and loop functions given above are consistent with the results of Refs.~\cite{He:2002pva,Grimus:2002ux,Lavoura:2003xp}.

\subsection{$Z$-boson form factors}
\label{sec:LFFFZ}

The $Z$-penguin diagrams displayed in Fig.~\ref{fig:photonZ} lead to the form factors:
\begin{equation}
F_{Z,L(R)}^{\alpha \beta} = F_{Z,(R)}^{\alpha \beta (1)} + F_{Z,L(R)}^{\alpha \beta (2)}\,, 
\end{equation}
where the $W^{\pm}$-boson contributions are
\begin{equation}
F_{Z,L}^{\alpha \beta (1)} = \sum_{i,j = 1}^{n_f} \mathbf{B}_{\beta i} \mathbf{B}^{*}_{\alpha j} \left[ \delta_{i j} F^{(1)}_{Z}(\lambda_i) + \bm{\mathcal{C}}_{i j} G^{(1)}_{Z}(\lambda_i, \lambda_j) + \bm{\mathcal{C}}_{i j}^{*} H_{Z}(\lambda_i, \lambda_j) \right]\;, \;
F_{Z,R}^{\alpha \beta (1)} = 0\,,
\end{equation}
with the matrix $\bm{\mathcal{C}}$ defined in Eq.~\eqref{eq:BC}, while the $H^{\pm}$ scalar contribution is
\begin{equation}
\begin{aligned}
F_{Z,L}^{\alpha \beta (2)} &= \frac{1}{M_W^2}\sum_{i,j=1}^{n_f} \left(\mathbf{B}\mathbf{N}_{\nu}\right)_{\beta i} \left(\mathbf{N}_{\nu}^{\dagger} \mathbf{B}^{\dagger}\right)_{j \alpha} \left[\bm{\mathcal{C}}_{i j}  \ G^{(2)}_{Z}(\omega_i,\omega_j) - \bm{\mathcal{C}}_{i j}^{*} G^{(3)}_{Z}(\omega_i,\omega_j)\right],\\
F_{Z,R}^{\alpha \beta (2)} &= \frac{1}{M_W^2} \sum_{i,j=1}^{n_f} \left(\mathbf{N}_{e}^{\dagger} \mathbf{B}\right)_{\beta i} \left(\mathbf{B}^{\dagger} \mathbf{N}_{e}\right)_{j \alpha} \left[ \delta_{i j} F^{(2)}_{Z}(\omega_i) + \bm{\mathcal{C}}_{i j} G^{(3)}_{Z}(\omega_i,\omega_j) - \bm{\mathcal{C}}_{i j}^{*}  G^{(2)}_{Z}(\omega_i,\omega_j)\right].
\end{aligned}
\end{equation}
The loop functions entering the above form factors are given by
\begin{align}
&F^{(1)}_{Z}(x)=-\frac{5 x}{2 (1-x)} - \frac{5 x^2}{2 (1-x)^2} \ln x, \ F^{(2)}_{Z}(x)=-\frac{x}{4(1-x)}-\frac{x^2 \ln x}{4 (1-x)^2} + \frac{1}{8},\nonumber \\
&G^{(1)}_{Z}(x,y) = \frac{- 1}{2(x-y)} \left[\frac{x^2 (1-y)}{(1-x)} \ln x - \frac{y^2 (1-x)}{(1-y)} \ln y \right],\ G^{(2)}_{Z}(x, y) = \frac{- \sqrt{x y}}{2 (x -y)} \left[ \frac{ x \ln x}{(1-x)} - \frac{ y \ln y}{(1-y)} \right],\\
&G^{(3)}_{Z}(x, y) = \frac{1}{4 (x -y)} \left[ \frac{ x^2 \ln x}{(1-x)} - \frac{ y^2 \ln y}{(1-y)} \right] + \frac{3}{8}, \ H_{Z}(x,y) = \frac{\sqrt{x y}}{4(x-y)} \left[\frac{x^2 - 4x}{(1-x)} \ln x - \frac{y^2 - 4y}{(1-y)} \ln y \right].\nonumber
\end{align}
The charged Higgs form factors and loop functions given above are consistent with the results of Ref.~\cite{Grimus:2002ux}.

\subsection{Semi-leptonic box form factors}
\label{sec:LFFFsemilep}

\allowdisplaybreaks[1]

Here we present the form factors and loop functions relevant for the amplitudes \eqref{eq:TAlepboxqq} corresponding to the semi-leptonic $u$-type and $d$-type diagrams presented in Fig.~\ref{fig:boxsemi}. We write \footnote{Although we use general $\alpha$ and $\beta$ indices,  for $\mu - e$ conversion one obviously has $\alpha=\mu$ and $\beta=e$.}
\begin{equation}
 B_{V,LL}^{\alpha \beta q q}  = B_{V,LL}^{\alpha \beta q q (1)} + B_{V,LL}^{\alpha \beta q q (2)},
\end{equation}
where $W^{\pm}$-boson contributes only to $B_{V,LL}^{\alpha \beta q q (1)}$, while the $H^{\pm}$ loops produces all types of form factors in~\eqref{eq:TAlepboxqq}, including $B_{V,LL}^{\alpha \beta q q (2)}$. We now present the various $B_{A,XY}^{\alpha \beta q q}$ for the $u$ and $d$-type diagrams shown in Fig.~\ref{fig:boxsemi}.\\

\noindent {\bf $\bm{u}$-type diagrams:} The only $W^{\pm}$-boson contribution is
\begin{equation}
B_{V,LL}^{\alpha \beta u u (1)} = \sum_{i=1}^{n_f} \ \sum_{d_{j}=d,s,b} \mathbf{V}_{u d_{j}} \mathbf{V}_{u d_{j}}^{*} \mathbf{B}_{\beta i} \mathbf{B}^{*}_{\alpha i} F^{(1)}_{\text{Box}}(\lambda_i,\lambda_{d_{j}})\,,
\label{eq:FBoxu}
\end{equation}
while the diagrams with $H^{\pm}$ lead to~
\footnote{\label{note:Nud} Since for our numerical analysis we have considered diagonal quark mass matrices and $\tan \beta =1$ we will have $\mathbf{N}_{u,d} \equiv 0$ and, therefore, the semi-leptonic box form factors involving $H^{\pm}$ in the loop will not contribute to the rates in that case. }
\begin{equation}
\begin{aligned}
&B_{V,LL}^{\alpha \beta u u (2)} = \frac{1}{ M_W^2 m_{H^{\pm}}^2} \sum_{i=1}^{n_f} \ \sum_{d_{j}=d,s,b} \Bigg\{ \left(\mathbf{B} \mathbf{N_{\nu}}\right)_{\beta i} \left(\mathbf{N^{\dagger}_{\nu}} \mathbf{B}^{\dagger} \right)_{i \alpha} \left(\mathbf{V} \mathbf{N}_{d} \right)_{u d_{j}} \left(\mathbf{N}^{\dagger}_{d} \mathbf{V}^{\dagger}\right)_{d_{j} u} \ F^{(3)}_{\text{Box}}(\omega_i, \omega_{d_{j}}) \\
&\quad\quad\quad\quad+m_{H^{\pm}}^2 \left[ \mathbf{V}_{u d_{j}}^{*} \mathbf{B}^{*}_{\alpha i} \left(\mathbf{B} \mathbf{N_{\nu}}\right)_{\beta i} \left(\mathbf{V} \mathbf{N}_{d}\right)_{u d_{j}} + \mathbf{V}_{u d_{j}} \mathbf{B}_{\beta i} \left(\mathbf{N}^{\dagger}_{\nu} \mathbf{B}^{\dagger} \right)_{i \alpha} \left(\mathbf{N}^{\dagger}_{d} \mathbf{V}^{\dagger} \right)_{d_{j} u} \right] H^{(1)}_{\text{Box}}(\lambda_i, \lambda_{d_{j}}, \lambda_{\pm}) \Bigg\},
\\
&B_{V,LR}^{\alpha \beta u u} = \frac{1}{M_W^2 m_{H^{\pm}}^2} \sum_{i=1}^{n_f} \ \sum_{d_{j}=d,s,b} \left(\mathbf{B} \mathbf{N_{\nu}}\right)_{\beta i} \left(\mathbf{N^{\dagger}_{\nu}} \mathbf{B}^{\dagger} \right)_{i \alpha} \left(\mathbf{N}^{\dagger}_{u} \mathbf{V}\right)_{u d_{j}} \left(\mathbf{V}^{\dagger} \mathbf{N}_{u}\right)_{d_{j} u} \ F^{(3)}_{\text{Box}}(\omega_i, \omega_{d_{j}}),\\
&B_{V,RL}^{\alpha \beta u u} =  \frac{1}{M_W^2 m_{H^{\pm}}^2} \sum_{i=1}^{n_f} \ \sum_{d_{j}=d,s,b}
\left(\mathbf{N}^{\dagger}_{e} \mathbf{B} \right)_{\beta i} \left(\mathbf{B}^{\dagger} \mathbf{N}_{e} \right)_{i \alpha} \left(\mathbf{V} \mathbf{N}_{d}\right)_{u d_{j}} \left(\mathbf{N}^{\dagger}_{d} \mathbf{V}^{\dagger}\right)_{d_{j} u} \ F^{(3)}_{\text{Box}}(\omega_i, \omega_{d_{j}}),\\
&B_{V,RR}^{\alpha \beta u u}=  \frac{1}{M_W^2 m_{H^{\pm}}^2} \sum_{i=1}^{n_f} \ \sum_{d_{j}=d,s,b} \left(\mathbf{N}^{\dagger}_{e} \mathbf{B} \right)_{\beta i} \left(\mathbf{B}^{\dagger} \mathbf{N}_{e} \right)_{i \alpha} \left(\mathbf{N}^{\dagger}_{u} \mathbf{V}\right)_{u d_{j}} \left(\mathbf{V}^{\dagger} \mathbf{N}_{u}\right)_{d_{j} u} \ F^{(3)}_{\text{Box}}(\omega_i, \omega_{d_{j}}),\\
&B_{S,LL}^{\alpha \beta u u} = \sum_{i=1}^{n_f} \sum_{d_{j}=d,s,b} \frac{\left(\mathbf{N}_{e}^{\dagger} \mathbf{B}\right)_{\beta i} \left(\mathbf{N}^{\dagger}_{u} \mathbf{V}\right)_{u d_{j}}}{M_W^2} \Bigg[\mathbf{V}_{u d_{j}}^{*} \mathbf{B}^{*}_{\alpha i} H^{(2)}_{\text{Box}}(\lambda_i, \lambda_{d_{j}}, \lambda_{\pm})+  \left( \mathbf{N}^{\dagger}_{\nu} \mathbf{B}^{\dagger}\right)_{i \alpha}  \left(\mathbf{N}^{\dagger}_{d} \mathbf{V}^{\dagger} \right)_{d_{j} u} \frac{F^{(4)}_{\text{Box}}(\omega_i,\omega_{d_{j}})}{m_{H^{\pm}}^2}\Bigg],\\
&B_{S,LR}^{\alpha \beta u u} = \frac{1}{ M_W^2 m_{H^{\pm}}^2} \sum_{i=1}^{n_f} \ \sum_{d_{j}=d,s,b} \left(\mathbf{N}^{\dagger}_{e} \mathbf{B} \right)_{\beta i} \left(\mathbf{N}^{\dagger}_{\nu} \mathbf{B}^{\dagger} \right)_{i \alpha} \left(\mathbf{V} \mathbf{N}_{d}\right)_{u d_{j}} \left(\mathbf{V}^{\dagger} \mathbf{N}_{u}\right)_{d_{j} u} \ F^{(4)}_{\text{Box}}(\omega_i, \omega_{d_{j}}),\\
&B_{S,RL}^{\alpha \beta u u} = \frac{1}{ M_W^2 m_{H^{\pm}}^2} \sum_{i=1}^{n_f} \ \sum_{d_{j}=d,s,b} \left(\mathbf{B} \mathbf{N_{\nu}}\right)_{\beta i} \left(\mathbf{B}^{\dagger} \mathbf{N}_{e}\right)_{i \alpha} \left(\mathbf{N}^{\dagger}_{u} \mathbf{V}\right)_{u d_{j}} \left(\mathbf{N}^{\dagger}_{d} \mathbf{V}^{\dagger} \right)_{d_{j} u} \ F^{(4)}_{\text{Box}}(\omega_i, \omega_{d_{j}}), \\
&
B_{S,RR}^{\alpha \beta u u} =  \sum_{i=1}^{n_f} \sum_{d_{j}=d,s,b} \frac{\left(\mathbf{B}^{\dagger} \mathbf{N}_{e}\right)_{i \alpha} \left(\mathbf{V}^{\dagger} \mathbf{N}_{u} \right)_{d_{j} u}}{M_W^2} \Bigg[\mathbf{V}_{u d_{j}} \mathbf{B}_{\beta i} H^{(2)}_{\text{Box}}(\lambda_i, \lambda_{d_{j}}, \lambda_{\pm}) + \left(\mathbf{B} \mathbf{N}_{\nu}\right)_{\beta i} \left(\mathbf{V} \mathbf{N}_{d}\right)_{u d_{j}} \frac{F^{(4)}_{\text{Box}}(\omega_i, \omega_{d_{j}})}{m_{H^{\pm}}^2} \Bigg],\\
\end{aligned}
\end{equation}
where $\lambda_{d_{j}} = m_{d_{j}}^2/M_W^2$ and $\omega_{d_{j}} = m_{d_{j}}^2/m_{H^{\pm}}^2$. $\mathbf{N}_{d}$ and $\mathbf{N}_{u}$ are defined in Eqs.~\eqref{eq:Nd}~and~\eqref{eq:Nu}, respectively.\\

\noindent {\bf $\bm{d}$-type diagrams:} The form factor for the $d$-type diagrams coming from the $W^{\pm}$-boson contribution is
\begin{equation}
  B_{V,LL}^{\alpha \beta d d (1)} = \sum_{i=1}^{n_f} \ \sum_{u_{j}=u,c,t} \mathbf{V}_{d u_{j}} \mathbf{V}_{d u_{j}}^{*} \mathbf{B}_{\beta i} \mathbf{B}^{*}_{\alpha i} F^{(2)}_{\text{Box}}(\lambda_i, \lambda_{u_{j}}),
  \label{eq:FBoxd}
\end{equation}
while for the $H^{\pm}$ loops:
\begin{equation}
\begin{aligned}
&\begin{aligned}
B_{V,LL}^{\alpha \beta d d (2)} &= -\frac{1}{M_W^2 m_{H^{\pm}}^2} \sum_{i=1}^{n_f} \sum_{u_{j}=u,c,t}  \Bigg\{\left(\mathbf{B} \mathbf{N_{\nu}}\right)_{\beta i} \left(\mathbf{N^{\dagger}_{\nu}} \mathbf{B}^{\dagger} \right)_{i \alpha}  \left(\mathbf{V}^{\dagger} \mathbf{N}_{u} \right)_{d u_j} \left(\mathbf{N}^{\dagger}_{u} \mathbf{V}\right)_{u_j d} \ F^{(3)}_{\text{Box}}(\omega_i, \omega_{u_{j}}) \\
&+ m_{H^{\pm}}^2 \left[\mathbf{V}_{d u_j} \mathbf{B}^{*}_{\alpha i} \left(\mathbf{B} \mathbf{N_{\nu}}\right)_{\beta i} \left( \mathbf{N}^{\dagger}_{u} \mathbf{V}\right)_{u_j d} + \mathbf{B}_{\beta i} \mathbf{V}_{d u_j}^{*} \left( \mathbf{N}^{\dagger}_{\nu} \mathbf{B}^{\dagger}\right)_{i \alpha} \left(\mathbf{V}^{\dagger} \mathbf{N}_{u}\right)_{d u_j} \right] H^{(1)}_{\text{Box}}(\lambda_i, \lambda_{u_{j}}, \lambda_{\pm})\Bigg\},
\end{aligned} \nonumber \\
&B_{V,LR}^{\alpha \beta d d} = - \frac{1}{M_W^2 m_{H^{\pm}}^2} \sum_{i=1}^{n_f} \sum_{u_{j}=u,c,t}  \left(\mathbf{B} \mathbf{N_{\nu}}\right)_{\beta i} \left(\mathbf{N^{\dagger}_{\nu}} \mathbf{B}^{\dagger} \right)_{i \alpha} \left(\mathbf{N}^{\dagger}_{d} \mathbf{V}^{\dagger}\right)_{d u_j} \left(\mathbf{V} \mathbf{N}_{d} \right)_{u_j d} \ F^{(3)}_{\text{Box}}(\omega_i, \omega_{u_{j}}),\\
&B_{V,RL}^{\alpha \beta d d}=- \frac{1}{M_W^2 m_{H^{\pm}}^2} \sum_{i=1}^{n_f} \ \sum_{u_{j}=u,c,t}  \left( \mathbf{N}^{\dagger}_{e} \mathbf{B}\right)_{\beta i} \left( \mathbf{B}^{\dagger} \mathbf{N}_{e} \right)_{i \alpha} \left(\mathbf{V}^{\dagger} \mathbf{N}_{u}\right)_{d u_j} \left(\mathbf{N}^{\dagger}_{u} \mathbf{V}\right)_{u_j d} \ F^{(3)}_{\text{Box}}(\omega_i, \omega_{u_{j}}),\\
&B_{V,RR}^{\alpha \beta d d} = - \frac{1}{M_W^2 m_{H^{\pm}}^2} \sum_{i=1}^{n_f} \ \sum_{u_{j}=u,c,t}  \left( \mathbf{N}^{\dagger}_{e} \mathbf{B}\right)_{\beta i} \left(\mathbf{B}^{\dagger} \mathbf{N}_{e} \right)_{i \alpha} \left(\mathbf{N}^{\dagger}_{d} \mathbf{V}^{\dagger}\right)_{d u_j} \left(\mathbf{V} \mathbf{N}_{d} \right)_{u_j d} \ F^{(3)}_{\text{Box}}(\omega_i, \omega_{u_{j}}),
\end{aligned}
\end{equation}

\begin{align}
&B_{S,LL}^{\alpha \beta d d} = \frac{1}{M_W^2 m_{H^{\pm}}^2} \sum_{i=1}^{n_f} \sum_{u_{j}=u,c,t}   \left(\mathbf{N}^{\dagger}_{e} \mathbf{B}\right)_{ \beta i} \left(\mathbf{N}^{\dagger}_{\nu} \mathbf{B}^{\dagger}\right)_{i \alpha}
\left(\mathbf{N}^{\dagger}_{d} \mathbf{V}^{\dagger} \right)_{d u_j} \left(\mathbf{N}^{\dagger}_{u} \mathbf{V} \right)_{u_j d} \ F^{(4)}_{\text{Box}}(\omega_i, \omega_{u_{j}}), \nonumber\\
&
B_{S,LR}^{\alpha \beta d d} = \sum_{i=1}^{n_f} \ \sum_{u_{j}=u,c,t}  \frac{\left( \mathbf{N}^{\dagger}_{e} \mathbf{B} \right)_{\beta i} \left(\mathbf{V} \mathbf{N}_{d}\right)_{u_j d}}{ M_W^2} \Bigg[\mathbf{V}_{d u_j} \mathbf{B}^{*}_{\alpha i} H^{(2)}_{\text{Box}}(\lambda_i, \lambda_{u_{j}}, \lambda_{\pm}) + \left(\mathbf{N^{\dagger}_{\nu}} \mathbf{B}^{\dagger} \right)_{i \alpha}  \left(\mathbf{V}^{\dagger} \mathbf{N}_{u} \right)_{d u_j} \frac{F^{(4)}_{\text{Box}}(\omega_i, \omega_{u_{j}})}{m_{H^{\pm}}^2} \Bigg], \nonumber\\
&
B_{S,RL}^{\alpha \beta d d} = \sum_{i=1}^{n_f} \  \sum_{u_{j}=u,c,t}  \frac{\left(\mathbf{B}^{\dagger} \mathbf{N}_{e} \right)_{i \alpha} \left(\mathbf{N}^{\dagger}_{d} \mathbf{V}^{\dagger} \right)_{d u_j}}{M_W^2}  \Bigg[\mathbf{V}_{d u_j}^{*} \mathbf{B}_{\beta i} H^{(2)}_{\text{Box}}(\lambda_i, \lambda_{u_{j}}, \lambda_{\pm})+ \left(\mathbf{B} \mathbf{N}_{\nu}  \right)_{\beta i} \left(\mathbf{N}^{\dagger}_{u} \mathbf{V} \right)_{u_j d} \frac{F^{(4)}_{\text{Box}}(\omega_i, \omega_{u_{j}})}{m_{H^{\pm}}^2} \Bigg],\nonumber\\
&B_{S,RR}^{\alpha \beta d d} = \frac{1}{M_W^2 m_{H^{\pm}}^2} \sum_{i=1}^{n_f} \ \sum_{u_{j}=u,c,t}  \left(\mathbf{B} \mathbf{N_{\nu}}\right)_{\beta i} \left(\mathbf{B}^{\dagger} \mathbf{N}_{e}\right)_{i \alpha} 
\left(\mathbf{V}^{\dagger} \mathbf{N}_{u}\right)_{d u_j} \left(\mathbf{V} \mathbf{N}_{d} \right)_{u_j d} F^{(4)}_{\text{Box}}(\omega_i, \omega_{u_{j}})\,,
\end{align}
with $\lambda_{u_{j}} = m_{u_{j}}^2/M_W^2$ and $\omega_{u_{j}} = m_{u_{j}}^2/m_{H^{\pm}}^2$. The loop functions entering the semi-leptonic box form factors are 
\begin{equation}
\begin{aligned}
&\begin{aligned}
   F^{(1)}_{\text{Box}}(x, y) &= \frac{1}{x-y} \left\{ \left(4 + \frac{x y}{4} \right) \left[ \frac{1}{1-x} + \frac{x^2}{(1-x)^2} \ln x -  \frac{1}{1-y} - \frac{y^2}{(1-y)^2} \ln y \right] \right.\\
   &\left. -2 x y \left[ \frac{1}{1-x} + \frac{x}{(1-x)^2} \ln x -  \frac{1}{1-y} - \frac{y}{(1-y)^2} \ln y \right] \right\},
\end{aligned} \\
&\begin{aligned}
   F^{(2)}_{\text{Box}}(x, y) &= - \frac{1}{x-y} \left\{ \left(1 + \frac{x y}{4} \right) \left[ \frac{1}{1-x} + \frac{x^2}{(1-x)^2} \ln x -  \frac{1}{1-y} - \frac{y^2}{(1-y)^2} \ln y \right] \right. \\
   &\left. -2 x y \left[ \frac{1}{1-x} + \frac{x}{(1-x)^2} \ln x -  \frac{1}{1-y} - \frac{y}{(1-y)^2} \ln y \right] \right\},
\end{aligned} \\
&F^{(3)}_{\text{Box}}(x, y) = \frac{1}{4 (x- y)} \left[\frac{1}{1-x} + \frac{x^2}{(1-x)^2} \ln x -  \frac{1}{1-y} - \frac{y^2}{(1-y)^2} \ln y \right], \\
&F^{(4)}_{\text{Box}}(x, y) = -\frac{\sqrt{x y}}{ (x- y)} \left[ \frac{1}{1-x} + \frac{x}{(1-x)^2} \ln x -  \frac{1}{1-y} - \frac{y}{(1-y)^2} \ln y \right], 
\end{aligned}
\end{equation}
\begin{equation}
\begin{aligned}
&H^{(1)}_{\text{Box}}(x, y, z) = -\frac{\sqrt{x y}}{4}\left[ \frac{x (x-4) \ln x}{(1-x)(x - y) (x - z)} + \frac{y (y-4) \ln y}{(1-y)(y - x) (y - z)} + \frac{z (z-4) \ln z}{(1-z)(z - x) (z - y)}\right], \\
&H^{(2)}_{\text{Box}}(x, y, z) = -\left[\frac{x^2 (1-y)\ln x}{(1-x)(x - y) (x - z)} + \frac{y^2 (1-x) \ln y}{(1-y)(y - x) (y - z)} + \frac{z (z-x y) \ln z}{(1-z)(z - x) (z - y)} \right], \\
&H^{(3)}_{\text{Box}}(x, y, z) = -\frac{1}{4}\left[ \frac{x^2 \ln x}{(1-x)(x - y) (x - z)} + \frac{y^2 \ln y}{(1-y)(y - x) (y - z)} + \frac{z^2 \ln z}{(1-z)(z - x) (z - y)}\right].
\end{aligned}
\end{equation}

\subsection{Leptonic box form factors}
\label{sec:LFFFlep}

Finally, we present the form factors in $\mathcal{A}_{\text{Box}}^{\alpha \beta \gamma \delta}$ of Eq.~\eqref{eq:TAlepbox}, corresponding to the leptonic one-loop box diagrams generically presented in Fig.~\ref{fig:boxleptonic}. As explained in Appendix~\ref{sec:clfv}, we consider all contributions, including those stemming from LNV and cross diagrams. The form factor $B_{V,LL}^{\alpha \beta \gamma \delta}$ has two types of contributions, coming from the $W^{\pm}$ boson and $H^{\pm}$ scalar. The remaining form factors are due to box diagrams with $H^{\pm}$. We write
\begin{equation}
 B_{V,LL}^{\alpha \beta \gamma \delta}  = B_{V,LL}^{\alpha \beta  \gamma \delta(1)} + B_{V,LL}^{\alpha \beta \gamma \delta(2)}.
\end{equation}
where the $W^{\pm}$ contribution is
\begin{equation} 
  B_{V,LL}^{\alpha \beta \gamma \delta (1)} = \sum_{i,j=1}^{n_f} \left[ \mathbf{B}^{*}_{\alpha i} \mathbf{B}_{\gamma i}^{*} \mathbf{B}_{\beta j} \mathbf{B}_{\delta j} G_{\text{Box}}(\lambda_i,\lambda_j) - \ \mathbf{B}^{*}_{\alpha i} \mathbf{B}_{\gamma j}^{*} \left(\mathbf{B}_{\beta i} \mathbf{B}_{\delta j} + \mathbf{B}_{\beta j} \mathbf{B}_{\delta i}\right) F^{(2)}_{\text{Box}}(\lambda_i, \lambda_j) \right]\,.
  \label{eq:FBoxl}
\end{equation}
For the box form factors with $H^{\pm}$ we have:
\begin{equation}
\begin{aligned}
&\begin{aligned}
B_{V,LL}^{\alpha \beta \gamma \delta (2)} &= \frac{1}{M_W^2} \sum_{i,j=1}^{n_f} \Bigg\{\frac{1}{m_{H^{\pm}}^2} \left(\mathbf{N}_{\nu}^{\dagger} \mathbf{B}^{\dagger}\right)_{i \alpha}  \left(\mathbf{B} \mathbf{N}_{\nu} \right)_{\beta j} \left(\mathbf{N}_{\nu}^{\dagger} \mathbf{B}^{\dagger}\right)_{i \gamma} \left(\mathbf{B} \mathbf{N}_{\nu}\right)_{\delta j} F^{(4)}_{\text{Box}}(\omega_i, \omega_j)  \\
&+\frac{1}{m_{H^{\pm}}^2} \left(\mathbf{N}_{\nu}^{\dagger} \mathbf{B}^{\dagger}\right)_{i \alpha} \left(\mathbf{N}_{\nu}^{\dagger} \mathbf{B}^{\dagger}\right)_{j \gamma} \left[\left(\mathbf{B} \mathbf{N}_{\nu} \right)_{\beta j}  \left(\mathbf{B} \mathbf{N}_{\nu}\right)_{\delta i} + (\beta \leftrightarrow \delta) \right] F^{(3)}_{\text{Box}}(\omega_i,\omega_j)\\
&+\left[\mathbf{B}^{*}_{\alpha i} \left(\mathbf{N}_{\nu}^{\dagger} \mathbf{B}^{\dagger}\right)_{i \gamma} + (\alpha \leftrightarrow \gamma) \right] \Big[ \mathbf{B}_{\beta j} \left(\mathbf{B} \mathbf{N}_{\nu}\right)_{\delta j} + (\beta \leftrightarrow \delta) \Big] \frac{J^{(2)}_{\text{Box}}(\lambda_i,\lambda_j,\lambda_{\pm})}{2} \\
& + \left[\mathbf{B}^{*}_{\alpha i} \left(\mathbf{N}_{\nu}^{\dagger} \mathbf{B}^{\dagger}\right)_{j \gamma}+ (\alpha \leftrightarrow \gamma) \right] \left[ \mathbf{B}_{\beta i} \left(\mathbf{B} \mathbf{N}_{\nu}\right)_{\delta j} +  (\beta \leftrightarrow \delta) \right] H^{(1)}_{\text{Box}}(\lambda_i,\lambda_j,\lambda_{\pm}) \Bigg\},
\end{aligned} \\
&\begin{aligned}
B_{V,LR}^{\alpha \beta \gamma \delta} &=- \frac{1}{M_W^2} \sum_{i,j=1}^{n_f} \Bigg\{\mathbf{B}_{\beta j} \mathbf{B}^{*}_{\alpha i} \Bigg[\left(\mathbf{N}_{e}^{\dagger} \mathbf{B}\right)_{\delta j} \left(\mathbf{B}^{\dagger} \mathbf{N}_{e}\right)_{i \gamma} J^{(1)}_{\text{Box}}(\lambda_i,\lambda_j,\lambda_{\pm})-\left(\mathbf{B}^{\dagger} \mathbf{N}_{e}\right)_{j \gamma} \left(\mathbf{N}_{e}^{\dagger} \mathbf{B}\right)_{\delta i} \frac{H^{(2)}_{\text{Box}}(\lambda_i,\lambda_j,\lambda_{\pm})}{2} \Bigg] \\
&- \frac{1}{m_{H^{\pm}}^2} \left(\mathbf{N}_{\nu}^{\dagger} \mathbf{B}^{\dagger}\right)_{i \alpha} \left(\mathbf{B}^{\dagger} \mathbf{N}_{e}\right)_{j \gamma} \Bigg[ \left(\mathbf{B} \mathbf{N}_{\nu}\right)_{\beta i} \left(\mathbf{N}_{e}^{\dagger} \mathbf{B}\right)_{\delta j} F^{(3)}_{\text{Box}}(\omega_i, \omega_j)+ \left(\mathbf{B} \mathbf{N}_{\nu}\right)_{\beta j} \left(\mathbf{N}_{e}^{\dagger} \mathbf{B}\right)_{\delta i} \frac{F^{(4)}_{\text{Box}}(\omega_i, \omega_j)}{2} \Bigg] \Bigg\},
\end{aligned}\\
&\begin{aligned}
B_{V,RL}^{\alpha \beta \gamma \delta} &= - \frac{1}{M_W^2} \sum_{i,j=1}^{n_f} \Bigg\{ \left(\mathbf{B}^{\dagger} \mathbf{N}_{e}\right)_{i \alpha} \left(\mathbf{N}_{e}^{\dagger} \mathbf{B}\right)_{\beta j} \Bigg[ \mathbf{B}_{\delta j} \mathbf{B}^{*}_{\gamma i} J^{(1)}_{\text{Box}}(\lambda_i,\lambda_j,\lambda_{\pm})- \mathbf{B}_{\delta i} \mathbf{B}^{*}_{\gamma j} \frac{H^{(2)}_{\text{Box}}(\lambda_i,\lambda_j,\lambda_{\pm})}{2} \Bigg]\\
&- \frac{1}{m_{H^{\pm}}^2} \left(\mathbf{B}^{\dagger} \mathbf{N}_{e}\right)_{i \alpha} \left(\mathbf{N}_{\nu}^{\dagger} \mathbf{B}^{\dagger}\right)_{j \gamma} \Bigg[ \left(\mathbf{N}_{e}^{\dagger} \mathbf{B}\right)_{\beta i} \left(\mathbf{B} \mathbf{N}_{\nu}\right)_{\delta j} F^{(3)}_{\text{Box}}(\omega_i, \omega_j)+ \left(\mathbf{N}_{e}^{\dagger} \mathbf{B}\right)_{\beta j} \left(\mathbf{B} \mathbf{N}_{\nu}\right)_{\delta i} \frac{F^{(4)}_{\text{Box}}(\omega_i, \omega_j)}{2} \Bigg] \Bigg\},
\end{aligned}\\
&\begin{aligned}
B_{V,RR}^{\alpha \beta \gamma \delta} &= \frac{1}{M_W^2 m_{H^{\pm}}^2} \sum_{i,j=1}^{n_f} \Bigg\{\left(\mathbf{B}^{\dagger} \mathbf{N}_{e}\right)_{i \alpha} \left(\mathbf{N}_{e}^{\dagger} \mathbf{B}\right)_{\beta j} \left(\mathbf{B}^{\dagger} \mathbf{N}_{e}\right)_{i \gamma} \left(\mathbf{N}_{e}^{\dagger} \mathbf{B}\right)_{\delta j} F^{(4)}_{\text{Box}}(\omega_i, \omega_j) \\
&+ \left(\mathbf{B}^{\dagger} \mathbf{N}_{e}\right)_{i \alpha} \left(\mathbf{B}^{\dagger} \mathbf{N}_{e}\right)_{j \gamma} \left[ \left(\mathbf{N}_{e}^{\dagger} \mathbf{B}\right)_{\beta j} \left(\mathbf{N}_{e}^{\dagger} \mathbf{B}\right)_{\delta i} + \left(\mathbf{N}_{e}^{\dagger} \mathbf{B}\right)_{\beta i} \left(\mathbf{N}_{e}^{\dagger} \mathbf{B}\right)_{\delta j} \right] F^{(3)}_{\text{Box}}(\omega_i, \omega_j) \Bigg\},
\end{aligned}\\
&\begin{aligned}
B_{S,LL}^{\alpha \beta \gamma \delta} &= -\frac{1}{M_W^2 m_{H^{\pm}}^2} \sum_{i,j=1}^{n_f} \Bigg\{ \Bigg[ \left(\mathbf{N}_{\nu}^{\dagger} \mathbf{B}^{\dagger}\right)_{i \alpha} \left(\mathbf{N}_{\nu}^{\dagger} \mathbf{B}^{\dagger}\right)_{j\gamma} \left(\left(\mathbf{N}_{e}^{\dagger} \mathbf{B}\right)_{\beta i} \left(\mathbf{N}_{e}^{\dagger} \mathbf{B}\right)_{\delta j} - \frac{1}{2} \left(\mathbf{N}_{e}^{\dagger} \mathbf{B}\right)_{\beta j} \left(\mathbf{N}_{e}^{\dagger} \mathbf{B}\right)_{\delta i} \right) \\
&+ \left(\mathbf{N}_{\nu}^{\dagger} \mathbf{B}^{\dagger}\right)_{i \alpha}  \left(\mathbf{N}_{e}^{\dagger} \mathbf{B}\right)_{\beta j} \left(\mathbf{N}_{\nu}^{\dagger} \mathbf{B}^{\dagger}\right)_{i \gamma} \left(\mathbf{N}_{e}^{\dagger} \mathbf{B}\right)_{\delta j} \Bigg] F^{(4)}_{\text{Box}}(\omega_i, \omega_j) \Bigg\},
\end{aligned} \\
&\begin{aligned}
B_{S,LR}^{\alpha \beta \gamma \delta} &= \frac{1}{M_W^2} \sum_{i,j=1}^{n_f} \Bigg\{\mathbf{B}_{\delta j} \mathbf{B}^{*}_{\alpha i} \Bigg[  \left(\mathbf{N}_{e}^{\dagger} \mathbf{B}\right)_{\beta j} \left(\mathbf{B}^{\dagger} \mathbf{N}_{e}\right)_{i \gamma}  2 J^{(1)}_{\text{Box}}(\lambda_i,\lambda_j,\lambda_{\pm})- \left(\mathbf{B}^{\dagger} \mathbf{N}_{e}\right)_{j \gamma} \left(\mathbf{N}_{e}^{\dagger} \mathbf{B}\right)_{\beta i} H^{(2)}_{\text{Box}}(\lambda_i,\lambda_j,\lambda_{\pm}) \Bigg]\\
&- \frac{1}{m_{H^{\pm}}^2} \left(\mathbf{N}_{\nu}^{\dagger} \mathbf{B}^{\dagger}\right)_{i \alpha} \left(\mathbf{B}^{\dagger} \mathbf{N}_{e}\right)_{j \gamma} \Bigg[ \left(\mathbf{B} \mathbf{N}_{\nu}\right)_{\delta i} \left(\mathbf{N}_{e}^{\dagger} \mathbf{B}\right)_{\beta j} 2 F^{(3)}_{\text{Box}}(\omega_i, \omega_j)+ \left(\mathbf{B} \mathbf{N}_{\nu}\right)_{\delta j} \left(\mathbf{N}_{e}^{\dagger} \mathbf{B}\right)_{\beta i} F^{(4)}_{\text{Box}}(\omega_i, \omega_j) \Bigg] \Bigg\},
\end{aligned}\\
&\begin{aligned}
B_{S,RL}^{\alpha \beta \gamma \delta} &= \frac{1}{M_W^2} \sum_{i,j=1}^{n_f} \Bigg\{ \left(\mathbf{B}^{\dagger} \mathbf{N}_{e}\right)_{i \alpha} \left(\mathbf{N}_{e}^{\dagger} \mathbf{B}\right)_{\delta j} \Bigg[\mathbf{B}_{\beta j} \mathbf{B}^{*}_{\gamma i} 2 J^{(1)}_{\text{Box}}(\lambda_i,\lambda_j,\lambda_{\pm})- \mathbf{B}_{\beta i} \mathbf{B}^{*}_{\gamma j} H^{(2)}_{\text{Box}}(\lambda_i,\lambda_j,\lambda_{\pm})\Bigg]\\
&- \frac{1}{m_{H^{\pm}}^2} \left(\mathbf{N}_{\nu}^{\dagger} \mathbf{B}^{\dagger}\right)_{j \gamma} \left(\mathbf{B}^{\dagger} \mathbf{N}_{e}\right)_{i \alpha} \Bigg[ \left(\mathbf{B} \mathbf{N}_{\nu}\right)_{\beta j} \left(\mathbf{N}_{e}^{\dagger} \mathbf{B}\right)_{\delta i} 2 F^{(3)}_{\text{Box}}(\omega_i, \omega_j)+\left(\mathbf{B} \mathbf{N}_{\nu}\right)_{\beta i} \left(\mathbf{N}_{e}^{\dagger} \mathbf{B}\right)_{\delta j} F^{(4)}_{\text{Box}}(\omega_i, \omega_j) \Bigg] \Bigg\},
\end{aligned}\\
&\begin{aligned}
B_{S,RR}^{\alpha \beta \gamma \delta} &= -\frac{1}{M_W^2 m_{H^{\pm}}^2} \sum_{i,j=1}^{n_f} \Bigg\{\Bigg[ \left(\mathbf{B}^{\dagger} \mathbf{N}_{e}\right)_{i \alpha} \left(\mathbf{B}^{\dagger} \mathbf{N}_{e}\right)_{j \gamma} \left( \left(\mathbf{B} \mathbf{N_{\nu}}\right)_{\beta i} \left(\mathbf{B} \mathbf{N_{\nu}}\right)_{\delta j} - \frac{1}{2} \left(\mathbf{B} \mathbf{N_{\nu}}\right)_{\beta j} \left(\mathbf{B} \mathbf{N_{\nu}}\right)_{\delta i} \right) \\
&+ \left(\mathbf{B}^{\dagger} \mathbf{N}_{e}\right)_{i \alpha} \left(\mathbf{B} \mathbf{N_{\nu}}\right)_{\beta j} \left(\mathbf{B}^{\dagger} \mathbf{N}_{e}\right)_{i \gamma} \left(\mathbf{B} \mathbf{N_{\nu}}\right)_{\delta j} \Bigg] F^{(4)}_{\text{Box}}(\omega_i, \omega_j) \Bigg\},
\end{aligned}\\
&\begin{aligned}
B_{T,LL}^{\alpha \beta \gamma \delta} = \frac{1}{M_W^2 m_{H^{\pm}}^2} \sum_{i,j=1}^{n_f} \left\{\left(\mathbf{N}_{\nu}^{\dagger} \mathbf{B}^{\dagger}\right)_{i \alpha}  \left(\mathbf{N}_{e}^{\dagger} \mathbf{B}\right)_{\beta j} \left[\left( \mathbf{N}_{\nu}^{\dagger} \mathbf{B}^{\dagger}\right)_{i \gamma} \left(\mathbf{N}_{e}^{\dagger} \mathbf{B}\right)_{\delta j}  + \frac{1}{2}  \left( \mathbf{N}_{\nu}^{\dagger} \mathbf{B}^{\dagger}\right)_{j \gamma} \left(\mathbf{N}_{e}^{\dagger} \mathbf{B}\right)_{\delta i} \right] \ \frac{F^{(4)}_{\text{Box}}(\omega_i, \omega_j)}{4} \right\},
\end{aligned}\\
&\begin{aligned}
B_{T,RR}^{\alpha \beta \gamma \delta} = \frac{1}{M_W^2 m_{H^{\pm}}^2} \sum_{i,j=1}^{n_f} \left\{ \left(\mathbf{B}^{\dagger} \mathbf{N}_{e}\right)_{i \alpha} \left(\mathbf{B} \mathbf{N_{\nu}}\right)_{\beta j}  \left[ \left(\mathbf{B}^{\dagger} \mathbf{N}_{e}\right)_{i \gamma} \left(\mathbf{B} \mathbf{N_{\nu}}\right)_{\delta j} + \frac{1}{2} \left(\mathbf{B}^{\dagger} \mathbf{N}_{e}\right)_{j \gamma} \left(\mathbf{B} \mathbf{N_{\nu}}\right)_{\delta i} \right]\ \frac{F^{(4)}_{\text{Box}}(\omega_i, \omega_j)}{4} \right\},
\end{aligned}
\end{aligned}
\end{equation}
where  $\mathbf{N}_{e}$ and $\mathbf{N}_{\nu}$ have been defined in Eqs.~\eqref{eq:Ne}~and~\eqref{eq:Nnu}, respectively. The loop functions relevant for the leptonic box form factors are
\begin{align}
&\begin{aligned}
G_{\text{Box}}(x, y) &= - \frac{\sqrt{x y}}{x-y} \left\{ \left(4 + x y \right) \left[ \frac{1}{1-x} + \frac{x}{(1-x)^2} \ln x -  \frac{1}{1-y} - \frac{y}{(1-y)^2} \ln y \right] \right.\\
&\left. -2 \left[ \frac{1}{1-x} + \frac{x^2}{(1-x)^2} \ln x -  \frac{1}{1-y} - \frac{y^2}{(1-y)^2} \ln y \right] \right\},
\end{aligned} \\
&J^{(1)}_{\text{Box}}(x, y, z) = - \frac{\sqrt{x y}}{4} \left[\frac{x (x+4) \ln x}{(1-x)(x - y) (x - z)} + \frac{y (y+4) \ln y}{(1-y)(y - x) (y - z)} + \frac{z (z+4) \ln z}{(1-z)(z - x) (z - y)}\right], \\
&J^{(2)}_{\text{Box}}(x, y, z) = -\left[\frac{x^2 (1+y)\ln x}{(1-x)(x - y) (x - z)} + \frac{y^2 (1+x) \ln y}{(1-y)(y - x) (y - z)} + \frac{z (z+x y) \ln z}{(1-z)(z - x) (z - y)}\right].
\end{align}

The contributions to the box form factors stemming from $H^{\pm}$ in the loop are consistent with the derivation performed in Ref.~\cite{Grimus:2002ux}. We refer the reader to Refs.~\cite{Ilakovac:2012sh,Abada:2014kba} where charged-Higgs contributions were considered in the context of low-scale seesaw SUSY. This enabled us to check the types of loop functions and of matrix and chiral structures that enter the resulting form factors. However, to the best of our knowledge, the box form factors and loop functions presented in a compact form in this work within the framework of a seesaw type model and 2HDM scalar sector have not been presented elsewhere.

%
%*******************************************************************************************************
%*******************************************************************************************************

\end{document}